\documentstyle[prl,eqsecnum,aps]{revtex}

\begin{document}
\author{Jian Qi Shen $^{1,}$$^{2}$ \footnote{E-mail address: jqshen@coer.zju.edu.cn}}
\address{$^{1}$  Centre for Optical
and Electromagnetic Research, State Key Laboratory of Modern
Optical Instrumentation, \\Zhejiang University,
Hangzhou Yuquan 310027, P.R. China\\
$^{2}$ Zhejiang Institute of Modern Physics and Department of
Physics, \\Zhejiang University, Hangzhou 310027, P.R. China}
\date{\today }
\title{Introduction to the theory of left-handed media}
\maketitle

\begin{abstract}
This paper is devoted to investigating the physically interesting
optical and electromagnetic properties, phenomena and effects of
wave propagation in the negative refractive index materials, which
is often referred to as the {\it left-handed media} in the
literature. This paper covers a wide range of subjects and related
topics of left-handed media such as many mathematical treatment of
fundamental effects ({\it e.g.}, the reflection and the refraction
laws on the interface between LH and RH media, the group velocity
and energy density in dispersive materials, the negative optical
refractive index resulting from a moving regular medium, the
reversal of Doppler effect in left-handed media, the reversal of
Cerenkov radiation in left-handed media, the optical refractive
index of massive particles and physical meanings of left-handed
media, the anti-shielding effect and negative temperature in
left-handed media, {\it etc.}), and their some applications to
certain areas ({\it e.g.}, three kinds of compact thin
subwavelength cavity resonators (rectangular, cylindrical,
spherical) made of left-handed media, the photon geometric phases
due to helicity inversions inside a periodical fiber made of
left-handed media, {\it etc.}).
\end{abstract}

\section{Introduction}
During the past 50 years, the design and fabrication of artificial
materials attracted extensive attention in various scientific and
technological areas\cite{Ziolkowski}. For example, in the 1950s
and 1960s artificial dielectrics were exploited for light-weight
microwave antenna lenses, and in the 1980s and 1990s artificial
chiral materials were investigated for microwave radar absorber
applications\cite{Jaggard}. During the last decades, a kind of
material termed photonic crystals, which is patterned with a
periodicity in dielectric constant and can therefore create a
range of forbidden frequencies called a photonic band gap focus
considerable attention of many researchers\cite{Yablonovitch}.
Such dielectric structure of crystals offers the possibility of
molding the flow of light inside media. Recently a number of
interest was captured by the artificial electric and magnetic
molecules\cite{Ziolkowski1997} and artificial electric and
magnetic materials, which, like many of the chiral materials, can
exhibit positive or negative permittivity or permeability
properties. More recently, a kind of artificial composite
metamaterials (the so-called {\it left-handed media}) having a
frequency band where the effective permittivity and the effective
permeability are simultaneously negative receives particular and
intensive attention of many authors both experimentally and
theoretically\cite{Smith,Klimov,Shelby,Ziolkowski2,Kong,Garcia,Lindell,Pendryprl,Zhang,Jianqi,Shen}.

In the above, we present the briefly history of artificial
materials, which was stated based on the point of view of pure
technology. It is shown in what background (or environment) such a
material with simultaneous negative permittivity and permeability
arises. However, in what follows we will discuss this problem from
the purely theoretical point of view.

In 1930s, Dirac predicted a new particle (positive electron) based
on his relativistic wave equation. In Dirac's theory, the energy
eigenvalue of Fermions satisfies the equation
$E^{2}=p^{2}c^{2}+m_{0}^{2}c^{4}$ with $p$ and $m_{0}$ denoting
the momentum and rest mass of the Dirac particle. It follows that
the energy eigenvalue $E=\pm\sqrt{p^{2}c^{2}+m_{0}^{2}c^{4}}$.
According to the convention of classical physics, the negative
energy solution may be thought of as a one that has no physical
meanings and therefore can be abandoned without any hesitation.
But in quantum mechanics, the completeness of solutions is a very
essential subject in solving wave equation. If we desert the
negative energy solution, we cannot guarantee the completeness
property of solutions of wave equation. So, the negative energy
solution should also be of physical meanings and deserves
consideration in quantum mechanics. Thus, the positive electron
was predicted by Dirac. Likewise, in Maxwell electrodynamics, we
may also experience such things. It is well known that in the
second-order Maxwellian equation, the relation between the squared
of optical refractive index $n$ and the permittivity $\epsilon$
and permeability $\mu$ is $n^{2}=\epsilon\mu$. If both $\epsilon$
and $\mu$ are positive, it follows that one can arrive at the
following case of $n$:
\begin{equation}
n=+\sqrt{\epsilon\mu},  \quad   n=-\sqrt{\epsilon\mu},  \quad
n=+\sqrt{(-\epsilon)(-\mu)},  \quad  n=-\sqrt{(-\epsilon)(-\mu)},
\label{ieq0}
\end{equation}
all of which agree with the second-order Maxwellian equation. But
for a time-harmonic electromagnetic wave, not all of them
satisfies the first-order Maxwellian equation. In order to clarify
this point, let us first consider the following two equations
\begin{equation}
{\bf k}\times {\bf E}=\mu \mu_{0}\omega {\bf H},  \qquad {\bf
k}\times {\bf H}=-\epsilon \epsilon_{0}\omega {\bf E},
\label{ieq1}
\end{equation}
for a time-harmonic wave, where $\mu_{0}$ and $\epsilon_{0}$ stand
for the electric permittivity and magnetic permeability in a free
space, respectively. Here the wave vector ${\bf k}$ may be
rewritten as ${\bf k}=n\frac{\omega}{c}\hat{{\bf k}}$, where the
unit vector $\hat{{\bf k}}$ is so defined that $\hat{{\bf k}}$,
${\bf E}$ and ${\bf H}$ form a right-handed system. Thus according
to Eq.(\ref{ieq1}), one can arrive at $\frac{n}{\mu
\mu_{0}c}\hat{{\bf k}}\times {\bf E}={\bf H}$. It is readily
verified that if $\mu>0$, then $n$ is positive, and while if
$\mu<0$, then $n$ is negative. In the same fashion, from the
equation ${\bf k}\times {\bf H}=-\epsilon \epsilon_{0}\omega {\bf
E}$, one can obtain $\frac{n}{\epsilon \epsilon_{0}c}\hat{{\bf
k}}\times {\bf H}=-{\bf E}$. Thus the similar result can also be
obtained: if $n>0$, then $\epsilon$ should be positive, and while
if $n<0$, then $\epsilon$ should be negative. This, therefore,
means that in Eq.(\ref{ieq0}), only the following two
possibilities
\begin{equation}
n=+\sqrt{\epsilon\mu},  \quad \quad   n=-\sqrt{(-\epsilon)(-\mu)}
\label{ieq2}
\end{equation}
may satisfy the first-order Maxwellian equation\footnote{Someone
shows how the negative index of refraction arises:
$n=\sqrt{\epsilon\mu}=\sqrt{(-\epsilon)(-\mu)}=\sqrt{-\epsilon}\sqrt{-\mu}=i^{2}\sqrt{\epsilon\mu}$.
We think that such a viewpoint may not be suitable for considering
the existence of negative index of refraction.}. These two
situations correspond to the two optical refractive indices of
electromagnetic wave propagating inside {\it right-handed media}
and {\it left-handed media}, respectively. Historically, in
1967\footnote{Note that, in the literature, some authors mentioned
the wrong year when Veselago suggested the {\it left-handed
media}. They claimed that Veselago proposed or introduced the
concept of {\it left-handed media} in 1968 or 1964. On the
contrary, the true history is as follows: Veselago's excellent
paper was first published in Russian in July, 1967 [Usp. Fiz. Nauk
{\bf 92}, 517-526 (1967)]. This original paper was translated into
English by W.H. Furry and published again in 1968 in the journal
of Sov. Phys. Usp.\cite{Veselago}. Unfortunately, Furry stated
erroneously in his English translation that the original version
of Veselago' work was first published in 1964.}, Veselago first
considered this peculiar medium having negative simultaneously
negative $\epsilon $ and $\mu $, and showed from Maxwellian
equations that such media exhibit a negative index of
refraction\cite{Veselago}. It follows from the Maxwell's curl
equations that the phase velocity of light wave propagating inside
this medium is pointed opposite to the direction of energy flow,
that is, the Poynting vector and wave vector of electromagnetic
wave would be antiparallel, {\it i.e.}, the vector {\bf {k}}, the
electric field {\bf {E}} and the magnetic field {\bf {H}} form a
left-handed system\footnote{Note that by using the mirror
reflection operation, a right-handed system may be changed into a
left-handed one. It is thus clearly seen that the permittivity and
permeability of the free vacuum in a mirror world may be negative
numbers.}; thus Veselago referred to such materials as
``left-handed'' media, and correspondingly, the ordinary/regular
medium in which {\bf {k}}, {\bf {E}} and {\bf {H}} form a
right-handed system may be termed the ``right-handed'' one. Other
authors call this class of materials ``negative-index media
(NIM)''\cite{Gerardin}, ``backward media (BWM)''\cite{Lindell},
``double negative media (DNM)''\cite{Ziolkowski2} and Veselago's
media. There exist a number of peculiar electromagnetic and
optical properties, for instance, many dramatically different
propagation characteristics stem from the sign change of the
optical refractive index and phase velocity, including reversal of
both the Doppler shift and Cerenkov radiation, anomalous
refraction, amplification of evanescent waves\cite{Pendryprl},
unusual photon tunneling\cite{Zhang}, modified spontaneous
emission rates and even reversals of radiation pressure to
radiation tension\cite{Klimov}.

In experiments, this artificial negative electric permittivity
media may be obtained by using the {\it array of long metallic
wires} (ALMWs)\cite{Pendry2}, which simulates the plasma behavior
at microwave frequencies, and the artificial negative magnetic
permeability media may be built up by using small resonant
metallic particles, {\it e.g.}, the {\it split ring resonators}
(SRRs), with very high magnetic
polarizability\cite{Pendry1,Pendry3,Maslovski}. A combination of
the two structures yields a left-handed medium. Recently, Shelby
{\it et al.} reported their first experimental realization of this
artificial composite medium, the permittivity and permeability of
which have negative real parts\cite{Shelby}. One of the potential
applications of negative refractive index materials is to
fabricate the so-called ``superlenses'' (perfect lenses):
specifically, a slab of such materials may has the power to focus
all Fourier components of a 2D image, even those that do not
propagate in a radiative manner\cite{Pendryprl,Hooft}.

In this paper, I will investigate theoretically the novel
phenomena, effects and properties of wave propagation in
left-handed media.

\section{The reflection and the refraction laws on the interface between LH and RH media}

The influence of the interface between LH and RH media on the
radiation propagation is governed by the reflection and the
refraction laws, which differ substantially from the laws in the
conventional (right-handed) media. When a wave in a RH material
hits an interface with a LH medium, it will have negative angle of
refraction, {\it i.e.}, the refracted wave will be on the same
side of the normal as the incident wave, which are peculiar to the
usual media. It is due to the fact that the refraction index is
negative for LH media. Moreover, the refracted wave can be {\it
absent} at all if the absolute values of the refraction indices of
LH and RH media are equal\cite{Veselago,Klimov}

\section{Negative permittivity and negative permeability}
\subsection{Negative permittivity resulting from ALMWs structure}
A 2-layer periodic array of conducting straight wires will give
rise to a negative permittivity. As observed, there is a single
gap in the propagation up to a cutoff frequency for modes with the
electric field polarized along the axis of the wire. Metals have a
characteristic response to electromagnetic radiation due to the
plasma resonance of the electron gas\cite{Pendry2}.

Here we consider the effective rest mass of electromagnetic wave
in superconducting media (or electron plasma) and
dispersive/absorptive materials. It is well known that the
Lagrangian density of electromagnetics is
\begin{equation}
\ell =-\frac{1}{4}F_{\mu \nu }F^{\mu \nu }-\frac{1}{\epsilon
_{0}c^{2}}\rho \varphi +\mu _{0}{\bf J}\cdot {\bf A}.
\label{eqB1}
\end{equation}
The equation of motion of an electron in superconducting media (
or electron plasma) acted upon by the external electromagnetic
wave is $m_{\rm e}\dot{{\bf v}}=e{\bf E}$ with ${\bf
E}=-\frac{\partial }{\partial t}{\bf A}$, where dot on the
velocity ${\bf v}$ denotes the derivative of ${\bf v}$ with
respect to time $t$ and ${\bf A}$ stands for the magnetic vector
potentials of the applied electromagnetic wave. One can therefore
arrive at

\begin{equation}
\frac{{\rm d}}{{\rm d}t}\left(m_{\rm e}{\bf v}+e{\bf A}\right) =0,
\label{eqB2}
\end{equation}
namely, the canonical momentum $m_{\rm e}{\bf v}+e{\bf A}$ of the
electron is conserved. Set $m_{\rm e}{\bf v}+e{\bf A}=0$ and then
$m_{\rm e}{\bf v}=-e{\bf A}$. So, the electric current density is
of the form
\begin{equation}
{\bf J}=Ne{\bf v}=-\frac{Ne^{2}}{m_{\rm e}}{\bf A} \label{eqB3}
\end{equation}
with $N$ being the electron number in per unit volume. It follows
from (\ref{eqB1}) that the interaction term $\mu _{0}{\bf J}\cdot
{\bf A}$ in Lagrangian density is rewritten as follows
\begin{equation}
\mu _{0}{\bf J}\cdot {\bf A}=-\frac{\mu _{0}Ne^{2}}{m_{\rm e}}{\bf
A}^{2}=-\frac{Ne^{2}}{\epsilon _{0}m_{\rm e}c^{2}}{\bf A}^{2},
\label{eqB4}
\end{equation}
where use is made of the relation $c^{2}=\frac{1}{\epsilon _{0}\mu
_{0}}$. It is apparently seen that since electron current acts as
the self-induced charge current, the interaction term $\mu
_{0}{\bf J}\cdot {\bf A}$ is therefore transformed into the mass
term $-\frac{m_{\rm eff}^{2}c^{2}}{\hbar ^{2}}{\bf A}^{2}$ of
electromagnetic fields. Hence, we have
\begin{equation}
\frac{m_{\rm eff}^{2}c^{2}}{\hbar ^{2}}=\frac{Ne^{2}}{\epsilon
_{0}m_{\rm e}c^{2}}   \label{eqB5}
\end{equation}
and the {\it effective mass} squared
\begin{equation}
m_{\rm eff}^{2}=\frac{\hbar ^{2}}{c^{2}}\left(
\frac{Ne^{2}}{\epsilon _{0}m_{\rm e}c^{2}}\right). \label{eqB6}
\end{equation}

The relation between the frequency $\omega $ and wave vector $k$
of lightwave propagating inside the dispersive medium may be
$\frac{\omega ^{2}}{k^{2}}=\frac{c^{2}}{n^{2}}$ with $n$ being the
optical refractive index. For highly absorptive media, the
corresponding modification of this method can also apply. So, for
convenience, here we are not ready to deal with the case of
absorptive materials. This dispersion relation $\frac{\omega
^{2}}{k^{2}}=\frac{c^{2}}{n^{2}}$ yields

\begin{equation}
\frac{\omega ^{2}}{k^{2}}=\frac{c^{2}}{1-\frac{c^{4}}{\hbar
^{2}}\frac{m_{\rm eff}^{2}}{\omega ^{2}}} \label{eq101}
\end{equation}
with the effective rest mass, $m_{\rm eff}$, of the light being so
defined that $m_{\rm eff}$, $n$ and $\omega $ together satisfy the
following relation

\begin{equation}
\frac{m_{\rm eff}^{2}c^{4}}{\hbar ^{2}}=(1-n^{2})\omega ^{2}.
                 \label{eqq102}
\end{equation}
This, therefore, implies that the particle velocity of light (
photons) is $v=nc$ and the phase velocity of light is $v_{\rm
p}=\frac{c}{n}=\frac{c^{2}}{v}$. Since this relation is familiar
to us, we thus show that the above calculation is self-consistent,
at least for the de Broglie wave (particle).

By using $n^{2}=\epsilon$, it follows from Eq.(\ref{eqB6}) and
(\ref{eqq102}) that
\begin{equation}
\epsilon (\omega)=1-\frac{\omega _{\rm p}^{2}}{\omega ^{2}}, \quad
 \omega _{\rm p}^{2}=\frac{Ne^{2}}{m_{\rm e}\epsilon _{0}}  \label{eqq103}
\end{equation}
with $\omega _{\rm p},N,e,m_{\rm e}$ and $\epsilon _{0}$ being the
electron plasma frequency, number of electrons per unit volume,
electron charge, electron mass and electric permittivity in a
vacuum, respectively.

Generally speaking, when taking into consideration the absorption
of media, the electric permittivity of ALMWs structures may be
rewritten as follows
\begin{equation}
\epsilon=1-\frac{\omega _{\rm p}^{2}}{\omega(\omega+i\gamma)}.
\label{permittivity}
\end{equation}
Appropriate choice for the frequency of wave will yield negative
permittivity.

An alternative approach to deriving the expression for the
permittivity of ALMWs structures can be seen in the Appendix A to
this paper.
\subsection{Negative permeability resulting from SRRs structure}

The {\it split ring resonator} (SRR) can be used for manufacturing
negative magnetic permeability media and hence designing
left-handed materials. SRR is often formed by two coupled
conducting rings printed on a dielectric slab of thickness with,
say, 0.216 mm\cite{Marques}. The structure of SRRs under
consideration may be referred to the figures in the
references\cite{Smith,Marques}. The physical mechanism to obtain
negative permeability may be as follows: a time-varying magnetic
field applied parallel to the axis of the rings induces currents
that, depending on the resonant properties of the unit, produce a
magnetic field that may either oppose or enhance the incident
field\cite{Smith}. The associated magnetic field pattern from the
SRR is dipolar. By having splits in the rings, the SRR unit can be
made resonant at wavelengths much larger than the diameter of
rings (the SRR particle size is about one tenth of that
wavelength\cite{Marques}). The purpose of the second split ring,
inside and whose split is oriented {\it opposite} to the first, is
to generate a large capacitance in the small gap region between
the rings, lowering the resonant frequency considerably and
concentrating the electric field\cite{Smith}.

In what follows we will derive the expression for the magnetic
permeability of SRRs structures. Assume that the planes of SRRs
are parallel to the \^{x}-\^{y} plane, and an external magnetic
field ${\bf B}_{z}=B_{z}^{\rm ext}\exp[i\omega t]\hat{{\bf z}}$
acting on the SRRs. So, an electromotive force ${\mathcal
E}=-i\omega\pi r^{2}_{0}B_{z}^{\rm ext}$ (with $\pi r^{2}_{0}$
being the plane area of a SRR particle) is induced along the rings
which is responsible for creating a current flow which produces a
total magnetic moment in the particle. The slot between the rings
acts as a distributed capacitance, which stores the same amount of
charge (but of opposite sign) at both sides of the
slot\cite{Marques}. According to the charge conservation law, one
can obtain
\begin{eqnarray}
\frac{\rm d}{r_{0}{\rm d}\phi}I_{\rm i}=-\frac{\rm d}{{\rm d}t}q_{\rm i}=-i\omega C(V_{\rm i}-V_{\rm o}),                  \nonumber \\
\frac{\rm d}{r_{0}{\rm d}\phi}I_{\rm o}=-\frac{\rm d}{{\rm
d}t}q_{\rm o}=-i\omega C(V_{\rm o}-V_{\rm i}),
\label{permeability}
\end{eqnarray}
where $\phi$ and $C$ denote the angle displacement of 2D polar
coordinate and the per unit length capacitance between the rings,
and $I_{\rm i}$ and $I_{\rm o}$ the currents flowing on both rings
(inner and outer). $q_{\rm i}$ and $q_{\rm o}$ are the per unit
length charge at the inner and outer rings, respectively. Note
that on the right-handed sides of Eqs. (\ref{permeability}),
$\frac{\rm d}{{\rm d}t}$ can be replaced with $i\omega$. It is
assumed that in the ring the current can be considered to be
homogeneous. This, therefore, implies that $\frac{\rm d}{r_{0}{\rm
d}\phi}I_{\rm i}=\frac{I_{\rm i}}{2\pi {r_{0}}}$. It is apparent
that $(V_{\rm i}-V_{\rm o})=2i\omega\left(LI+\pi r^{2}_{0}B^{\rm
ext}\right)$, where $L$ is the total inductance of the SRR. Thus,
we have
\begin{equation}
I=4\pi \omega^{2}r_{0}C\left(LI+\pi r^{2}_{0}B^{\rm ext}\right)
\end{equation}
and consequently
\begin{equation}
I=\frac{4\pi^{2}\omega^{2}r_{0}^{3}C}{1-4\pi\omega^{2}LCr_{0}}B^{\rm
ext}.
\end{equation}
Thus the moment per unit volume is
\begin{equation}
M=NI\pi r_{0}^{2}=N\frac{\pi r_{0}^{2}\mu_{0}^{-1}\omega^{2}(4\pi
Cr_{0}\cdot\mu_{0}\pi
r_{0}^{2})}{1-\frac{\omega^{2}}{\omega^{2}_{0}}}B^{\rm ext},
\label{induc}
\end{equation}
where $\omega^{2}_{0}=\frac{1}{4\pi LCr_{0}}$ and $N$ is the SRR
unit number per unit volume. It can be easily verified\footnote{In
accordance with Amp\`{e}re's circuital law, the magnetic induction
$B$ inside a long solenoid with $N$ circular loops carrying a
current $I$ in the region remote from the ends is axial, uniform
and equal to $\mu_{0}$ times the number of Amp\`{e}re-turns per
meter $nI$, {\it i.e.}, $B=\mu_{0}nI$, where $n=\frac{N}{l}$ with
$l$ being the solenoid length. So, the magnetic flux is
$\mu_{0}nI\pi r_{0}^{2}l$ (here $\pi r_{0}^{2}$ is the
cross-section area of this long solenoid), and the electromotive
force ${\mathcal E}=-\mu_{0}n\pi r_{0}^{2}l\frac{{\rm d}I}{{\rm d
}t}$. By the aid of ${\mathcal E}=-L\frac{{\rm d}I}{{\rm d }t}$,
one can arrive at the total inductance of the solenoid, {\it
i.e.}, $L=\mu_{0}n\pi r_{0}^{2}l$. Thus the per unit length
inductance of the solenoid is $\frac{L}{2\pi
r_{0}nl}=\frac{\mu_{0}r_{0}}{2}$.} that the per unit length
inductance of ring with radius $r_{0}$ is
$\frac{\mu_{0}r_{0}}{2}$. So, the total inductance of the SRR is
$\mu_{0}\pi r_{0}^{2}$. It follows from Eq.(\ref{induc}) that
\begin{equation}
M=\frac{N\pi
r_{0}^{2}\mu_{0}^{-1}\frac{\omega^{2}}{\omega^{2}_{0}}}{1-\frac{\omega^{2}}{\omega^{2}_{0}}}B^{\rm
ext}.                                             \label{MMM}
\end{equation}
Let $B^{\rm int}$, $H^{\rm int}$ and $H^{\rm ext}$ stand for the
magnetic induction and field strength of internal and external
fields, respectively\footnote{$B^{\rm int}$ refers to the field
influenced by the moment.}. Since all the currents (such as free
current, polarization current and magnetization current) will
contribute to magnetic induction $B$, while the magnetic filed $H$
is produced only by the free current, we may have $B^{\rm int}\neq
B^{\rm ext}$, $H^{\rm int}=H^{\rm ext}$. Because of $B^{\rm
int}=\mu_{0}\mu_{\rm r}H^{\rm int}=\mu_{0}\mu_{\rm r}H^{\rm
ext}=\mu_{\rm r}B^{\rm ext}$, we can obtain $B^{\rm
ext}=\frac{B^{\rm int}}{\mu_{\rm r}}$. So, it follows that
\begin{equation}
M=\frac{N\pi
r_{0}^{2}\mu_{0}^{-1}\frac{\omega^{2}}{\omega^{2}_{0}}}{1-\frac{\omega^{2}}{\omega^{2}_{0}}}\frac{B^{\rm
int}}{\mu_{\rm r}}.
\end{equation}
Insertion of $M=\frac{1}{\mu_{0}}\frac{\mu_{\rm r}-1}{\mu_{\rm
r}}B^{\rm int}$ yields
\begin{equation}
\mu_{\rm r}=1+\frac{N\pi
r_{0}^{2}\frac{\omega^{2}}{\omega^{2}_{0}}}{1-\frac{\omega^{2}}{\omega^{2}_{0}}}=1-\frac{F\omega^{2}}{\omega^{2}-\omega^{2}_{0}}.
\label{expre}
\end{equation}
In general, when taking account of the dissipation factor of SRR,
the expression (\ref{expre}) may be rewritten as
\begin{equation}
\mu_{\rm r}=1-\frac{F\omega^{2}}{\omega^{2}-\omega^{2}_{0}+i\omega
\Gamma }.                    \label{permeability2}
\end{equation}
Appropriate choice for the frequency of wave will yield negative
permeability.

Readers may be referred to Appendix B for the derivation of the
expression for the permeability of SRRs structures.

In addition, someone may argue that the magnetization $M$ should
be $M=\frac{1}{\mu_{0}}\frac{\mu_{\rm r}-1}{\mu_{\rm r}}B^{\rm
ext}$ (rather than $M=\frac{1}{\mu_{0}}\frac{\mu_{\rm
r}-1}{\mu_{\rm r}}B^{\rm int}$), where the relative permeability
$\mu_{\rm r}$ is defined. Thus it follows from Eq.(\ref{MMM}) that
\begin{equation}
\frac{N\pi
r_{0}^{2}\mu_{0}^{-1}\frac{\omega^{2}}{\omega^{2}_{0}}}{1-\frac{\omega^{2}}{\omega^{2}_{0}}}=\frac{1}{\mu_{0}}\frac{\mu_{\rm
r}-1}{\mu_{\rm r}},
\end{equation}
and in consequence
\begin{equation}
\mu_{\rm r}=\frac{1}{1-\frac{N\pi
r_{0}^{2}\omega^{2}}{\omega^{2}_{0}-\omega^{2}}}=\frac{\omega^{2}_{0}-\omega^{2}}{\omega^{2}_{0}-(1+N\pi
r_{0}^{2})\omega^{2}}=\frac{\omega^{2}_{0}-(1+N\pi
r_{0}^{2})\omega^{2}+N\pi
r_{0}^{2}\omega^{2}}{\omega^{2}_{0}-(1+N\pi r_{0}^{2})\omega^{2}},
\end{equation}
which can be rewritten as
\begin{equation}
\mu_{\rm r}=1+\frac{\frac{N\pi r_{0}^{2}}{1+N\pi
r_{0}^{2}}\omega^{2}}{\frac{\omega^{2}_{0}}{1+N\pi
r_{0}^{2}}-\omega^{2}}=1-\frac{F'\omega^{2}}{\omega^{2}-\omega'^{2}_{0}},
\end{equation}
which is also of the form analogous to (\ref{expre}).

\subsection{Other design to obtain negative optical refractive index}

Pendry {\it et al.} used the {\it array of long metallic wires}
(ALMWs)\cite{Pendry2} that simulates the plasma behavior at
microwave frequencies, and the {\it split ring resonators} (SRRs)
that gives rise to the artificial negative magnetic permeability
to realize the left-handed materials. But there are alternative
approaches to the realization of such media. For example,
Podolskiy {\it et al.} investigated the electromagnetic field
disrtribution for the thin metal nanowires by using the discrete
dipole approximation, and considered the plasmon polariton modes
in wires by using numerical simulation\cite{Podolskiy}. In the
paper\cite{Podolskiy}, they described a material comprising pairs
of nanowires parallel to each other and showed that such materials
can have a negative refraction index in the near IR and the
visible spectral range. Consider a thin layer of material,
composed from pairs of nanowires parallel to each other. The
length of individual nanowires is $2b_{1}$, their diameter is
$2b_{2}$, and the distance between the nanowires in the pair is
$d$. The needles are assumed to be embedded into host with
dielectric constant. Podolskiy {\it et al.} considered the case of
closely placed long nanowires so that $b_{2}\ll d\ll
b_{1}$\cite{Podolskiy}. The incident wave propagates normal to the
composite surface so that the electric field is parallel to the
nanowires, while the magnetic field is perpendicular to the
nanowire pairs. Podolskiy {\it et al.} found the effective
dielectric constant and magnetic permeability for the 2D
nanoneedle composite film are
\begin{equation}
\epsilon=1+\frac{4p}{b_{1}b_{2}d}\frac{d_{E}}{E},   \qquad
\mu=1+\frac{4p}{b_{1}b_{2}d}\frac{m_{H}}{H},
\end{equation}
where $p$ and $m_{H}$ are the surface metal concentration and the
moment of the individual two-needle system, respectively. Note
that the resonance position (and therefore the spectral range
where the material refractive index is negative) is determined by
parameters $b_{1}$, $b_{2}$ and $d$. By varying the these
parameters the negative-refraction spectral range can be moved to
the visible part of the spectrum\cite{Podolskiy}.

It should be noted that the structure of a combination of two
lattices ({\it i.e.}, a lattice of infinitely long parallel wires
and a lattice of the split ring resonators) is not isotropic. Such
structures can be described with negative permittivity and
permeability only if the propagation direction is orthogonal to
the axes of wires\cite{Simovski}. More recently, Simovski and He
presented a analytical model for a rectangular lattice of
isotropic scatters with electric and magnetic resonances. Here
each isotropic scatterer is formed by putting appropriately 6
$\Omega$-shaped perfectly conducting particles on the faces of a
cubic unit cell\cite{Simovski}. They derived a self-consistent
dispersion equation and used it to calculate correctly the
effective permittivity and permeability in the frequency band
where the lattice can be homogenized. The frequency range in which
both the effective permittivity and permeability are negative
corresponds to the mini-band of backward waves within the resonant
band of the individual isotropic scatterer. For the detailed
analysis of the electromagnetic properties of materials formed by
the rectangular lattice of isotropic cubic unit cells of $\Omega$
particles, authors may be referred to Simovski and He's recently
published paper\cite{Simovski}.

\section{Wave propagation in dispersive materials}

\subsection{Group velocity and energy density in dispersive
materials}

In a dispersive (isotropic) medium, the group velocity of an
electromagnetic wave is
\begin{equation}
v_{\rm g}=\frac{{\rm d}\omega}{{\rm d}k}=
\frac{1}{\sqrt{\epsilon_{0}\mu_{0}}}\left(\frac{1}{\sqrt{\epsilon\mu}+\omega\frac{\rm
d }{{\rm d}\omega}\sqrt{\epsilon\mu}}\right),
\end{equation}
where $k=\sqrt{\epsilon\mu}\sqrt{\epsilon_{0}\mu_{0}}\omega$. In
what follows we will calculate the energy flow velocity, {\it
i.e.}, $\frac{{\bf E}\times{\bf H}}{w_{\rm em}}$. The magnitude of
Poynting vector may be given by
\begin{equation}
|{\bf E}\times{\bf
H}|=\sqrt{\frac{\epsilon\epsilon_{0}}{\mu\mu_{0}}}E^{2}.
\end{equation}
Assume that the energy flow velocity equals the group velocity and
we will obtain
\begin{equation}
\frac{\sqrt{\frac{\epsilon\epsilon_{0}}{\mu\mu_{0}}}E^{2}}{w}=\frac{1}{\sqrt{\epsilon_{0}\mu_{0}}}\left(\frac{1}{\sqrt{\epsilon\mu}+\omega\frac{\rm
d }{{\rm d}\omega}\sqrt{\epsilon\mu}}\right). \label{above}
\end{equation}
The right-handed side of Eq.(\ref{above}) can be rewritten as
\begin{equation}
v_{\rm
g}=\frac{1}{\sqrt{\epsilon_{0}\mu_{0}}}\left(\frac{1}{\sqrt{\epsilon\mu}+\frac{1}{2}\frac{\omega}{\sqrt{\epsilon\mu}}\frac{\rm
d}{{\rm d}\omega}(\epsilon\mu)}\right)
=\frac{\sqrt{\frac{\epsilon\epsilon_{0}}{\mu\mu_{0}}}E^{2}}{\epsilon\epsilon_{0}E^{2}\left\{1+\frac{1}{2}\frac{\omega}{\epsilon\mu}\left[\frac{{\rm
d }(\epsilon\mu)}{{\rm d}\omega}\right]\right\}}. \label{above2}
\end{equation}
The comparison between (\ref{above}) and (\ref{above2}) shows that
the energy density of electromagnetic system takes the form
\begin{equation}
w=\epsilon\epsilon_{0}E^{2}\left\{1+\frac{1}{2}\frac{\omega}{\epsilon\mu}\left[\frac{{\rm
d }(\epsilon\mu)}{{\rm d}\omega}\right]\right\},
\end{equation}
which can also be rewritten as
\begin{equation}
w=\frac{1}{2}\left[\frac{\rm d}{{\rm
d}\omega}(\omega\epsilon)+\frac{\rm d}{{\rm
d}\omega}(\omega\mu)\frac{\epsilon}{\mu}\right]\epsilon_{0}E^{2}.
\label{above3}
\end{equation}
By using the expression
$H=\sqrt{\frac{\epsilon\epsilon_{0}}{\mu\mu_{0}}}E$, $\frac{\rm
d}{{\rm d}\omega}(\omega\mu)\frac{\epsilon}{\mu}\epsilon_{0}E^{2}$
is therefore $\frac{\rm d}{{\rm d}\omega}(\omega\mu)\mu_{0}H^{2}$.
So, it follows from Eq.(\ref{above3}) that
\begin{equation}
w=\frac{1}{2}\left[\frac{\rm d}{{\rm
d}\omega}(\omega\epsilon)\epsilon_{0}E^{2}+\frac{\rm d}{{\rm
d}\omega}(\omega\mu)\mu_{0}H^{2}\right], \label{above4}
\end{equation}
which is just the electromagnetic energy density of dispersive
linear materials.

Note that according to the Kramers-Kronig relation
\begin{equation}
{\rm
Re}\epsilon(\omega)=1+\frac{1}{\pi}P\int^{+\infty}_{-\infty}\frac{{\rm
Im}\epsilon(\omega')}{\omega'-\omega}{\rm d}\omega',    \quad {\rm
Im}\epsilon(\omega)=-\frac{1}{\pi}P\int^{+\infty}_{-\infty}\frac{\left[{\rm
Re}\epsilon(\omega')-1\right]}{\omega'-\omega}{\rm d}\omega',
\end{equation}
the negative refractive index materials is sure to be the
dispersive (and also absorptive) media. So we should make use of
Eq.(\ref{above4}) when calculating the energy density of
electromagnetic materials. It can be easily seen that by
substituting the expressions for the permittivity and permeability
(\ref{permittivity}) and (\ref{permeability2}) of left-handed
media into Eq.(\ref{above4}), the energy density is always
positive.

It is well known that in the ``non-dispersive'' medium, the
electromagnetic energy density is $w=\frac{1}{2}\left({\bf E
}\cdot{\bf D}+{\bf H}\cdot{\bf B}\right)$. How can we understand
the relation between $w$ and the electromagnetic potential energy
(and kinetic energy) of dipoles? Does $\frac{1}{2}{\bf E
}\cdot{\bf D}$ contain the potential energy $-{\bf P}\cdot{\bf
E}$, or does it contain the kinetic energy of dipoles?

Consider the following equation
\begin{equation}
\ddot{p}+\omega_{\rm r}^{2}p=\epsilon_{0}\omega_{\rm p}^{2}E
\label{dipole}
\end{equation}
that governs the electric dipole $p$ in the presence of an
electric field $E=E_{0}\exp\left[\frac{1}{i}\omega t\right]$. The
solution to Eq.(\ref{dipole}) is
$p=p_{0}\exp\left[\frac{1}{i}\omega t\right]$, where
\begin{equation}
p_{0}=\frac{\epsilon_{0}\omega^{2}_{\rm
p}}{-\omega^{2}+\omega^{2}_{\rm r}}E_{0},   \quad
E_{0}=\frac{-\omega^{2}+\omega^{2}_{\rm
r}}{\epsilon_{0}\omega^{2}_{\rm p}}p_{0}.
\end{equation}
Here $\omega^{2}_{\rm r}=\frac{K}{m}$, $\omega^{2}_{\rm
p}=\frac{e^{2}}{m\epsilon_{0}}$. Thus the kinetic and potential
energies (purely mechanical energy) of the dipole oscillator
(experience mechanical vibration) are written

\begin{eqnarray}
E_{\rm k}&=&\frac{1}{2}m\dot{x}^{2}=-\frac{1}{2}m\omega^{2}x^{2},                 \nonumber \\
E_{\rm p}&=&\frac{1}{2}Kx^{2}=\frac{1}{2}m\omega^{2}_{\rm r}x^{2},
\end{eqnarray}
where $x=\frac{p}{e}$. So, the total mechanical energy of the
dipole oscillator is

\begin{equation}
E_{\rm mecha \ tot}=E_{\rm k}+E_{\rm
p}=\frac{1}{2}\frac{m}{e^{2}}\left(\omega^{2}_{\rm
r}-\omega^{2}\right)p^{2}=\frac{1}{2}\frac{1}{\epsilon_{0}\omega^{2}_{\rm
p}}\left(\omega^{2}_{\rm r}-\omega^{2}\right)p^{2}.
\label{answer}
\end{equation}
The electric potential energy of the dipole oscillator is
\begin{equation}
V=-pE=-\frac{-\omega^{2}+\omega^{2}_{\rm
r}}{\epsilon_{0}\omega^{2}_{\rm p}}p^{2}.
\end{equation}
This therefore means that the relation between the mechanical
energy and the electric potential energy of the dipole oscillator
is
\begin{equation}
V=-2E_{\rm mecha \ tot}.                 \label{answer2}
\end{equation}
Thus the total energy of the dipole oscillator reads
\begin{equation}
E_{\rm tot}=E_{\rm mecha \ tot}+V=-\frac{1}{2}pE. \label{tot}
\end{equation}
What is the relationship between $E_{\rm tot}$ and $\frac{1}{2}pE$
in $\frac{1}{2}ED$ with $D=\epsilon_{0}E+p$? Now I will briefly
discuss this problem: in fact, $\frac{1}{2}pE$ in $\frac{1}{2}ED$
is just the total mechanical energy, $E_{\rm mecha \ tot}$, of the
dipole oscillator, while the electric potential energy has always
been involved\footnote{Electric potential energy is involved in
the electric field energy density. This may be discussed as
follows: consider a self-interacting electric system, by using the
Gaussian theorem $\nabla^{2}\phi=-\frac{\rho}{\epsilon_{0}}$, we
may obtain the total potential energy $V=\frac{1}{2}\int \rho\phi
{\rm d}{\bf x}=-\frac{1}{2}\epsilon_{0}\int \phi\nabla^{2}\phi{\rm
d}{\bf x}=-\frac{1}{2}\epsilon_{0}\int
\left[\nabla\cdot\left(\phi\nabla\phi\right)-\left(\nabla\phi\right)^{2}\right]{\rm
d}{\bf x}=\frac{1}{2}\epsilon_{0}\int E^{2}{\rm d}{\bf x}$.} in
$\frac{1}{2}\epsilon_{0}E^{2}$. This can also be understood as
follows: when the incident wave with electric energy density
$\frac{1}{2}\epsilon_{0}E^{2}$ in a vacuum propagates inside a
medium, the electric energy density will be changed into
$\frac{1}{2}ED$. Where does the energy difference $\frac{1}{2}pE$
come? The answer lies in Eq.(\ref{tot}), which shows that the
total energy (mechanical energy and electric potential energy) of
the dipole oscillator is just the energy difference
$-\frac{1}{2}pE$.

It should be emphasized that when we take account of the
electromagnetic energy density in the dispersive material, we
should not neglect the terms
$\frac{1}{2}\left[\epsilon_{0}\omega\frac{{\rm d }\epsilon}{{\rm
d}\omega}{\bf E}^{2}+\mu_{0}\omega\frac{{\rm d }\mu}{{\rm
d}\omega}{\bf H}^{2}\right]$. We think that in Ruppin's
work\cite{Rup}, such a dispersive term should be added to his
obtained expression for the electromagnetic energy density.

\subsection{Wave propagation in dispersive materials}
Now let us consider the wave propagation in a dispersive material,
where the electromagnetic field equation that governs the wave
propagation is
\begin{equation}
\nabla \times{\mathcal E}+\frac{\partial}{\partial
t}\left(\mu_{0}\mu{\mathcal H}\right)=0,   \quad \nabla
\times{\mathcal H}-\frac{\partial}{\partial
t}\left(\epsilon_{0}\epsilon{\mathcal E}\right)=0.
\end{equation}
Here the electric and magnetic field strengths are written in
terms of quasi-monochromatic field components as
follows\cite{Prade}
\begin{eqnarray}
{\mathcal E}&=&\exp\left[-i(\omega_{0}t-\beta_{0}z)\right]\int \exp[-i\alpha(t-\beta' z)]{\bf E}({\bf x}, \omega_{0}+\alpha){\rm d}\alpha,                  \nonumber \\
{\mathcal H}&=&\exp\left[-i(\omega_{0}t-\beta_{0}z)\right]\int
\exp[-i\alpha(t-\beta' z)]{\bf H}({\bf x}, \omega_{0}+\alpha){\rm
d}\alpha,
\end{eqnarray}
where $\omega=\omega_{0}+\alpha$, $\beta'=\frac{{\rm d}\beta}{{\rm
d}\omega}|_{\omega_{0}}$ with ${\omega_{0}}$ being the central
frequency. Assuming the wave vector of electromagnetic wave is
parallel to \^{z}-direction, it follows that
\begin{eqnarray}
& & i\beta_{0}{\rm e}_{z}\times{\bf E}+i\beta' \alpha {\rm e}_{z}\times{\bf E}+\nabla\times{\bf E}-i\mu_{0}\omega_{0}\mu(\omega_{0}){\bf H}-i\mu_{0}\alpha\frac{\partial}{\partial \omega}(\omega\mu){\bf H}=0,                \nonumber \\
& & i\beta_{0}{\rm e}_{z}\times{\bf H}+i\beta' \alpha {\rm
e}_{z}\times{\bf H}+\nabla\times{\bf
H}+i\epsilon_{0}\omega_{0}\epsilon(\omega_{0}){\bf
E}+i\epsilon_{0}\alpha\frac{\partial}{\partial
\omega}(\omega\epsilon){\bf E}=0,
\end{eqnarray}
and consequently (multiplying the above two equations by ${\bf H}$
and ${\bf E} $, respectively)
\begin{eqnarray}
& & i\beta_{0}{\bf H}\cdot\left({\rm e}_{z}\times{\bf E}\right)+i\beta' \alpha {\bf H}\cdot\left({\rm e}_{z}\times{\bf E}\right)+{\bf H}\cdot\left(\nabla\times{\bf E}\right)-i\mu_{0}\omega_{0}\mu(\omega_{0}){\bf H}^{2}-i\mu_{0}\alpha\frac{\partial}{\partial \omega}(\omega\mu){\bf H}^{2}=0,                       \nonumber \\
& & i\beta_{0}{\bf E}\cdot\left({\rm e}_{z}\times{\bf
H}\right)+i\beta' \alpha {\bf E}\cdot\left({\rm e}_{z}\times{\bf
H}\right)+{\bf E}\cdot\left(\nabla\times{\bf
H}\right)+i\epsilon_{0}\omega_{0}\epsilon(\omega_{0}){\bf
E}^{2}+i\epsilon_{0}\alpha\frac{\partial}{\partial
\omega}(\omega\epsilon){\bf E}^{2}=0.   \label{multi}
\end{eqnarray}
Subtracting the second equation from the first one in
(\ref{multi}), one can arrive at
\begin{eqnarray}
& & 2i\left[\beta_{0}{\bf H}\cdot\left({\bf e}_{z}\times{\bf
E}\right)-\epsilon_{0}\omega_{0}\epsilon(\omega_{0}){\bf
E}^{2}\right]+\nabla\cdot\left({\bf E}\times{\bf H}\right)                  \nonumber \\
&+& 2i\alpha\left[\beta'{\bf H}\cdot\left({\bf e}_{z}\times{\bf
E}\right)-\frac{1}{2}\mu_{0}\frac{\partial}{\partial
\omega}(\omega\mu){\bf
H}^{2}-\frac{1}{2}\epsilon_{0}\frac{\partial}{\partial
\omega}(\omega\epsilon){\bf E}^{2}\right]=0.    \label{inte}
\end{eqnarray}
It is readily verified that by using the relations
$\beta_{0}=\sqrt{\epsilon\mu}\frac{\omega}{c}$ and $|{\bf
H}|=\sqrt{\frac{\epsilon\epsilon_{0}}{\mu\mu_{0}}}|{\bf E}|$, one
can prove that the following relation
\begin{equation}
\beta_{0}{\bf H}\cdot\left({\bf e}_{z}\times{\bf
E}\right)-\epsilon_{0}\omega_{0}\epsilon(\omega_{0}){\bf E}^{2}=0
\end{equation}
holds. Thus integrating the equation (\ref{inte}) and using $\int
\nabla\cdot\left({\bf E}\times{\bf H}\right){\rm d}{\bf x}=0$, we
will obtain\cite{Prade}
\begin{equation}
\frac{1}{\beta'}=\frac{\int^{+\infty}_{-\infty}S_{z}{\rm d
}z}{\int^{+\infty}_{-\infty}\left[\frac{1}{2}\mu_{0}\frac{\partial}{\partial
\omega}(\omega\mu){\bf
H}^{2}+\frac{1}{2}\epsilon_{0}\frac{\partial}{\partial
\omega}(\omega\epsilon){\bf E}^{2}\right]{\rm d }z}.
\label{density}
\end{equation}
Hence we obtain the expression for group velocity of wave
propagation in the dispersive materials, {\it i.e.},
\begin{equation}
v_{\rm g}=\frac{1}{\beta'}.
\end{equation}
It follows from (\ref{density}) that the electromagnetic energy
density $w$ in dispersive media takes the form
\begin{equation}
w=\frac{1}{2}\epsilon_{0}\frac{\partial}{\partial
\omega}(\omega\epsilon){\bf
E}^{2}+\frac{1}{2}\mu_{0}\frac{\partial}{\partial
\omega}(\omega\mu){\bf H}^{2}.
\end{equation}

\section{Negative optical refractive index resulting from a moving regular medium}
Apart from the above artificial fabrications, does there exist
another alternative to the negative index of refraction? In this
section, I will demonstrate that a moving regular medium with a
velocity in a suitable region (associated with its rest refractive
index $n$)\footnote{The rest refractive index is just the one
measured by the observer fixed at the regular/ordinary medium. In
what follows, we will derive the relativistic transformation for
the optical refractive index (tensor) of a moving medium.}
possesses a negative index of refraction, which may be a
physically interesting effect and might deserve further
discussion.

Suppose we have a regular anisotropic medium with the rest
refractive index tensor $\hat{n}$, which is moving relative to an
inertial frame K with speed ${\bf v}$ in the arbitrary directions.
The rest refractive index tensor $\hat{n}$ can be written as $
\hat{n}={\rm diag}\left[n_{1}, n_{2}, n_{3}\right] $ with
$n_{i}>0, i=1,2,3$. Thus the wave vector of a propagating
electromagnetic wave with the frequency $\omega$ reads $ {\bf
k}=\frac{\omega}{c}\left(n_{1}, n_{2}, n_{3}\right) \label{eqq1} $
 measured by the observer fixed at this medium. In this sense,
one can define a 3-D vector ${\bf n}=\left(n_{1}, n_{2},
n_{3}\right)$, and the wave vector ${\bf k}$ may be rewritten as
${\bf k}={\bf n}\frac{\omega}{c}$. Now we analyze the phase
$\omega t-{\bf k}\cdot{\bf x}$ of the above time-harmonic
electromagnetic wave under the following Lorentz transformation
\begin{equation}
{\bf x}'=\gamma\left({\bf x}-{\bf v}t\right), \quad
t'=\gamma\left(t-\frac{{\bf v}\cdot{\bf x}}{c^{2}}\right),
\label{eq1}
\end{equation}
where $\left({\bf x}', t'\right)$ and $\left({\bf x}, t\right)$
respectively denote the spacetime coordinates of the initial frame
K and the system of moving medium, the spatial origins of which
coincide when $t=t'=0$. Here the relativistic factor
$\gamma=\left(1-\frac{v^{2}}{c^{2}}\right)^{-\frac{1}{2}}$. Thus
by using the transformation (\ref{eq1}), the phase $\omega t-{\bf
k}\cdot{\bf x}$ observed inside the moving medium may be rewritten
as the following form by using the spacetime coordinates of K
\begin{equation}
\omega t-{\bf k}\cdot{\bf x}=\gamma\omega\left(1-\frac{{\bf
n}\cdot{\bf v }}{c}\right)t'-\gamma\omega\left(\frac{{\bf n
}}{c}-\frac{{\bf v}}{c^{2}}\right)\cdot{\bf x}',
\end{equation}
the term on the right-handed side of which is just the expression
for the wave phase in the initial frame K. Hence, the frequency
$\omega'$ and the wave vector ${\bf k}'$ of the observed wave we
measure in the initial frame K are given
\begin{equation}
\omega'=\gamma\omega\left(1-\frac{{\bf n}\cdot{\bf v }}{c}\right),
\quad        {\bf k}'=\gamma\omega\left(\frac{{\bf n
}}{c}-\frac{{\bf v}}{c^{2}}\right),             \label{eq2}
\end{equation}
respectively.

To gain some insight into the meanings of the expression
(\ref{eq2}), let us consider the special case of a boost (of the
Lorentz transformation (\ref{eq1})) in the
$\hat{x}_{1}$-direction, in which the medium velocity relative to
K along the positive $\hat{x}_{1}$-direction is $v$. In the
meanwhile, we assume that the wave vector of the electromagnetic
wave is also parallel to the positive $\hat{x}_{1}$-direction. If
the {\it rest} refractive index of the medium in the
$\hat{x}_{1}$-direction is $n$, its (moving) refractive index in
the same direction measured by the observer fixed at the initial
frame K is of the form
\begin{equation}
n'=\frac{ck'}{\omega'}=\frac{n-\frac{v}{c}}{1-\frac{nv}{c}},
\label{eq3}
\end{equation}
which is a relativistic formula for the addition of ``refractive
indices'' ($-\frac{v}{c}$ provides an effective index of
refraction).

Note that here $n'$ observed in the frame K may be negative. If,
for example, when $n>1$, the medium moves at
\begin{equation}
c>v>\frac{c}{n}                           \label{eq5}
\end{equation}
(and in consequence, $k'>0$, $\omega'<0$); or when $0<n<1$, the
medium velocity with respect to K satisfies
\begin{equation}
c>v>nc                                         \label{eq6}
\end{equation}
(and consequently, $k'<0$, $\omega'>0$), then $n'$ is negative.
The former case where $k'>0$, $\omega'<0$ is of physical interest.
In the paper\cite{Shenanti}, it was shown that the photon in the
negative refractive index medium behaves like its {\it
anti-particle}, which, therefore, implies that we can describe the
wave propagation in both left- and right- handed media in a
unified way by using a complex vector field\cite{Shenanti}.

In the above, even though we have shown that in some certain
velocity regions, the moving medium possesses a negative $n'$,
such a moving medium cannot be surely viewed as a left-handed
medium. For this point, it is necessary to take into account the
problem as to whether the vector {\bf {k}}, the electric field
{\bf {E}} and the magnetic field {\bf {H}} of the electromagnetic
wave form a left-handed system or not\footnote{This requirement is
the standard definition of a left-handed medium. Apparently, it is
seen that the two concepts {\it left-handed medium} and {\it
negative refractive index medium} is not completely equivalent to
each other.}. For simplicity and without the loss of generality,
we choose the electric and magnetic fields of the electromagnetic
wave in the medium system as

\begin{equation}
{\bf E}=\left(0, E_{2}, 0\right),   \quad   {\bf
B}=\left(0,0,B_{3}\right).
\end{equation}
In the regular medium (right-handed medium with $n>0$), the wave
vector (along the positive $\hat{x}_{1}$-direction with the
modulus $k$) and such ${\bf E}$ and ${\bf B}$ (or ${\bf H}$) form
a right-handed system\footnote{Thus if $E_{2}>0$ is chosen to be
positive, then $B_{3}$ is also positive. In what follows, we will
adopt this case ({\it i.e.}, $E_{2}>0$ and $B_{3}>0$) without the
loss of generality.}. In the frame of reference K, according to
the Lorentz transformation, one can arrive at

\begin{equation}
E'_{2}=\gamma\left(E_{2}-vB_{3}\right)       \quad
B'_{3}=\gamma\left(B_{3}-\frac{v}{c^{2}}E_{2}\right). \label{eq7}
\end{equation}
For convenience, here we will think of the medium as an isotropic
one, {\it i.e.}, the refractive index $n'$, permittivity
$\epsilon'$ and the permeability $\mu'$ of the moving medium are
all the scalars rather than the tensors. First we consider the
case of Eq.(\ref{eq6}) where $k'<0$, $\omega'>0$. In order to let
the moving medium with a negative index of refraction be a real
left-handed one, the wave vector ($k'<0$), $E'_{2}$ and $H'_{3}$
of the electromagnetic wave measured in K should form a
left-handed system. Because of $k'<0$ and
$B'_{3}=\mu'\mu_{0}H'_{3}$ with $\mu_{0}$ being the magnetic
permeability in a vacuum, if $\mu'$ is negative\footnote{In the
left-handed medium, the electric permittivity and the magnetic
permeability are simultaneously negative.}, then only the case of
$E'_{2}<0$ and $B'_{3}>0$ (and vice versa) will lead to the
left-handed system\footnote{Since $k'<0$, in order to form a
left-handed system, the signs of $E'_{2}$ and $H'_{3}$ should not
be opposite to each other. So, the signs of $E'_{2}$ and $B'_{3}$
should be opposite due to $\mu'<0$.} formed by the wave vector
($k'<0$), $E'_{2}$ and $H'_{3}$. So, it follows from
Eq.(\ref{eq7}) that when the medium velocity relative to K is in
the range
\begin{equation}
\frac{c^{2}B_{3}}{E_{2}}<v<\frac{E_{2}}{B_{3}}      \quad
\left({\rm if}  \quad  E^{2}_{2}>c^{2}B^{2}_{3}\right),
\label{eq8}
\end{equation}
or
\begin{equation}
\frac{c^{2}B_{3}}{E_{2}}>v>\frac{E_{2}}{B_{3}}      \quad
\left({\rm if}  \quad  E^{2}_{2}<c^{2}B^{2}_{3}\right),
\label{eq9}
\end{equation}
then the wave vector, electric field and magnetic field measured
in K will truly form a left-handed system.

In conclusion, if the medium velocity satisfies both
Eq.(\ref{eq6}) and Eq.(\ref{eq8}) or (\ref{eq9}), then the moving
medium will be a left-handed one observed from the observer fixed
in the initial frame K.

As far as the case of Eq.(\ref{eq5}) where $k'>0$, $\omega'<0$ is
concerned, it is also possible for the moving material to become a
left-handed medium. But the case of $\omega'<0$ has no the
practical counterpart (at least in the real situation where the
artificial composite materials is designed). So, we would not
further discuss the case of $\omega'<0$.

Although the subject of the present note seems to be somewhat
trivial, it is helpful for understanding the properties of wave
propagation (such as causality problem, the problem of energy
propagating outwards and backwards from source\cite{Smith2}, {\it
etc.}) in the left-handed medium.

\section{The reversal of Doppler effect in left-handed media}
As stated in the Introduction, the reversals of many optical and
electromagnetic effects arise due to the negative index of
refraction. In what follows we will consider these unusual
phenomena.

Let us assume that the medium is fixed at the initial frame K and
the light source is moving inside the medium at velocity $-{\bf
v}$ with respect to the rest frame K. It follows from
Eq.(\ref{eq2}) ({\it i.e.},
$\omega'=\gamma\omega\left(1-\frac{{\bf n}\cdot{\bf v
}}{c}\right)$) that if here we consider a one-dimensional case in
the regular (right-handed) medium with the refractive index $n=1$,
then the frequency we measure in the rest frame K is
$\omega'=\gamma\omega\left(1-\frac{v}{c}\right)$. Since the
relativistic factor
$\gamma=\left(1-\frac{v^{2}}{c^{2}}\right)^{-\frac{1}{2}}$, we may
have

\begin{equation}
\omega'=\omega\sqrt{\frac{c-v}{c+v}},         \label{Doppler1}
\end{equation}
which is the normal Doppler effect. However, if the medium is a
left-handed material with $n=-1$, then it follows from
$\omega'=\gamma\omega\left(1-\frac{{\bf n}\cdot{\bf v
}}{c}\right)$ that the frequency measured in K is
$\omega'=\gamma\omega\left(1+\frac{v}{c}\right)$, which can be
rewritten as
\begin{equation}
\omega'=\omega\sqrt{\frac{c+v}{c-v}}.      \label{Doppler2}
\end{equation}
It is apparently seen that it is an abnormal Doppler effect
compared with Eq.(\ref{Doppler1}).

\section{The reversal of Cerenkov radiation in left-handed media}
Electromagnetic radiation, usually bluish light, emitted by a beam
of high-energy charged particles passing through a transparent
medium at s speed greater than the speed of light in that medium.
This effect is similar to that of a sonic boom when an object
moves faster than the speed of sound. In this case the radiation
is a shock wave set up in the electromagnetic field.

In the theory of Cerenkov radiation, the Fourier component of the
magnetic fields produced by the moving charged particle is
\begin{equation}
B_{\omega}=\frac{i\omega ne}{4\pi
\epsilon_{0}c^{3}}\frac{\exp(ikR)}{R}\sin \theta \delta
\left(\frac{\omega}{v}-\frac{\omega n}{c}\cos \theta\right),
\end{equation}
where $\theta$is the angle between the particle speed ${\bf v}$
and the direction of radiation. Let us first assume that the
refractive index $n>0$. It is well known that when the charged
particle velocity $v$ satisfies $v>\frac{c}{n}$, then in the
direction of $\cos \theta=\frac{c}{nv}$, the magnetic field
$B_{\omega}$ is nonvanishing. If the refractive index $n<0$, then
the radiation direction may change, {\it i.e.}, $\theta
\rightarrow \theta+\pi$. Additionally, the reversal of the
direction of $B_{\omega}$ will also arise, which results from the
coefficient $\frac{i\omega ne}{4\pi \epsilon_{0}c^{3}}$ of
$B_{\omega}$ (because of the coefficient involving $n$).
\section{Optical Refractive Index of Massive Particles and Physical Meanings of Left-handed Media}
A number of novel electromagnetic and optical properties in
left-handed media result from the negative index of refraction.
But what is the physical meanings of the negative index of
refraction? Although in the literature many researchers have
investigated the negative refractive index of left-handed media by
a variety of means of applied electromagnetism, classical optics,
materials science as well as condensed matter physics, less
attention than it deserves is paid to the physical meanings of
negative refractive index. In this section we will study this
fundamental problem from the purely physical point of view.

Before we consider the negative refractive index of left-handed
media, we first deal with the ``optical refractive index" problem
of de Broglie wave, the results of which will be helpful in
discussing the former problem.

For the case of de Broglie particle moving in a force field, we
can also take into account its ``optical refractive index" $n$.
According to the Einstein- de Broglie relation, the dispersion
relation of the de Broglie particle with rest mass $m_{0}$ in a
scalar potential field $V\left({\bf x}\right) $ agrees with
\begin{equation}
\left(\omega -\phi \right)
^{2}=k^{2}c^{2}+\frac{m_{0}^{2}c^{4}}{\hbar ^{2}}, \label{eeeeq1}
\end{equation}
where $\phi$ is defined to be $\phi=\frac{V}{\hbar }$, and $k$,
$\hbar$ and $c$ stand for the wave vector, Plank constant and
speed of light in a vacuum, respectively. It follows from
Eq.(\ref{eeeeq1}) that
\begin{equation}
\omega ^{2}\left[ 1-\frac{m_{0}^{2}c^{4}}{\hbar ^{2}\omega
^{2}}-2\frac{\phi }{\omega }+\left(\frac{\phi }{\omega }\right)
^{2}\right] =k^{2}c^{2},     \label{eeeeq2}
\end{equation}
which yields
\begin{equation}
\frac{\omega
^{2}}{k^{2}}=\frac{c^{2}}{1-\frac{m_{0}^{2}c^{4}}{\hbar ^{2}\omega
^{2}}-2\frac{\phi }{\omega }+\left(\frac{\phi }{\omega }\right)
^{2}}.   \label{eeeeq3}
\end{equation}
Compared Eq.(\ref{eeeeq3}) with the dispersion relation
$\frac{\omega ^{2}}{k^{2}}=\frac{c^{2}}{n^{2}}$, one can arrive at
\begin{equation}
n^{2}=1-\frac{m_{0}^{2}c^{4}}{\hbar ^{2}\omega ^{2}}-2\frac{\phi
}{\omega }+\left(\frac{\phi }{\omega }\right) ^{2}, \label{eeeeq4}
\end{equation}
which is the square of ``optical refractive index" of de Broglie
wave in the presence of a potential field $\phi $.

It is of physical interest to discuss the Fermat's principle of de
Broglie wave in a potential field $V$. One can readily verify that
the variation $\delta \int cn^{2}{\rm d}t=0$ can serve as the
mathematical expression for the Fermat's principle of de Broglie
particle (for the derivation of this expression, the readers may
be referred to the the paper\cite{Shenanti}). If, for example, by
using the weak-field and low-motion approximation where $\left|
\frac{\phi }{\omega }\right| \ll 1$, $\frac{v^{2}}{c^{2}}\ll 1$
and $\left(\frac{\phi }{\omega }\right) ^{2}$ in Eq.(\ref{eeeeq4})
can therefore be ignored, $1-\frac{m_{0}^{2}c^{4}}{\hbar
^{2}\omega ^{2}}\simeq \frac{v^{2}}{c^{2}}$, then $n^{2}$ is
approximately equal to
\begin{equation}
n^{2}\simeq \frac{2}{mc^{2}}\left(\frac{1}{2}mv^{2}-V\right).
\label{eeeeq5}
\end{equation}
Thus in the non-relativistic case, Fermat's principle, $\delta
\int cn^{2}{\rm d}t=0$, is equivalent to the following well-known
action principle $\delta \int L({\bf x},{\bf v}){\rm d}t=0$
(differing only by a constant coefficient $\frac{2}{mc^{2}}$),
where the function $L({\bf x},{\bf v})=\frac{1}{2}mv^{2}-V({\bf
x})$ denotes the Lagrangian of massive particle in a potential
field $V({\bf x})$. Hence the above formulation for the ``optical
refractive index" of the de Broglie particle inside a potential
field is said to be self-consistent.

According to Eq.(\ref{eeeeq4}), there are two square roots ({\it
i.e.}, positive and negative roots, $n_{+}$, $n_{-}$) for the
refractive index of de Broglie wave. It is well known that a
subatomic particle has its corresponding antiparticle that has the
same mass as it but opposite values of some other property or
properties. Indeed, it is verified that the particle and its
antiparticle possess these two square roots, respectively. If the
refractive index of particle is positive, then that of its
antiparticle will acquire a minus sign, which may be seen by
utilizing the charge conjugation transformation:
$\omega\rightarrow-\omega$, $\phi\rightarrow-\phi$. If, for
example, we take into consideration the refractive index of
electron and positron by regarding the positron as the negative
energy solution to Dirac's equation, then we can classify the
solutions of Dirac's equation into the following four cases:

(i) \quad  $E=E_{+}$, \quad $h=\hbar k$, \quad $n=n_{+}$;

(ii) \quad  $E=E_{+}$, \quad $h=-\hbar k$, \quad $n=n_{+}$;

(iii) \quad $E=E_{-}$, \quad $h=\hbar k$,\quad  $n=n_{-}$;

(iv) \quad   $E=E_{-}$,  \quad  $h=-\hbar k$, \quad   $n=n_{-}$,
\\
where $h$, $E_{+}$ and $E_{-}$ respectively represent the
helicity, positive and negative energy corresponding to the
electron and positron. It is apparent that the relationship
between the cases (i) and (iv) is just the charge conjugation
transformation. This transformation also relates the case (ii) to
(iii).

Compare the case of wave propagation in left-handed media with the
results presented above, we can conclude that the negative optical
refractive index in left-handed media corresponds to the
antiparticles of photons (the so-called {\it antiphotons}).
However, as is well known, there exist no {\it antiphotons} in
free space. The theoretical reason for this is that the
four-dimensional vector potentials $A_{\mu}$ with $\mu=0,1,2,3$
are always taking the real numbers. But in a dispersive and
absorptive medium, as an effective medium theory\cite{Nicorovici},
the vector potentials $A_{\mu}$ of which may probably take the
complex numbers. Such complex vector field theory has been
considered previously\cite{Lurie}. Here the Lagrangian density is
written ${\mathcal L}=-\frac{1}{2}F^{\ast}_{\mu\nu}F_{\mu\nu}$,
where the electromagnetic field tensors
$F_{\mu\nu}=\partial_{\mu}A_{\nu}-\partial_{\nu}A_{\mu}$,
$F^{\ast}_{\mu\nu}=\partial_{\mu}A^{\ast}_{\nu}-\partial_{\nu}A^{\ast}_{\mu}$
with $A^{\ast}_{\mu}$ being the complex conjugation of $A_{\mu}$.
The complex four-dimensional vector potentials $A_{\mu}$ and
$A^{\ast}_{\mu}$ characterize the propagating behavior of both
photons and antiphotons.

Now a problem left to us is that does the complex vector field
theory apply well in investigating the light propagation inside
the negative refractive index medium. Detailed analysis will show
that this theory is truly applicable to the consideration of
optical index of refraction and wave propagation in left-handed
media. This will be proved in what follows by using another
mathematical form
\begin{equation}
\frac{i}{c_{n}}\frac{\partial}{\partial t}{\bf M}=\nabla\times{\bf
M},  \quad       \nabla\cdot{\bf M}=0         \label{eeeeq6}
\end{equation}
for the Maxwellian equations, where ${\bf M}=\sqrt{\epsilon}{\bf
E}+i\sqrt{\mu}{\bf H}$, $c_{n}=1/\sqrt{\epsilon\mu}$, $\epsilon$
and $\mu$ are absolute dielectric constant and magnetic
conductivity of the medium, respectively. It can be easily
verified that the Maxwellian equations can be rewritten as
Eq.(\ref{eeeeq6}).

If $\epsilon>0$ and $\mu>0$ are assumed, then the light
propagation inside the isotropic linear right-handed medium
(regular medium) is governed by Eq.(\ref{eeeeq6}). For the
time-harmonic electromagnetic wave in the right-handed medium,
Eq.(\ref{eeeeq6}) can be rewritten
\begin{equation}
i\frac{\omega}{c_{n}}{\bf M}=-{\bf k}\times{\bf M} \label{eeeeq7}
\end{equation}
with $\omega$ being the frequency of electromagnetic wave.

Now let us take the complex conjugation of the two sides of
Eq.(\ref{eeeeq7}), and the result is of the form
\begin{equation}
i\frac{\omega}{c_{n}}{\bf M^{\ast}}={\bf k}\times{\bf M^{\ast}},
\label{eeeeq8}
\end{equation}
where the field ${\bf M^{\ast}}$ is expressed in terms of both
electric and magnetic fields, ${\bf E^{\ast}}$ and ${\bf
H^{\ast}}$, {\it i.e.}, ${\bf M^{\ast}}=\sqrt{\epsilon}{\bf
E^{\ast}}-i\sqrt{\mu}{\bf H^{\ast}}$. Thus, we have
\begin{equation}
i\frac{\omega}{c_{n}}[\sqrt{\epsilon}{\bf
E^{\ast}}-i\sqrt{\mu}{\bf H^{\ast}}]={\bf
k}\times[\sqrt{\epsilon}{\bf E^{\ast}}-i\sqrt{\mu}{\bf H^{\ast}}].
\label{eeeeq9}
\end{equation}
Multiplying the two sides of Eq.(\ref{eeeeq9}) by the imaginary
unit $i$, one can arrive at
\begin{equation}
i\frac{\omega}{c_{n}}[\sqrt{-\epsilon}{\bf
E^{\ast}}-i\sqrt{-\mu}{\bf H^{\ast}}]={\bf
k}\times[\sqrt{-\epsilon}{\bf E^{\ast}}-i\sqrt{-\mu}{\bf
H^{\ast}}], \label{eeeeq10}
\end{equation}
where the magnitude of $c_{n}$ ({\it i.e.},
$1/\sqrt{\epsilon\mu}$) does not change but it can be rewritten as
a new form $c_{n}=1/\sqrt{(-\epsilon)(-\mu)}$. According to
Eq.(\ref{eeeeq10}), one can obtain the following relations
\begin{equation}
{\bf k}\times{\bf E^{\ast}}=\omega(-\mu){\bf H^{\ast}},   \quad
{\bf k}\times{\bf H^{\ast}}=-\omega(-\epsilon){\bf E^{\ast}}.
\label{eeeeq11}
\end{equation}
Note that here $\epsilon>0$, $\mu>0$. Thus it follows from
Eq.(\ref{eeeeq11}) that the wave vector ${\bf {k}}$, the electric
field ${\bf {E^{\ast}}}$ and the magnetic field ${\bf {H^{\ast}}}$
of the electromagnetic wave in this medium, through which the
fields ${\bf E^{\ast}}$ and ${\bf H^{\ast}}$ propagate, form a
left-handed system. This, therefore, means that here the Poynting
vector ${\bf S}$ is pointed opposite to the direction of the wave
vector of electromagnetic wave. For this reason, it is expected
that the reversal of wave vector, instantaneous helicity
inversion\cite{Shenpla} and anomalous refraction will take place
when an incident lightwave from the right-handed medium travels to
the interfaces between left- and right- handed media. Since the
direction of phase velocity and energy flow of lightwave
propagating in the left-handed medium would be antiparallel, the
change of wave vector ${\bf k}$ will truly occur during the light
propagation through the interfaces. The inversion of ${\bf k}$
means that its magnitude acquires a minus sign, namely, the
optical index of refraction $n$ becomes negative since the
magnitude of ${\bf k}$ is defined to be $k=n\omega/c$. In view of
the above discussion, it is concluded that Eq.(\ref{eeeeq10})
governs the propagation of time-harmonic planar electromagnetic
wave, where the electromagnetic fields are characterized by ${\bf
E^{\ast}}$ and ${\bf H^{\ast}}$.

Thus by making use of the complex vector field theory we obtain a
formulation which can treat the wave propagation in both right-
and left- handed media. One of the advantages of the present
formulation is that the minus sign of negative refractive index
can be placed into the theoretical calculations by using the above
unified approach rather than {\it by hand}. Such unified approach
will be useful in dealing with the electromagnetic wave
propagation and photonic band gap structures in the so-called
left-handed medium photonic crystals, which are artificial
materials patterned with a periodicity in negative and positive
indices of refraction.
\section{Anti-shielding effect and negative temperature in left-handed media}
In linear media the electric and magnetic polarizations (per unit
volume) is respectively

\begin{equation}
{\bf P}=(\epsilon -1)\epsilon _{0}{\bf E}, \quad {\bf M}=(\mu
-1){\bf H}.
\end{equation}
The potential energy density of electric and magnetic dipoles in
electromagnetic fields reads

\begin{equation}
U_{e}=-{\bf P}\cdot {\bf E}=-(\epsilon -1)\epsilon _{0}{\bf
E}^{2},\quad U_{m}=-\mu _{0}{\bf M}\cdot {\bf H}=-(\mu -1)\mu
_{0}{\bf H}^{2}   \label{teqq51}
\end{equation}
with $\mu _{0}$ being the permeability of free space. It is
apparent that in left-handed media where both $\epsilon $ and $\mu
$ are negative, the electric and magnetic polarizations, ${\bf P}$
and ${\bf M}$, are opposite to the electric and magnetic fields,
${\bf E}$ and ${\bf H}$, thus in general, the potential energy
density, both $U_{e}$ and $U_{m}$, are positive, which differs
from the cases in the conventional media ({\it i.e.}, the
right-handed media). This, therefore, means that left-handed media
possess the anti-shielding effect, which results in many anomalous
electromagnetic and optical behaviors\cite{Klimov}, as is
mentioned above.

May the electric and magnetic dipole systems be brought into the
state of negative absolute temperature when the electromagnetic
wave is propagating in left-handed media? We show that under some
restricted conditions, these dipole systems may possess a negative
temperature indeed. The detailed analysis is given in what
follows:

According to Ramsey's claim\cite{Ramsey} that the essential
requirement for a thermodynamical system to be capable of a
negative temperature is that the elements of the thermodynamical
system must be in thermodynamical equilibrium among themselves in
order that the system can be described by a temperature at all.
For the left-handed media, this may be understood in the sense:
the electric and magnetic dipole moments produced by ALMWs and
SRRs should be coupled respectively to each other, and this
dipole-dipole interactions must be strong and therefore the
thermodynamical equilibriums between themselves may be brought
about in a short time, which is characterized by the relaxation
time $\tau _{1}$. But here both ALMWs and SRRs are macroscopic
elements of left-handed medium ({\it e.g.}, the length scale of
SRRs is mm), which implies that there exists a problem as to
whether the concept of thermodynamical equilibrium is applicable
to these macroscopic elements or not. This problem, however, is
not the subject of this section, and it will be published in a
separate paper; furthermore, if, for example, we can find a kind
of molecular-type SRRs ({\it i.e.}, the C- or $\rm {\Omega}$-
like\cite{Saadoun} curved organic molecule that possesses the
delocalized $\pi $ bond and can therefore provide a molecular
electric current induced by magnetic field of lightwave) for
manufacturing the left-handed media, then the dipole system of
this media can be appropriately described by a
temperature\cite{Ramsey}. So in our tentative consideration in
this paper, we can study the anti-shielding effect and negative
temperature of left-handed media from the phenomenological point
of view, ignoring the detailed information about the elemental
atoms which constitute the left-handed media.

In order to generate a state of negative absolute temperature, the
following requirement for a thermodynamical system should also be
satisfied: the system must be thermally isolated from all other
systems, namely, the thermal equilibrium time ($\tau _{1}$) among
the elements of the negative-temperature system must be short
compared to the time during which appreciable energy is lost to or
gained from other systems\cite{Ramsey}. In left-handed media, the
dipole-lattice interaction causes heat to flow from the dipole
system (negative temperature) to the lattice system (positive
temperature). If the relaxation time ($\tau _{2}$) during which
this interaction brings the dipole-lattice system into
thermodynamical equilibrium is much greater than $\tau _{1},$ then
it may be said that the dipole system is thermally isolated from
the lattice system, and the dipole system is therefore capable of
a negative temperature. In addition to this condition, there
exists another restriction, {\it i.e.}, the state of negative
temperature should be built up within half period of
electromagnetic wave propagating inside the left-handed medium,
{\it i.e.}, the relaxation time $\tau _{1}$ should be much less
than $\frac{1}{f}$ with $f$ being the frequency of electromagnetic
wave, which means that this thermodynamical process caused by
dipole-wave interaction can be considered a quasi-stationary
process and in consequence the state of negative temperature of
dipole system due to dipole-dipole interaction may adiabatically
achieve establishment in every instantaneous interval during one
period of electromagnetic wave. If this condition is not
satisfied, say, the relaxation time $\tau _{1}$ is several times
as long as $\frac{1}{f}$, then the negative temperature of dipole
systems cannot be achieved since the thermodynamical equilibrium
among dipoles induced by the oscillating electromagnetic fields
${\bf E}$ and ${\bf H}$ may lose its possibility, namely, the
concept of thermodynamical equilibrium is not applicable to this
non-stationary process; in fact, it should be treated by using
non-equilibrium statistical mechanics. In view of above
discussions, we conclude that the principal conditions for the
electric and magnetic dipole systems of left-handed media
possessing a negative temperature may be ascribed to the following
two essential requirements

\begin{equation}
\tau _{1}\ll \tau _{2},\quad \tau _{1}\ll \frac{1}{f}.
\end{equation}

Generally speaking, the first restriction, $\tau _{1}\ll \tau
_{2}$, is readily achieved, since in electromagnetic media the
dipole-dipole interaction are often much stronger than the
dipole-lattice interaction. For example, in some nuclear spin
systems such as the pure crystal of LiF where the negative
temperature was first realized in experiments, the relaxation time
$\tau _{1}$ of spin-spin process is approximately the period of
the Larmor precession of one nucleus in the field of its neighbor
and is of the order of $10^{-5}$ second while the relaxation time
that depends strongly upon the interaction between the spin system
and the crystal lattice is several minutes\cite{Ramsey}. In the
recent work of designing the structures of more effective
left-handed media, use is made of the SRRs of high electromagnetic
polarizability\cite{Smith}. This leads to the very strong
dipole-dipole interaction and may achieve the thermodynamical
equilibrium of dipole system with shorter relaxation time. If this
interaction is rather strong and therefore agrees with the second
restriction, $\tau _{1}\ll \frac{1}{f}$ with $f$ being in the
region of GHz\cite{Pendry2,Pendry1,Pendry3}, then the dipole
systems of left-handed media may be said to be in the state of
negative absolute temperature.

Since the most conspicuous feature of left-handed media may be the
electromagnetic anti-shielding effect, the previous considerations
concerning the superluminal light propagation in the instantly
varying electric fields based on the viewpoint of effective rest
mass can also be applied to the incident electromagnetic wave
propagating inside left-handed media. As is stated, in left-handed
media many electromagnetic effects such as Doppler shift,
Cherenkov radiation and wave refraction are inverted compared with
those in ordinary right-handed media. In the similar fashion, it
is believed that this superluminal light propagation may also
probably occur in left-handed media. Moreover, in some
electromagnetic materials, there exists another superluminal pulse
propagation that also results from the anti-shielding
effect\cite{Ziolkowski3,Ziolkowski4} (but rather than from the
statistical fluctuations in negative temperature), since in this
media the designed equivalent permittivity and permeability based
on the so-called two-time-derivative Lorentz model are
simultaneously smaller than the values in free
space\cite{Ziolkowski3,Ziolkowski4}. Ziolkowski has studied the
superluminal pulse propagation and consequent superluminal
information exchange in this media, and demonstrated that they do
not violate the principle of causality\cite {Ziolkowski}. We
suggest that all these potential superluminality effects of light
propagation in anti-shielding media should be further
investigated, particularly in experiments.

\section{Frequency-independent effective rest mass of photons in the 2TDLM model}
\subsection{Extracting frequency-independent effective mass from
refractive index squared}

In this framework of definition of frequency-independent effective
rest mass, the photon should be regarded as acted upon by a
hypothetical force field, namely, it behaves like a massive de
Broglie particle moving in a force field. It has been verified
previously, for the case of de Broglie particle in a force field,
one can also consider its ``optical refractive index" $n$.

In what follows, an approach to frequency-independent effective
mass extracted from the optical refractive index squared
$n^{2}\left(\omega\right) $ is presented. It is assumed that
$n^{2}\left(\omega\right) $ can be rewritten as the following
series expansions
\begin{equation}
n^{2}\left(\omega \right) =\sum_{k}\frac{a_{-k}}{\omega ^{k}},
\end{equation}
then one can arrive at $n^{2}\left(\omega \right) \omega
^{2}=\sum_{k}\frac{a_{-k}}{\omega ^{k-2}}$, namely,
\begin{equation}
n^{2}\left(\omega \right) \omega ^{2}= ...+\frac{a_{-3}}{\omega
}+a_{-2}+a_{-1}\omega +a_{0}\omega ^{2}+a_{+1}\omega ^{3}+...\quad
. \label{jeq37}
\end{equation}
Compared (\ref{jeq37}) with (\ref{eeeeq4}), it is apparent that
the frequency-independent effective rest mass squared $m_{\rm
eff}^{2}$ of the photon can be read off from $n^{2}\left(\omega
\right)$, namely, the constant term on the right-handed side on
(\ref{jeq37}) is related only to $m_{\rm eff}^{2}$, {\it i.e.},
\begin{equation}
a_{-2}=-\frac{m_{\rm eff}^{2}c^{4}}{\hbar ^{2}}.
\end{equation}

Thus it follows from Eq.(\ref{eeeeq4}) that after we extract the
term $\frac{{\rm const.}}{\omega^{2}}$, the retained terms
${\mathcal R}$ in $n^{2}(\omega)-1$ belongs to $-2\frac{\phi
}{\omega }+\left(\frac{\phi }{\omega }\right) ^{2}$. If the
effective potential $\phi$ can be rewritten in a form
$\phi=\sum_{k}\phi_{k}\omega^{k}$ with $\phi_{k}$ being
$\omega$-independent, then one can obtain the expansion
coefficients $\phi_{k}$ of $\phi$. It should be noted that here
the terms in $-2\frac{\phi }{\omega }+\left(\frac{\phi }{\omega
}\right) ^{2}$ that takes the form $\frac{{\rm
const.}}{\omega^{2}}$ has already been involved in
$\frac{a_{-2}}{\omega^{2}}$. We will clarify further this point in
what follows. It is apparently seen that by using the form of
series expansion of $\phi$, the terms $-2\frac{\phi }{\omega }$
and $\left(\frac{\phi }{\omega }\right) ^{2}$ in Eq.(\ref{eeeeq4})
can be rewritten
\begin{equation}
-2\frac{\phi }{\omega}=-2\frac{\phi_{-1}}{\omega^{2}}-2\sum_{k\neq
-1}\phi_{k}\omega^{k-1}                 \label{jeqq39}
\end{equation}
and
\begin{equation}
\left(\frac{\phi }{\omega }\right)
^{2}=\left(\frac{\phi_{0}^{2}+2\sum_{k>0}\phi_{k}\phi_{-k}}{\omega^{2}}\right)+\sum_{k\neq
0 }\phi^{2}_{k}\omega^{2k-2}+2\sum_{k\neq -l,
k>l}\phi_{k}\phi_{l}\omega^{k+l-2}.             \label{jeqq40}
\end{equation}
Note that the coefficient factors of $\omega^{-2}$, {\it i.e.},
$-2\phi_{-1}$ in Eq.(\ref{jeqq39}) and
$\phi_{0}^{2}+2\sum_{k>0}\phi_{k}\phi_{-k}$ in Eq.(\ref{jeqq40})
contributes to the photon effective rest mass, that is, it is
difficult to distinguish the effect of these coefficient factors
from the effective rest mass. So we can attribute the effective
rest mass term to the coefficient factors of $\omega^{-2}$,
namely, $a_{-2}$ contains
$-2\phi_{-1}+\phi_{0}^{2}+2\sum_{k>0}\phi_{k}\phi_{-k}$. After the
term $\frac{{\rm const.}}{\omega^{2}}$ is extracted from
$n^{2}(\omega)-1$, the retained term ${\mathcal R}$ equals the
summation of those terms in Eq.(\ref{jeqq39}) and (\ref{jeqq40})
that does not take the form $\frac{{\rm const.}}{\omega^{2}}$,
{\it i.e.},
\begin{equation}
{\mathcal R}=-2\sum_{k\neq -1}\phi_{k}\omega^{k-1}+\sum_{k\neq 0
}\phi^{2}_{k}\omega^{2k-2}+2\sum_{k\neq -l,
k>l}\phi_{k}\phi_{l}\omega^{k+l-2}.               \label{jeqq41}
\end{equation}

To conclude, both the photon effective rest mass $m_{\rm eff}$ and
the effective potential $\phi$ are defined in the above
formulation. In a free space, since $n^{2}-1=0$ and the vacuum
contributes no effective mass to photons. However, in a dispersive
medium, if $n^{2}(\omega)-1$ is regarded as the contribution of
the interaction between light and medium, then the term
$\frac{{\rm const.}}{\omega^{2}}$ in $n^{2}(\omega)-1$ will result
in an $\omega$-independent photon effective rest mass. Other terms
retained in $n^{2}(\omega)-1$ can be ascribed to the effective
potential $\phi$, which is so defined that its expansion
coefficients $\phi_{k}$ satisfy Eq.(\ref{jeqq41}).

Generally speaking, the optical refractive index $n^{2}\left(
\omega \right) $ is often of complicated form, particularly for
the metamaterials. So, we cannot extract frequency-independent
$m_{\rm eff}^{2}$ immediately from $n^{2}\left(\omega \right) $.
In the following, this problem is resolved by calculating the
limit value of $n^{2}\left(\omega \right) \omega ^{2}$, where
$\omega $ tends to $\infty $ or zero. We first consider the large
frequency behavior of $n^{2}\left(\omega \right) \omega ^{2}$. It
follows from (\ref{jeq37}) that $n^{2}\left(\omega \right) \omega
^{2}$ at high frequencies ({\it i.e.}, $\omega \rightarrow \infty
$) is of the form
\begin{equation}
n^{2}\left(\omega \right) \omega ^{2}{\rightarrow
}a_{-2}+a_{-1}\omega +a_{0}\omega ^{2}+a_{+1}\omega
^{3}+a_{+2}\omega ^{4}+...\quad .
\end{equation}
This, therefore, means that $n^{2}\left(\omega \right) \omega
^{2}$ at high frequencies ($\omega \rightarrow \infty $) can be
split into the following two terms
\begin{equation}
n^{2}\left(\omega \right) \omega ^{2}{\rightarrow }\left({\rm
constant \quad term}\right) +\left({\rm divergent \quad
term}\right).
\end{equation}
So, we can obtain $m_{\rm eff}^{2}$ from the following formula
\begin{equation}
\lim_{\omega \rightarrow \infty }\left[ {n^{2}\left(\omega \right)
\omega ^{2}-\left({\rm divergent \quad
term}\right)}\right]=-\frac{m_{\rm eff}^{2}c^{4}}{\hbar ^{2}}.
\label{jeq311}
\end{equation}
Note that the constant term in $n^{2}\left(\omega \right) \omega
^{2}$ can also be extracted by considering the low-frequency
(zero-frequency) behavior of $n^{2}\left(\omega \right) $. It is
certain that in this case the obtained constant term is the same
as (\ref{jeq311}). Apparently $n^{2}\left(\omega \right) \omega
^{2}$ at low-frequencies ({\it i.e.}, $\omega \rightarrow 0$)
behaves like
\begin{equation}
n^{2}\left(\omega \right) \omega ^{2}\rightarrow
...+\frac{a_{-5}}{\omega ^{3}}+\frac{a_{-4}}{\omega
^{2}}+\frac{a_{-3}}{\omega }+a_{-2}, \label{jeq312}
\end{equation}
where the sum term ($...+\frac{a_{-5}}{\omega
^{3}}+\frac{a_{-4}}{\omega ^{2}}+\frac{a_{-3}}{\omega }$) is
divergent if $\omega$ approaches zero. Thus, subtraction of the
divergent term from $n^{2}\left(\omega \right) \omega ^{2}$ yields
\begin{equation}
\lim_{\omega \rightarrow 0}\left[ n^{2}\left(\omega \right) \omega
^{2}-\left({\rm divergent \quad term}\right) \right]
=-\frac{m_{\rm eff}^{2}c^{4}}{\hbar ^{2}}.       \label{jeq313}
\end{equation}

In the next section, we will calculate the frequency-independent
effective rest mass of photons in the {\it two time derivative
Lorentz material} (2TDLM) model\cite{Ziolkowski3,Ziolkowski4} by
making use of the method presented above.

\subsection{Frequency-independent
effective rest mass of photons in 2TDLM model}

One of the metamaterials models proposed by Ziolkowski and
Auzanneau is called the {\it two time derivative Lorentz material}
(2TDLM) model\cite{Ziolkowski3,Ziolkowski4}, which is a
generalization of the standard Lorentz material model and
encompasses the permittivity and permeability material responses
experimentally obtained, which may be seen, for example, in the
reference of Smith {\it et al.}\cite{Smith}. According to the
definition of the 2TDLM model, the frequency-domain electric and
magnetic susceptibilities are given\cite{Ziolkowski}
\begin{equation}
\chi ^{e}\left(\omega \right)=\frac{\left(\omega _{\rm
p}^{e}\right) ^{2}\chi _{\alpha }^{e}+i\omega \omega _{\rm
p}^{e}\chi _{\beta }^{e}-\omega ^{2}\chi _{\gamma }^{e}}{-\omega
^{2}+i\omega \Gamma ^{e}+\left(\omega _{0}^{e}\right) ^{2}}, \quad
\chi ^{m}\left(\omega \right)=\frac{\left(\omega _{\rm
p}^{m}\right) ^{2}\chi _{\alpha }^{m}+i\omega \omega _{\rm
p}^{m}\chi _{\beta }^{m}-\omega ^{2}\chi _{\gamma }^{m}}{-\omega
^{2}+i\omega \Gamma ^{m}+\left(\omega _{0}^{m}\right) ^{2}},
\label{jeq41}
\end{equation}
where $\chi _{\alpha }^{e,m}$, $\chi _{\beta }^{e,m}$ and $\chi
_{\gamma }^{e,m}$ represent, respectively, the coupling of the
electric (magnetic) field and its first and second time
derivatives to the local electric (magnetic) dipole motions.
$\omega _{\rm p}$, $\Gamma ^{e}$, $\Gamma ^{m}$ and $\omega _{0}$
can be viewed as the plasma frequency, damping frequency and
resonance frequency of the electric (magnetic) dipole oscillators,
respectively. So, the refractive index squared in the 2TDLM model
reads
\begin{equation}
n^{2}\left(\omega \right) =1+\chi ^{e}\left(\omega \right) +\chi
^{m}\left(\omega \right) +\chi ^{e}\left(\omega \right) \chi
^{m}\left(\omega \right).                \label{jeq42}
\end{equation}
In order to obtain $m_{\rm eff}^{2}$ from $n^{2}\left(\omega
\right)$, according to the formulation in the previous section,
the high-frequency ($\omega \rightarrow \infty $) behavior of
$\chi ^{e}\left(\omega \right) \omega ^{2}$ should first be taken
into consideration and the result is
\begin{equation}
\chi ^{e}\left(\omega \right) \omega ^{2}\rightarrow -i\omega
\omega _{\rm p}^{e}\chi _{\beta }^{e}+\omega ^{2}\chi _{\gamma
}^{e}.                   \label{jeq43}
\end{equation}
Note that $-i\omega \omega _{\rm p}^{e}\chi _{\beta }^{e}+\omega
^{2}\chi _{\gamma }^{e}$ is the divergent term of $\chi ^{e}\left(
\omega \right) \omega ^{2}$ at large frequencies. Subtracting the
divergent term from $\chi ^{e}\left(\omega \right) \omega ^{2}$,
we consider again the high-frequency behavior ($\omega \rightarrow
\infty $) and obtain another divergent term $i\omega ^{3}\Gamma
^{e}\chi _{\gamma }^{e}$, {\it i.e.},
\begin{eqnarray}
& & \chi ^{e}\left(\omega \right) \omega ^{2}-\left(-i\omega
\omega _{\rm p}^{e}\chi _{\beta }^{e}+\omega ^{2}\chi _{\gamma
}^{e}\right)                                      \nonumber  \\
&=& \frac{\omega ^{2}\left(\omega _{\rm p}^{e}\right) ^{2}\chi
_{\alpha }^{e}-\omega ^{2}\omega _{\rm p}^{e}\Gamma ^{e}\chi
_{\beta }^{e}+i\omega \omega _{\rm p}^{e}\left(\omega
_{0}^{e}\right) ^{2}\chi _{\beta }^{e}-i\omega ^{3}\Gamma ^{e}\chi
_{\gamma }^{e}-\omega ^{2}\left(\omega _{0}^{e}\right) ^{2}\chi
_{\gamma }^{e}}{-\omega ^{2}+i\omega \Gamma ^{e}+\left(\omega
_{0}^{e}\right) ^{2}}\rightarrow i\omega ^{3}\Gamma ^{e}\chi
_{\gamma }^{e}.                 \label{jeq44}
\end{eqnarray}
Thus, after subtracting all the divergent terms from $\chi
^{e}\left(\omega \right) \omega ^{2}$, we obtain
\begin{eqnarray}
&& \lim_{\omega \rightarrow \infty }\left[ \chi ^{e}\left(\omega
\right) \omega ^{2}-\left(-i\omega \omega _{\rm p}^{e}\chi _{\beta
}^{e}+\omega ^{2}\chi _{\gamma }^{e}\right) -i\omega
^{3}\Gamma ^{e}\chi _{\gamma }^{e}\right]           \nonumber   \\
&=&-\left[ \left(\omega _{\rm p}^{e}\right) ^{2}\chi _{\alpha
}^{e}-\omega _{\rm p}^{e}\Gamma ^{e}\chi _{\beta }^{e}-\left(
\omega _{0}^{e}\right) ^{2}\chi _{\gamma }^{e}+\left(\Gamma
^{e}\right) ^{2}\chi _{\gamma }^{e}\right]. \label{jeq45}
\end{eqnarray}
In the same fashion, we obtain
\begin{eqnarray}
& & \lim_{\omega \rightarrow \infty }\left[ \chi ^{m}\left(\omega
\right) \omega ^{2}-\left(-i\omega \omega _{\rm p}^{m}\chi _{\beta
}^{m}+\omega ^{2}\chi _{\gamma }^{m}\right) -i\omega
^{3}\Gamma ^{m}\chi _{\gamma }^{m}\right]         \nonumber   \\
&=&-\left[ \left(\omega _{\rm p}^{m}\right) ^{2}\chi _{\alpha
}^{m}-\omega _{\rm p}^{m}\Gamma ^{m}\chi _{\beta }^{m}-\left(
\omega _{0}^{m}\right) ^{2}\chi _{\gamma }^{m}+\left(\Gamma
^{m}\right) ^{2}\chi _{\gamma }^{m}\right] \label{jeq46}
\end{eqnarray}
for the magnetic susceptibility $\chi ^{m}\left(\omega \right) $.
In what follows we continue to consider the term $\chi ^{e}\left(
\omega \right) \chi ^{m}\left(\omega \right) $ on the right-handed
side of (\ref{jeq42}). In the similar manner, the
infinite-frequency limit of $\chi ^{e}\left(\omega \right) \chi
^{m}\left(\omega \right) \omega ^{2}$ is given as follows
\begin{eqnarray}
\lim_{\omega \rightarrow \infty }\chi ^{e}\left(\omega \right)
\chi ^{m}\left(\omega \right) \omega ^{2}&=&-\chi _{\gamma
}^{m}\left[ \left(\omega _{\rm p}^{e}\right) ^{2}\chi _{\alpha
}^{e}-\omega _{\rm p}^{e}\Gamma ^{e}\chi _{\beta }^{e}-\left(
\omega _{0}^{e}\right) ^{2}\chi _{\gamma }^{e}+\left(\Gamma
^{e}\right)
^{2}\chi _{\gamma }^{e}\right]                               \nonumber \\
&-&\chi _{\gamma }^{e}\left[ \left(\omega _{\rm p}^{m}\right)
^{2}\chi _{\alpha }^{m}-\omega _{\rm p}^{m}\Gamma ^{m}\chi _{\beta
}^{m}-\left(\omega _{0}^{m}\right) ^{2}\chi _{\gamma }^{m}+\left(
\Gamma ^{m}\right) ^{2}\chi _{\gamma }^{m}\right]. \label{jeq47}
\end{eqnarray}
Hence, insertion of (\ref{jeq45})-(\ref{jeq47}) into
(\ref{jeq311}) yields
\begin{eqnarray}
\frac{m_{\rm eff}^{2}c^{4}}{\hbar ^{2}}&=&\left(1+\chi _{\gamma
}^{m}\right) \left[ \left(\omega _{\rm p}^{e}\right) ^{2}\chi
_{\alpha }^{e}-\omega _{\rm p}^{e}\Gamma ^{e}\chi _{\beta
}^{e}-\left(\omega _{0}^{e}\right) ^{2}\chi _{\gamma }^{e}+\left(
\Gamma ^{e}\right) ^{2}\chi _{\gamma
}^{e}\right]                \nonumber \\
&+&\left(1+\chi _{\gamma }^{e}\right)\left[ \left(\omega _{\rm
p}^{m}\right) ^{2}\chi _{\alpha }^{m}-\omega _{\rm p}^{m}\Gamma
^{m}\chi _{\beta }^{m}-\left(\omega _{0}^{m}\right) ^{2}\chi
_{\gamma }^{m}+\left(\Gamma ^{m}\right) ^{2}\chi _{\gamma
}^{m}\right].                \label{jeq48}
\end{eqnarray}
Thus the $\omega$-independent effective rest mass squared $m_{\rm
eff}^{2}$ is extracted from $n^{2}\left(\omega \right)$.
Apparently, $m_{\rm eff}^{2}$ in Eq.(\ref{jeq48}) does not depend
on $\omega$. The remainder on the right-handed side of Eq.
(\ref{eeeeq4}) is $n^{2}\left(\omega \right) -1+\frac{m_{\rm
eff}^{2}c^{4}}{\hbar ^{2}\omega ^{2}}$, which is equal to $\left(
\frac{\phi }{\omega }\right) ^{2}-2\frac{\phi }{\omega }$. For
simplicity, we set
\begin{equation}
\Delta \left(\omega \right) =n^{2}\left(\omega \right)
-1+\frac{m_{\rm eff}^{2}c^{4}}{\hbar ^{2}\omega ^{2}},
\label{jeq4111}
\end{equation}
and then from the equation
\begin{equation}
\left(\frac{\phi }{\omega }\right) ^{2}-2\frac{\phi }{\omega
}=\Delta \left(\omega \right),
\end{equation}
one can readily obtain the expression for the hypothetical
potential field $\phi \left(\omega \right)$ is $\phi \left(\omega
\right) =\omega \left[ 1- \sqrt{1+\Delta \left(\omega \right)
}\right]$, or
\begin{equation}
V\left(\omega \right) =\hbar \omega \left[ 1- \sqrt{1+\Delta
\left(\omega \right) }\right],  \label{jeq411}
\end{equation}
where $\phi =\frac{V}{\hbar }$.

Many metamaterials which have complicated expressions for electric
and magnetic susceptibilities such as (\ref{jeq41}) are generally
rarely seen in nature and often arises from the artificial
manufactures. In the following, we will consider the photon
effective rest mass in one of the artificial composite
metamaterials, {\it i.e.}, the left-handed medium. In general, the
dielectric parameter $ \epsilon \left(\omega \right)$ and magnetic
permeability $\mu \left(\omega \right)$ in the isotropic
homogeneous left-handed medium are of the form
\begin{equation}
 \epsilon \left(\omega \right)
=1-\frac{\omega _{\rm p}^{2}}{\omega \left(\omega +i\gamma \right)
},\quad \mu \left(\omega \right) =1-\frac{F\omega ^{2}}{\omega
^{2}-\omega _{0}^{2}+i\omega \Gamma } \label{jeq412}
\end{equation}
with the plasma frequency $\omega _{\rm p}$ and the magnetic
resonance frequency $\omega _{0}$ being in the GHz region.
$\gamma$ and $\Gamma$ stand for the electric and magnetic damping
parameters, respectively. Compared (\ref{jeq412}) with the
electric and magnetic susceptibilities (\ref{jeq41}), one can
arrive at
\begin{eqnarray}
\chi _{\alpha }^{e}=1, \quad \omega _{\rm p}^{e}=\omega _{\rm p},
\quad \chi _{\beta }^{e}=\chi _{\gamma }^{e}=0, \quad \omega
_{0}^{e}=0, \quad  \Gamma ^{e}=-\gamma
,                                         \nonumber \\
 \chi _{\gamma }^{m}=-F, \quad \chi _{\alpha }^{m}=\chi _{\beta
}^{m}=0, \quad \omega _{0}^{m}=\omega _{0}, \quad \Gamma
^{m}=-\Gamma.    \label{jeq413}
\end{eqnarray}
So, it is readily verified with the help of (\ref{jeq48}) that the
$\omega$-independent effective rest mass of photons inside
left-handed media is written as follows
\begin{equation}
\frac{m_{\rm eff}^{2}c^{4}}{\hbar ^{2}}=\left(1-F\right) \omega
_{\rm p}^{2}+F\left(\omega _{0}^{2}-\Gamma ^{2}\right).
\label{jeq414}
\end{equation}

It follows that Eq.(\ref{jeq414}) is a restriction on the
electromagnetic parameters $F, \omega _{\rm p},  \omega _{0},
\Gamma $ and $F$ in the electric permittivity and magnetic
permeability (\ref{jeq412}), since $m_{\rm eff}^{2} \geq 0$.
Insertion of the experimentally chosen values of the
electromagnetic parameters in the literature into Eq.
(\ref{jeq414}) shows that this restriction condition is satisfied.
For instance, in Ruppin's work\cite{Ruppin}, $F, \omega _{\rm p},
\omega _{0}, \gamma, \Gamma $ and $F$ were chosen $F=0.56$,
$\omega _{\rm p}=10.0 {\rm GHz}$, $\omega _{0}=4.0 {\rm GHz}$,
$\gamma=0.03\omega _{\rm p}$, $ \Gamma=0.03\omega _{0}$, which
indicates that the inequality $m_{\rm eff}^{2} \geq 0$ is
satisfied.

\subsection{Some applications of frequency-independent
effective rest mass of photons}

In this section we consider the applications of the concept of
frequency-independent effective rest mass of photons to some
optical and electromagnetic phenomena and effects.

(i) As was stated above, since the frequency-independent effective
rest mass of photons in electromagnetic media is related close to
the coupling parameters of the electric (magnetic) fields to the
electric (magnetic) dipole systems, and the damping and resonance
frequency of the electric (magnetic) dipole oscillators, it
contains the information on the interaction between light and
materials. It is shown that in certain frequency ranges this
effective rest mass governs the propagation behavior of
electromagnetic wave in media, and by using this concept one can
therefore consider the wave propagation somewhat conveniently. For
example, in the left-handed media whose permittivity and
permeability are expressed by (\ref{jeq412}), the potential field
$V(\omega)$ approaches
$-\hbar\sqrt{F(\omega_{0}^{2}-\omega^{2}_{\rm p})}$, which is
constant, as the frequency tends to zero. This, therefore, means
that at low frequencies the light propagation is governed mainly
by the frequency-independent effective rest mass of photons.

Likewise, in the case of the potential field $V(\omega)$ being
vanishing where the frequency squared of electromagnetic wave is
\begin{equation}
\Omega^{2}=\frac{\omega^{2}_{0}\pm \sqrt{4\omega^{2}_{\rm
p}\omega_{0}^{2}-3\omega_{0}^{4}}}{2},       \label{j51}
\end{equation}
the photon can be regarded as a ``free'' massive particle with the
rest mass being expressed in (\ref{jeq414}). In the frequency
region near $\Omega$, sine the photon is a ``quasi-free''
particle, the light propagation behavior and properties is rather
simple compared with other frequency ranges in which $V(\omega)$
is nonvanishing or large. For the left-handed media, the relevant
frequency parameters $\omega_{0}$ and $\omega_{\rm p}$ of which is
about GHZ, the frequency $\Omega$ in Eq.(\ref{j51}) that leads to
$V(\Omega)=0$ is just in the frequency band (GHZ) where the
permittivity and permeability are simultaneously negative. It
follows that when the effects and phenomena associated with
negative index of refraction occur in left-handed media, the wave
propagation ({\it e.g.}, scattering properties) and dispersion
properties of the light are not as complicated as those in other
frequency ranges, since in the negative-index frequency band, the
optical refractive index squared simulates nearly the plasma
behavior, particularly in the region near $\Omega$.

(ii) Historically, possibility of a light pulse with speed greater
than that ($c$) in a vacuum has been extensively investigated by
many authors\cite
{Ziolkowski,Akulshin,Bolda,Mitchell,Zhou,Nimtz1,Wang}. Ziolkowski
has studied the superluminal pulse propagation and consequent
superluminal information exchange in the 2TDLM media, and
demonstrated that they do not violate the principle of
causality\cite {Ziolkowski}. He showed that in the 2TDLM model,
both the phase and group velocities, {\it i.e.},
\begin{equation}
\lim_{\omega\rightarrow \infty}v(\omega)\sim
\frac{c}{|1+\chi_{\gamma}|}
\end{equation}
of light at high frequencies will exceed the speed ($c$) of light
in a vacuum, so long as $-2<\chi_{\gamma}<0$, where
$\chi_{\gamma}$ denotes the coupling coefficients of the second
time derivatives of electric (magnetic) fields to the local
electric (magnetic) dipole motions\cite {Ziolkowski}.

Although in the 2TDLM model the medium will exhibit a possibility
of superluminal speeds of wave propagation, the photon velocity
(rather than group velocity of light) is by no means larger than
$c$. This may be explained as follows: according to
Eq.(\ref{jeq411}), the kinetic energy of a photon is ${\mathcal
E}=\hbar \omega\sqrt{1+\Delta(\omega)}$. If the light frequency
$\omega$ tends to infinity, then it follows from
Eq.(\ref{jeq4111}) that ${\mathcal E}\rightarrow n\hbar\omega$.
Since the photon momentum $p=\frac{n\hbar\omega}{c}$, the velocity
$v$ of the photon with a frequency-independent rest mass
approaches $c$, {\it i.e.}, the particle (photon) velocity
$v=\frac{p}{{\mathcal E}}\rightarrow c$. Thus we show that in the
artificial composite metamaterials such as 2TDLM media the photon
velocity does not exceed the speed of light in a vacuum, no matter
whether the phase (and group) velocity of light inside it is a
superluminal speed or not.

(iii) The frequency-independent rest mass of photons is useful in
discussing the plus and minus signs of phase shifts in light
scattering inside media. The phase shifts and its signs contain
the information on the scattering between wave and potentials.
According to Eq.(\ref{jeq411}), if $\Delta(\omega)$ is positive
(negative), then the interaction energy $V(\omega)$ between light
and media is negative (positive). For this reason, it is possible
for us to determine the signs (plus or minus) of phase shifts in
light scattering process by calculating $\Delta(\omega)$ and the
frequency-independent rest mass $m_{\rm eff}$. Now with the
development in design and fabrication of left-handed materials,
the zero-index ($\epsilon\sim 0, \mu\sim 0$) materials receive
attention in materials science and applied
electromagnetism\cite{Hu}. For this kind of media with the optical
refractive index $n\sim 0$, from Eq.(\ref{jeq4111}) it follows
that
\begin{equation}
\Delta(\omega)\simeq \frac{m_{\rm eff}^{2}c^{4}}{\hbar ^{2}\omega
^{2}}-1,
\end{equation}
{\it i.e.}, $\Delta(\omega)$ is approximately equal to
$\frac{m_{\rm eff}^{2}c^{4}}{\hbar ^{2}\omega ^{2}}-1$ and we can
determine the sign by comparing the light frequency with the
photon effective rest mass $m_{\rm eff}$. This, therefore, means
that if the frequency $\omega$ is larger than $\frac{m_{\rm
eff}c^{2}}{\hbar}$, then $\Delta<0$ and consequently the phase
shift of light wave due to scattering acquires a minus sign; and
if the light frequency $\omega$ is less than $\frac{m_{\rm
eff}c^{2}}{\hbar}$, then $\Delta>0$ and consequently the phase
shift acquires a plus sign.

\section{Scattering problem of double-layered sphere containing left-handed media}

The extinction properties of a sphere (single-layered) with
negative permittivity and permeability is investigated by Ruppin.
Since recently Wang and Asher developed a novel method to
fabricate nanocomposite ${\rm SiO}_{2}$ spheres ($\sim 100$ nm)
containing homogeneously dispersed Ag quantum dots ($2\sim 5$ nm
)\cite{Wang}, which may has potential applications to design and
fabrication of photonic crystals, it is believed that the
absorption and transmittance of double-layered sphere deserves
consideration. In this paper, both the Mie coefficients and the
electromagnetic field distributions in the two-layered sphere
 containing left-handed media irradiated by a planar
 electromagnetic wave are presented.
\subsection{Methods}
It is well known that in the absence of electromagnetic sources,
the characteristic vectors such as ${\bf E}$, ${\bf B}$, ${\bf
D}$, ${\bf H}$ and Hertz vector in the isotropic homogeneous media
agree with the same differential equation
\begin{equation}
\nabla\nabla\cdot{\bf C}-\nabla\times\nabla\times{\bf C}+k^{2}{\bf
C}=0.
\end{equation}
The three independent vector solutions to the above equation is
\cite{Stratton}
\begin{equation}
{\bf L}=\nabla\psi,   \quad   {\bf M}=\nabla\times{\bf a}\psi,
\quad             {\bf N}=\frac{1}{k}\nabla\times{\bf M}
\end{equation}
with ${\bf a}$ being a constant vector, where the scalar function
$\psi$ satisfies $\nabla^{2}\psi+k^{2}\psi=0$. It is verified that
the vector solutions ${\bf M}$, ${\bf N}$ and ${\bf L}$ possess
the following mathematical properties
\begin{equation}
{\bf M}={\bf L}\times{\bf a}=\frac{1}{k}\nabla\times{\bf N}, \quad
{\bf L}\cdot{\bf M}=0,    \quad   \nabla\times{\bf L}=0,    \quad
\nabla\cdot{\bf L}=\nabla^{2}\psi=-k^{2}\psi, \quad
\nabla\cdot{\bf M}=0,  \quad       \nabla\cdot{\bf N}=0.
\end{equation}
Set ${\bf M}={\bf m}\exp(-i\omega t)$ and ${\bf N}={\bf
n}\exp(-i\omega t)$, the vector wave functions ${\bf M}$ and ${\bf
N}$ can be expressed in terms of the following spherical vector
wave functions\cite{Stratton}
\begin{eqnarray}
{\bf m}_{^{e}_{o}mn}&=&\mp\frac{m}{\sin
\theta}z_{n}(kr)P^{m}_{n}(\cos \theta){^{\sin}_{\cos}}m\phi{\bf
i}_{2}-z_{n}(kr)\frac{\partial P^{m}_{n}(\cos \theta)}{\partial
\theta}{^{\cos}_{\sin}}m\phi{\bf i}_{3},    \nonumber   \\
{\bf n}_{^{e}_{o}mn}&=&\frac{n(n+1)}{kr}z_{n}(kr)P^{m}_{n}(\cos
\theta){^{\cos}_{\sin}}m\phi{\bf
i}_{1}+\frac{1}{kr}\frac{\partial}{\partial
r}[rz_{n}(kr)]\frac{\partial}{\partial \theta}P^{m}_{n}(\cos
\theta){^{\cos}_{\sin}}m\phi{\bf i}_{2}
                                       \nonumber   \\
&\mp&\frac{m}{kr\sin \theta }\frac{\partial}{\partial
r}[rz_{n}(kr)]P^{m}_{n}(\cos \theta){^{\sin}_{\cos}}m\phi{\bf
i}_{3}.
\end{eqnarray}
In what follows we treat the Mie coefficients of double-layered
sphere irradiated by a plane wave.

\subsubsection{Definitions}
We consider a double-layered sphere with interior radius $a_{1}$
and external radius $a_{2}$ having relative permittivity
(permeability) $\epsilon_{1}$ ($\mu_{1}$) and $\epsilon_{2}$ (
$\mu_{2}$), respectively, placed in a medium having the relative
permittivity $\epsilon_{0}$ and permeability $\mu_{0}$. Suppose
that the double-layered sphere irradiated by the following plane
wave with the electric amplitude $E_{0}$ along the \^{z}-direction
of Cartesian coordinate system\cite{Stratton}
\begin{eqnarray}
{\bf E}_{\rm i}&=&{\bf a}_{x}E_{0}\exp(ik_{0}z-i\omega
t)=E_{0}\exp(-i\omega
t)\sum^{\infty}_{n=1}i^{n}\frac{2n+1}{n(n+1)}\left({\bf
m}^{(1)}_{o1n}-i{\bf n}^{(1)}_{e1n}\right),  \nonumber \\
{\bf H}_{\rm i}&=&{\bf
a}_{y}\frac{k_{0}}{\mu_{0}\omega}E_{0}\exp(ik_{0}z-i\omega
t)=-\frac{k_{0}}{\mu_{0}\omega}E_{0}\exp(-i\omega
t)\sum^{\infty}_{n=1}i^{n}\frac{2n+1}{n(n+1)}\left({\bf
m}^{(1)}_{e1n}+i{\bf n}^{(1)}_{o1n}\right),
\end{eqnarray}
where
\begin{eqnarray}
{\bf m}^{(1)}_{^{o}_{e}1n}&=&\pm
\frac{1}{\sin\theta}j_{n}(k_{0}r)P^{1}_{n}(\cos\theta){^{\cos}_{\sin}\phi}{\bf
i}_{2}-j_{n}(k_{0}r)\frac{\partial P^{1}_{n}}{\partial
\theta}{^{\sin}_{\cos}\phi}{\bf i}_{3},                     \nonumber \\
{\bf
n}^{(1)}_{^{o}_{e}1n}&=&\frac{n(n+1)}{k_{0}r}j_{n}(k_{0}r)P^{1}_{n}(\cos\theta){^{\sin}_{\cos}\phi}{\bf
i}_{1}+\frac{1}{k_{0}r}\left[k_{0}rj_{n}(k_{0}r)\right]'\frac{\partial
P^{1}_{n}}{\partial \theta}{^{\sin}_{\cos}\phi}{\bf i}_{2}\pm
\frac{1}{k_{0}r\sin\theta}\left[k_{0}rj_{n}(k_{0}r)\right]'P^{1}_{n}(\cos\theta){^{\cos}_{\sin}\phi}{\bf
i}_{3}                      \label{mieeq26}
\end{eqnarray}
with the primes denoting differentiation with respect to their
arguments and $k_{0}$ being
$\sqrt{\epsilon_{0}\mu_{0}}\frac{\omega}{c}$. In the region
$r<a_{1}$, the wave function is expanded as the following series
\begin{eqnarray}
{\bf E}_{\rm t}&=&E_{0}\exp(-i\omega
t)\sum^{\infty}_{n=1}i^{n}\frac{2n+1}{n(n+1)}\left(a^{\rm
t}_{n}{\bf
m}^{(1)}_{o1n}-ib^{\rm t}_{n}{\bf n}^{(1)}_{e1n}\right),                        \nonumber \\
{\bf H}_{\rm t}&=&-\frac{k_{1}}{\mu_{1}\omega}E_{0}\exp(-i\omega
t)\sum^{\infty}_{n=1}i^{n}\frac{2n+1}{n(n+1)}\left(b^{\rm
t}_{n}{\bf m}^{(1)}_{e1n}+ia^{\rm t}_{n}{\bf
n}^{(1)}_{o1n}\right).
\end{eqnarray}
Note that here the propagation constant in ${\bf m}^{(1)}_{o1n}$
and ${\bf n}^{(1)}_{e1n}$ which have been defined in
(\ref{mieeq26}) is replaced with $k_{1}$, {\it i.e.},
$\sqrt{\epsilon_{1}\mu_{1}}\frac{\omega}{c}$.
 In the region $a_{1}<r<a_{2}$, one may expand the electromagnetic
 wave amplitude as
\begin{eqnarray}
{\bf E}_{\rm m}&=&E_{0}\exp(-i\omega
t)\sum^{\infty}_{n=1}i^{n}\frac{2n+1}{n(n+1)}\left(a^{\rm
m}_{n}{\bf m}^{(1)}_{o1n}-ib^{\rm m}_{n}{\bf
n}^{(1)}_{e1n}+a^{\bar{\rm m}}_{n}{\bf m}^{(3)}_{o1n}-ib^{\bar{\rm
m}}_{n}{\bf n}^{(3)}_{e1n}\right),
\nonumber \\
{\bf H}_{\rm m}&=&-\frac{k_{2}}{\mu_{2}\omega}E_{0}\exp(-i\omega
t)\sum^{\infty}_{n=1}i^{n}\frac{2n+1}{n(n+1)}\left(b^{\rm
m}_{n}{\bf m}^{(1)}_{e1n}+ia^{\rm m}_{n}{\bf
n}^{(1)}_{o1n}+b^{\bar{\rm m}}_{n}{\bf m}^{(3)}_{e1n}+ia^{\bar{\rm
m}}_{n}{\bf n}^{(3)}_{o1n}\right).
\end{eqnarray}
Note that here in order to obtain ${\bf m}^{(3)}_{e1n}$ and ${\bf
n}^{(3)}_{o1n}$, the spherical Bessel functions $j_{n}$ in the
spherical vector function ${\bf m}^{(1)}_{e1n}$ and ${\bf
n}^{(1)}_{o1n}$ is replaced by the Hankel functions $h^{(1)}_{n}$,
namely, the explicit expressions for ${\bf m}^{(3)}_{e1n}$ and
${\bf n}^{(3)}_{o1n}$ can also be obtained from (\ref{mieeq26}),
so long as we replace the spherical Bessel functions $j_{n}$ in
(\ref{mieeq26}) with the Hankel functions $h^{(1)}_{n}$.
Apparently, here the propagation constant in ${\bf m}^{(1)}_{o1n}$
and ${\bf n}^{(1)}_{e1n}$ should be replaced with $k_{2}$, {\it
i.e.}, $\sqrt{\epsilon_{2}\mu_{2}}\frac{\omega}{c}$.

In the region $r>a_{2}$ the reflected wave is written
\cite{Stratton}
\begin{eqnarray}
{\bf E}_{\rm r}&=&E_{0}\exp(-i\omega
t)\sum^{\infty}_{n=1}i^{n}\frac{2n+1}{n(n+1)}\left(a^{\rm
r}_{n}{\bf
m}^{(3)}_{o1n}-ib^{\rm r}_{n}{\bf n}^{(3)}_{e1n}\right),                        \nonumber \\
{\bf H}_{\rm r}&=&-\frac{k_{0}}{\mu_{0}\omega}E_{0}\exp(-i\omega
t)\sum^{\infty}_{n=1}i^{n}\frac{2n+1}{n(n+1)}\left(b^{\rm
r}_{n}{\bf m}^{(3)}_{e1n}+ia^{\rm r}_{n}{\bf
n}^{(3)}_{o1n}\right),
\end{eqnarray}
where the propagation vector in ${\bf m}^{(3)}_{e1n}$ and ${\bf
n}^{(3)}_{o1n}$ is replaced by $k_{0}$, {\it i.e.},
$\sqrt{\epsilon_{0}\mu_{0}}\frac{\omega}{c}$.
\subsubsection{Boundary conditions}
At the boundary $r=a_{1}$, the boundary condition is given as
follows
\begin{equation}
{\bf i}_{1}\times{\bf E}_{\rm m}={\bf i}_{1}\times{\bf E}_{\rm t},
\quad {\bf i}_{1}\times{\bf H}_{\rm m}={\bf i}_{1}\times{\bf
H}_{\rm t},
\end{equation}
where the unit vectors of Cartesian coordinate system agree with
${\bf i}_{1}\times{\bf i}_{2}={\bf i}_{3}$, ${\bf i}_{2}\times{\bf
i}_{3}={\bf i}_{1}$, ${\bf i}_{3}\times{\bf i}_{1}={\bf i}_{2}$.

It follows from ${\bf i}_{1}\times{\bf E}_{\rm m}={\bf
i}_{1}\times{\bf E}_{\rm t}$ that
\begin{eqnarray}
a^{\rm m}_{n}j_{n}(N_{2}\rho_{1})&+&a^{\bar{\rm
m}}_{n}h^{(1)}_{n}(N_{2}\rho_{1})=a^{\rm
t}_{n}j_{n}(N_{1}\rho_{1}),
\nonumber  \\
N_{1}b^{\rm
m}_{n}\left[N_{2}\rho_{1}j_{n}(N_{2}\rho_{1})\right]'&+&N_{1}b^{\bar{\rm
m}}_{n}\left[N_{2}\rho_{1}h^{(1)}_{n}(N_{2}\rho_{1})\right]'=N_{2}b^{\rm
t}_{n}\left[N_{1}\rho_{1}j_{n}(N_{1}\rho_{1})\right]',
\end{eqnarray}
where $\rho_{1}=k_{0}a_{1}$, $N_{1}=\frac{k_{1}}{k_{0}}$.

In the similar manner, it follows from ${\bf i}_{1}\times{\bf
H}_{\rm m}={\bf i}_{1}\times{\bf H}_{\rm t}$ that
\begin{eqnarray}
\mu_{1}\left\{a^{\rm
m}_{n}\left[N_{2}\rho_{1}j_{n}(N_{2}\rho_{1})\right]'+a^{\bar{\rm
m}}_{n}\left[N_{2}\rho_{1}h^{(1)}_{n}(N_{2}\rho_{1})\right]'\right\}&=&\mu_{2}\left\{a^{\rm
t}_{n}\left[N_{1}\rho_{1}j_{n}(N_{1}\rho_{1})\right]'\right\},
\nonumber  \\
N_{2}\mu_{1}\left[b^{\rm m}_{n}j_{n}(N_{2}\rho_{1})+b^{\bar{\rm
m}}_{n}h^{(1)}_{n}(N_{2}\rho_{1})\right]&=&N_{1}\mu_{2}b^{\rm
t}_{n}j_{n}(N_{1}\rho_{1}).
\end{eqnarray}

At the boundary $r=a_{2}$, in the same fashion, it follows from
${\bf i}_{1}\times({\bf E}_{\rm i}+{\bf E}_{\rm r})={\bf
i}_{1}\times{\bf E}_{\rm m}$ that
\begin{eqnarray}
j_{n}(\rho_{2})+a^{\rm r}_{n}h^{(1)}_{n}(\rho_{2})&=&a^{\rm
m}_{n}j_{n}(N_{2}\rho_{2})+a^{\bar{\rm
m}}_{n}h^{(1)}_{n}(N_{2}\rho_{2}),
\nonumber  \\
N_{2}\left[\rho_{2}j_{n}(\rho_{2})\right]'+N_{2}b^{\rm
r}_{n}\left[\rho_{2}h^{(1)}_{n}(\rho_{2})\right]'&=&b^{\rm
m}_{n}\left[N_{2}\rho_{2}j_{n}(N_{2}\rho_{2})\right]'+b^{\bar{\rm
m}}_{n}\left[N_{2}\rho_{2}h^{(1)}_{n}(N_{2}\rho_{2})\right]',
\end{eqnarray}
where $\rho_{2}=k_{0}a_{2}$, $N_{2}=\frac{k_{2}}{k_{0}}$.

 In the meanwhile, it follows from
${\bf i}_{1}\times({\bf H}_{\rm i}+{\bf H}_{\rm r})={\bf
i}_{1}\times{\bf H}_{\rm m}$ that
\begin{eqnarray}
\mu_{2}\left\{a^{\rm
r}_{n}\left[\rho_{2}h^{(1)}_{n}(\rho_{2})\right]'+\left[\rho_{2}j_{n}(\rho_{2})\right]'\right\}&=&\mu_{0}\left\{a^{\rm
m}_{n}\left[N_{2}\rho_{2}j_{n}(N_{2}\rho_{2})\right]'+a^{\bar{\rm
m}}_{n}\left[N_{2}\rho_{2}h^{(1)}_{n}(N_{2}\rho_{2})\right]'\right\},
\nonumber  \\
\mu_{2}b^{\rm
r}_{n}h^{(1)}_{n}(\rho_{2})+\mu_{2}j_{n}(\rho_{2})&=&N_{2}\mu_{0}b^{\rm
m}_{n}j_{n}(N_{2}\rho_{2})+N_{2}\mu_{0}b^{\bar{\rm \rm
m}}_{n}h^{(1)}_{n}(N_{2}\rho_{2}).
\end{eqnarray}
\subsubsection{Calculation of Mie coefficients}

{\bf Mie coefficient} $a^{\rm r}_{n}$:

According to the above boundary conditions, one can arrive at the
following matrix equation
\begin{equation}
\left(\begin{array}{cccc}
-j_{n}(N_{1}\rho_{1})  & j_{n}(N_{2}\rho_{1}) & h^{(1)}_{n}(N_{2}\rho_{1})& 0 \\
-\mu_{2}\left[N_{1}\rho_{1}j_{n}(N_{1}\rho_{1})\right]' & \mu_{1}\left[N_{2}\rho_{1}j_{n}(N_{2}\rho_{1})\right]'& \mu_{1}\left[N_{2}\rho_{1}h^{(1)}_{n}(N_{2}\rho_{1})\right]' & 0  \\
 0 & j_{n}(N_{2}\rho_{2})&h^{(1)}_{n}(N_{2}\rho_{2}) &  -h^{(1)}_{n}(\rho_{2})   \\
 0 & \mu_{0}\left[N_{2}\rho_{2}j_{n}(N_{2}\rho_{2})\right]' &\left[N_{2}\rho_{2}h^{(1)}_{n}(N_{2}\rho_{2})\right]' &
 -\mu_{2}\left[\rho_{2}h^{(1)}_{n}(\rho_{2})\right]'
 \end{array}
 \right)\left(\begin{array}{cccc}
 a^{\rm t}_{n} \\
 a^{\rm m}_{n}\\
 a^{\bar{\rm m}}_{n}\\
  a^{\rm r}_{n}
 \end{array}
 \right)=\left(\begin{array}{cccc}
 0\\
 0\\
 j_{n}(\rho_{2})\\
 \mu_{2}\left[\rho_{2}j_{n}(\rho_{2})\right]'
 \end{array}
 \right).           \label{mieeq215}
\end{equation}

For convenience, the $4\times4$ matrix on the left handed side of
(\ref{mieeq215}) is denoted by
\begin{equation}
A=\left(\begin{array}{cccc}
A_{11} & A_{12} & A_{13}& 0 \\
A_{21} & A_{22}& A_{23}& 0  \\
 0 & A_{32} & A_{33} & A_{34}\\
 0&A_{42}&A_{43}&A_{44}
 \end{array}
 \right)
\end{equation}
whose determinant is
\begin{eqnarray}
{\rm det}
A&=&A_{11}A_{22}A_{33}A_{44}-A_{11}A_{22}A_{34}A_{43}-A_{11}A_{32}A_{23}A_{44}+A_{11}A_{42}A_{23}A_{34}
\nonumber  \\
&-&A_{21}A_{12}A_{33}A_{44}+A_{21}A_{12}A_{34}A_{43}+A_{21}A_{32}A_{13}A_{44}-A_{21}A_{42}A_{13}A_{34}.
\end{eqnarray}

So, the Mie coefficient $a^{\rm r}_{n}$ is of the form
\begin{equation} a^{\rm
r}_{n}=\frac{{\rm det} A_{\rm r}}{{\rm det} A}
\end{equation}
with
\begin{equation}
 A_{\rm r}=\left(\begin{array}{cccc}
A_{11} & A_{12} & A_{13}& 0 \\
A_{21} & A_{22}& A_{23}& 0  \\
 0 & A_{32} & A_{33} & j_{n}(\rho_{2})\\
 0&A_{42}&A_{43}&\mu_{2}\left[\rho_{2}j_{n}(\rho_{2})\right]'
 \end{array}
 \right)
\end{equation}
whose determinant is
\begin{eqnarray}
{\rm det} A_{\rm r
}&=&A_{11}A_{22}A_{33}\mu_{2}\left[\rho_{2}j_{n}(\rho_{2})\right]'-A_{11}A_{22}j_{n}(\rho_{2})A_{43}-A_{11}A_{32}A_{23}\mu_{2}\left[\rho_{2}j_{n}(\rho_{2})\right]'+A_{11}A_{42}A_{23}j_{n}(\rho_{2})
\nonumber   \\
&-&A_{21}A_{12}A_{33}\mu_{2}\left[\rho_{2}j_{n}(\rho_{2})\right]'+A_{21}A_{12}j_{n}(\rho_{2})A_{43}+A_{21}A_{32}A_{13}\mu_{2}\left[\rho_{2}j_{n}(\rho_{2})\right]'-A_{21}A_{42}A_{13}j_{n}(\rho_{2}).
\end{eqnarray}
\\ \\

{\bf Mie coefficient} $b^{\rm r}_{n}$:

According to the above boundary conditions, one can arrive at the
following matrix equation
 \begin{equation}
\left(\begin{array}{cccc}
-N_{2}\left[N_{1}\rho_{1}j_{n}(N_{1}\rho_{1})\right]' & N_{1}\left[N_{2}\rho_{1}j_{n}(N_{2}\rho_{1})\right]' & N_{1}\left[N_{2}\rho_{1}h^{(1)}_{n}(N_{2}\rho_{1})\right]' & 0 \\
-N_{1}\mu_{2}j_{n}(N_{1}\rho_{1}) & N_{2}\mu_{1}j_{n}(N_{2}\rho_{1}) & N_{2}\mu_{1}h^{(1)}_{n}(N_{2}\rho_{1})& 0  \\
 0 & \left[N_{2}\rho_{2}j_{n}(N_{2}\rho_{2})\right]' & \left[N_{2}\rho_{2}h^{(1)}_{n}(N_{2}\rho_{2})\right]'&
 -N_{2}\left[\rho_{2}h^{(1)}_{n}(\rho_{2})\right]'   \\
 0& N_{2}\mu_{0}j_{n}(N_{2}\rho_{2})& N_{2}\mu_{0}h^{(1)}_{n}(N_{2}\rho_{2})&
 -\mu_{2}h^{(1)}_{n}(\rho_{2})
 \end{array}
 \right)\left(\begin{array}{cccc}
b^{\rm t}_{n} \\
 b^{\rm m}_{n}\\
 b^{\bar{\rm m}}_{n}\\
  b^{\rm r}_{n}
 \end{array}
 \right)=\left(\begin{array}{cccc}
 0\\
 0\\
 N_{2}\left[\rho_{2}j_{n}(\rho_{2})\right]' \\
 \mu_{2}j_{n}(\rho_{2})
 \end{array}
 \right).                           \label{mieeq221}
\end{equation}

For convenience, the $4\times4$ matrix on the left handed side of
(\ref{mieeq221}) is denoted by
\begin{equation}
B=\left(\begin{array}{cccc}
B_{11} & B_{12} & B_{13}& 0 \\
B_{21} & B_{22}& B_{23}& 0  \\
 0 & B_{32} & B_{33} & B_{34}\\
 0&B_{42}&B_{43}&B_{44}
 \end{array}
 \right)
\end{equation}
whose determinant is
\begin{eqnarray}
{\rm det}
B&=&B_{11}B_{22}B_{33}B_{44}-B_{11}B_{22}B_{34}B_{43}-B_{11}B_{32}B_{23}B_{44}+B_{11}B_{42}B_{23}B_{34}
\nonumber  \\
&-&B_{21}B_{12}B_{33}B_{44}+B_{21}B_{12}B_{34}B_{43}+B_{21}B_{32}B_{13}B_{44}-B_{21}B_{42}B_{13}B_{34}.
\end{eqnarray}

So, the Mie coefficient $b^{\rm r}_{n}$ is of the form

\begin{equation}
b^{\rm r}_{n}=\frac{{\rm det} B_{\rm r}}{{\rm det} B}
\end{equation}
with
\begin{equation}
B_{\rm r}=\left(\begin{array}{cccc}
B_{11} & B_{12} & B_{13}& 0 \\
B_{21} & B_{22}& B_{23}& 0  \\
 0 & B_{32} & B_{33} &  N_{2}\left[\rho_{2}j_{n}(\rho_{2})\right]' \\
 0&B_{42}&B_{43}&\mu_{2}j_{n}(\rho_{2})
 \end{array}
 \right)
\end{equation}
whose determinant is

\begin{eqnarray}
{\rm det} B_{\rm r
}&=&B_{11}B_{22}B_{33}\mu_{2}j_{n}(\rho_{2})-B_{11}B_{22}N_{2}\left[\rho_{2}j_{n}(\rho_{2})\right]'B_{43}-B_{11}B_{32}B_{23}\mu_{2}j_{n}(\rho_{2})+B_{11}B_{42}B_{23}N_{2}\left[\rho_{2}j_{n}(\rho_{2})\right]'
 \nonumber   \\
&-&B_{21}B_{12}B_{33}\mu_{2}j_{n}(\rho_{2})+B_{21}B_{12}N_{2}\left[\rho_{2}j_{n}(\rho_{2})\right]'B_{43}+B_{21}B_{32}B_{13}\mu_{2}j_{n}(\rho_{2})-B_{21}B_{42}B_{13}N_{2}\left[\rho_{2}j_{n}(\rho_{2})\right]'.
\end{eqnarray}
\subsection{Reduced to the case of single-layered sphere}
In order to see whether the above Mie coefficients in the case of
double-layered sphere is correct or not, we consider the reduction
problem of double-layered case to the single-layered one when the
following reduction conditions are satisfied: $A_{11}=-A_{12}$,
$A_{21}=-A_{22}$ $N_{1}=N_{2}$, $\rho_{1}=\rho_{2}$,
$\mu_{1}=\mu_{2}$.

By lengthy calculation, we obtain
\begin{eqnarray}
{\rm
det}A&=&\left(A_{11}A_{23}-A_{21}A_{13}\right)\left(A_{42}A_{34}-A_{32}A_{44}\right)
\nonumber \\
&=&\left(A_{11}A_{23}-A_{21}A_{13}\right)\left\{j_{n}(N_{2}\rho_{2})\mu_{2}\left[\rho_{2}h^{(1)}_{n}(\rho_{2})\right]'-\mu_{0}\left[N_{2}\rho_{2}j_{n}(N_{2}\rho_{2})\right]'h^{(1)}_{n}(\rho_{2})\right\}
\end{eqnarray}
and
\begin{equation}
{\rm det}A_{\rm
r}=\left(A_{11}A_{23}-A_{21}A_{13}\right)\left\{\mu_{0}\left[N_{2}\rho_{2}j_{n}(N_{2}\rho_{2})\right]'j_{n}(\rho_{2})-j_{n}(N_{2}\rho_{2})\mu_{2}\left[\rho_{2}j_{n}(\rho_{2})\right]'\right\}.
\end{equation}
Thus
\begin{equation}
a^{\rm
r}_{n}=\frac{\mu_{0}\left[N_{2}\rho_{2}j_{n}(N_{2}\rho_{2})\right]'j_{n}(\rho_{2})-j_{n}(N_{2}\rho_{2})\mu_{2}\left[\rho_{2}j_{n}(\rho_{2})\right]'}{j_{n}(N_{2}\rho_{2})\mu_{2}\left[\rho_{2}h^{(1)}_{n}(\rho_{2})\right]'-\mu_{0}\left[N_{2}\rho_{2}j_{n}(N_{2}\rho_{2})\right]'h^{(1)}_{n}(\rho_{2})},
\end{equation}
which is just the Mie coefficient $a^{\rm r}_{n}$ of
single-layered sphere\cite{Stratton}.

 In the same fashion, when the reduction conditions $B_{11}=-B_{12}$, $B_{21}=-B_{22}$
$N_{1}=N_{2}$, $\rho_{1}=\rho_{2}$, $\mu_{1}=\mu_{2}$ are
satisfied, one can arrive at
\begin{eqnarray}
{\rm
det}B&=&\left(B_{11}B_{23}-B_{21}B_{13}\right)\left(B_{42}B_{34}-B_{32}B_{44}\right)
\nonumber \\
&=&\left(B_{11}B_{23}-B_{21}B_{13}\right)\left\{\left[N_{2}\rho_{2}j_{n}(N_{2}\rho_{2})\right]'\mu_{2}h^{(1)}_{n}(\rho_{2})-N^{2}_{2}\mu_{0}j_{n}(N_{2}\rho_{2})\left[\rho_{2}h^{(1)}_{n}(\rho_{2})\right]'\right\}
\end{eqnarray}
and
\begin{equation}
{\rm det}B_{\rm
r}=\left(B_{11}B_{23}-B_{21}B_{13}\right)\left\{N^{2}_{2}\mu_{0}j_{n}(N_{2}\rho_{2})\left[\rho_{2}j_{n}(\rho_{2})\right]'-\left[N_{2}\rho_{2}j_{n}(N_{2}\rho_{2})\right]'\mu_{2}j_{n}(\rho_{2})\right\}.
\end{equation}
So,
\begin{equation}
b^{\rm
r}_{n}=\frac{N^{2}_{2}\mu_{0}j_{n}(N_{2}\rho_{2})\left[\rho_{2}j_{n}(\rho_{2})\right]'-\left[N_{2}\rho_{2}j_{n}(N_{2}\rho_{2})\right]'\mu_{2}j_{n}(\rho_{2})}{\left[N_{2}\rho_{2}j_{n}(N_{2}\rho_{2})\right]'\mu_{2}h^{(1)}_{n}(\rho_{2})-N^{2}_{2}\mu_{0}j_{n}(N_{2}\rho_{2})\left[\rho_{2}h^{(1)}_{n}(\rho_{2})\right]'},
\end{equation}
which is just the Mie coefficient $b^{\rm r}_{n}$ of
single-layered sphere\cite{Stratton}.

This, therefore, means that the Mie coefficients of double-layered
sphere presented here is right.
\subsection{All the Mie coefficients inside the double-layered sphere}
\subsubsection{Mie coefficients $a^{\rm t}_{n}$, $a^{\rm m}_{n}$ and $a^{\bar{\rm m}}_{n}$}
The $4\times4$ matrix $A_{\rm t}$ is
\begin{equation}
A_{\rm t}=\left(\begin{array}{cccc}
0 & A_{12} & A_{13}& 0 \\
0 & A_{22}& A_{23}& 0  \\
  j_{n}(\rho_{2})& A_{32} & A_{33} & A_{34}\\
 \mu_{2}\left[\rho_{2}j_{n}(\rho_{2})\right]'&A_{42}&A_{43}&A_{44}
 \end{array}
 \right)
\end{equation}
whose determinant is
\begin{equation}
{\rm det}A_{\rm
t}=j_{n}(\rho_{2})A_{12}A_{23}A_{44}-j_{n}(\rho_{2})A_{22}A_{13}A_{44}-\mu_{2}\left[\rho_{2}j_{n}(\rho_{2})\right]'A_{12}A_{23}A_{34}+\mu_{2}\left[\rho_{2}j_{n}(\rho_{2})\right]'A_{22}A_{13}A_{34}.
\end{equation}
So,
\begin{equation}
a^{\rm t}_{n}=\frac{{\rm det} A_{\rm t}}{{\rm det} A}.
\end{equation}

The $4\times4$ matrix $A_{\rm m}$ is
\begin{equation}
A_{\rm m}=\left(\begin{array}{cccc}
A_{11} & 0& A_{13}& 0 \\
A_{21} & 0& A_{23}& 0  \\
  0 & j_{n}(\rho_{2}) & A_{33} & A_{34}\\
 0 & \mu_{2}\left[\rho_{2}j_{n}(\rho_{2})\right]'& A_{43}& A_{44}
 \end{array}
 \right)
\end{equation}
whose determinant is
\begin{equation}
{\rm det}A_{\rm
m}=-A_{11}j_{n}(\rho_{2})A_{23}A_{44}+A_{11}\mu_{2}\left[\rho_{2}j_{n}(\rho_{2})\right]'A_{23}A_{34}+A_{21}j_{n}(\rho_{2})A_{13}A_{44}-A_{21}\mu_{2}\left[\rho_{2}j_{n}(\rho_{2})\right]'A_{13}A_{34}.
\end{equation}
So,
\begin{equation}
a^{\rm m}_{n}=\frac{{\rm det} A_{\rm m}}{{\rm det} A}.
\end{equation}

The $4\times4$ matrix $A_{\bar{\rm m}}$ is
\begin{equation}
A_{\bar{\rm m}}=\left(\begin{array}{cccc}
A_{11} & A_{12}& 0& 0 \\
A_{21} & A_{22}& 0& 0  \\
  0 & A_{32} & j_{n}(\rho_{2})  & A_{34}\\
 0 & A_{42}& \mu_{2}\left[\rho_{2}j_{n}(\rho_{2})\right]'& A_{44}
 \end{array}
 \right)
\end{equation}
whose determinant is
\begin{equation}
{\rm det}A_{\bar{\rm m}}=A_{11}A_{22}j_{n}(\rho_{2})
A_{44}-A_{11}A_{22}A_{34}\mu_{2}\left[\rho_{2}j_{n}(\rho_{2})\right]'-A_{21}A_{12}j_{n}(\rho_{2})A_{44}+A_{21}A_{12}A_{34}\mu_{2}\left[\rho_{2}j_{n}(\rho_{2})\right]'.
\end{equation}
So,
\begin{equation}
a^{\bar{\rm m}}_{n}=\frac{{\rm det}A_{\bar{\rm m}}}{{\rm det} A}.
\end{equation}

It is readily verified that under the reduction conditions
$A_{11}=-A_{12}$, $A_{21}=-A_{22}$, $N_{1}=N_{2}$,
$\rho_{1}=\rho_{2}$, $\mu_{1}=\mu_{2}$, one can arrive at
\begin{equation}
{\rm det}A_{\rm t}={\rm det}A_{\rm m}, \quad a^{\rm t}_{n}=a^{\rm
m}_{n},  \quad {\rm det}A_{\bar{\rm m}}=0,     \quad   a^{\bar{\rm
m}}_{n}=0,
\end{equation}
which means that the above Mie coefficient can be reduced to the
those of single-layered sphere.
\subsubsection{Mie coefficients $b^{\rm t}_{n}$, $b^{\rm m}_{n}$ and $b^{\bar{\rm m}}_{n}$}

The $4\times4$ matrix $B_{\rm t}$ is
\begin{equation}
B_{\rm t}=\left(\begin{array}{cccc}
0 & B_{12} & B_{13}& 0 \\
0 & B_{22}& B_{23}& 0  \\
  N_{2}\left[\rho_{2}j_{n}(\rho_{2})\right]' & B_{32} & B_{33} & B_{34}\\
 \mu_{2}j_{n}(\rho_{2})&B_{42}&B_{43}&B_{44}
 \end{array}
 \right)
\end{equation}
whose determinant is
\begin{equation}
{\rm det}B_{\rm t}=N_{2}\left[\rho_{2}j_{n}(\rho_{2})\right]'
B_{12}B_{23}B_{44}-N_{2}\left[\rho_{2}j_{n}(\rho_{2})\right]'
B_{22}B_{13}B_{44}-\mu_{2}j_{n}(\rho_{2})
B_{12}B_{23}B_{34}+\mu_{2}j_{n}(\rho_{2})B_{22}B_{13}B_{34}.
\end{equation}
So,
\begin{equation}
b^{\rm t}_{n}=\frac{{\rm det} B_{\rm t}}{{\rm det} B}.
\end{equation}
\\ \\
The $4\times4$ matrix $B_{\rm m}$ is
\begin{equation}
B_{\rm m}=\left(\begin{array}{cccc}
B_{11} & 0& B_{13}& 0 \\
B_{21} & 0& B_{23}& 0  \\
  0 & N_{2}\left[\rho_{2}j_{n}(\rho_{2})\right]'  & B_{33} & B_{34}\\
 0 & \mu_{2}j_{n}(\rho_{2})& B_{43}& B_{44}
 \end{array}
 \right)
\end{equation}
whose determinant is
\begin{equation}
{\rm det}B_{\rm
m}=-B_{11}N_{2}\left[\rho_{2}j_{n}(\rho_{2})\right]'
B_{23}B_{44}+B_{11}\mu_{2}j_{n}(\rho_{2})
B_{23}B_{34}+B_{21}N_{2}\left[\rho_{2}j_{n}(\rho_{2})\right]'
B_{13}B_{44}-B_{21}\mu_{2}j_{n}(\rho_{2})B_{13}B_{34}.
\end{equation}
So,
\begin{equation}
b^{\rm m}_{n}=\frac{{\rm det} B_{\rm m}}{{\rm det} B}.
\end{equation}
\\ \\
The $4\times4$ matrix $B_{\bar{\rm m}}$ is
\begin{equation}
B_{\bar{\rm m}}=\left(\begin{array}{cccc}
B_{11} & B_{12}& 0& 0 \\
B_{21} & B_{22}& 0& 0  \\
  0 & B_{32} & N_{2}\left[\rho_{2}j_{n}(\rho_{2})\right]'   & B_{34}\\
 0 & B_{42}& \mu_{2}j_{n}(\rho_{2})& B_{44}
 \end{array}
 \right)
\end{equation}
whose determinant is
\begin{equation}
{\rm det}B_{\bar{\rm
m}}=B_{11}B_{22}N_{2}\left[\rho_{2}j_{n}(\rho_{2})\right]'
B_{44}-B_{11}B_{22}B_{34}\mu_{2}j_{n}(\rho_{2})
-B_{21}B_{12}N_{2}\left[\rho_{2}j_{n}(\rho_{2})\right]'
B_{44}+B_{21}B_{12}B_{34}\mu_{2}j_{n}(\rho_{2}).
\end{equation}
So,
\begin{equation}
b^{\bar{\rm m}}_{n}=\frac{{\rm det}B_{\bar{\rm m}}}{{\rm det} B}.
\end{equation}
\\ \\

It is readily verified that under the reduction conditions
$B_{11}=-B_{12}$, $B_{21}=-B_{22}$,   $N_{1}=N_{2}$,
$\rho_{1}=\rho_{2}$, $\mu_{1}=\mu_{2}$, one can arrive at
\begin{equation}
{\rm det}B_{\rm t}={\rm det}B_{\rm m}, \quad b^{\rm t}_{n}=b^{\rm
m}_{n},  \quad {\rm det}B_{\bar{\rm m}}=0,     \quad   b^{\bar{\rm
m}}_{n}=0,
\end{equation}
which means that the above Mie coefficient can be reduced to the
those of single-layered sphere.

\subsection{Discussion of some typical cases with left-handed media involved}
In this subsection we briefly discuss several cases with
left-handed media involved.

(i) If the following conditions $h^{(1)}_{n}(N_{2}\rho_{1})=0$,
$\left[h^{(1)}_{n}(N_{2}\rho_{1})\right]'=0$,
$h^{(1)}_{n}(N_{2}\rho_{2})=-h^{(1)}_{n}(\rho_{2})$,
$\left[h^{(1)}_{n}(N_{2}\rho_{2})\right]'=-\left[h^{(1)}_{n}(\rho_{2})\right]'$,
$N_{2}=\mu_{2}=\epsilon_{2}=-1$ and $\mu_{0}=+1$ are satisfied,
then it is easily verified that

\begin{equation}
{\rm det}A=0,   \quad     {\rm det}B=0     \quad  {\rm and} \quad
a^{\rm r}_{n}=\infty,    \quad     b^{\rm r}_{n}=\infty.
\end{equation}
Thus in this case the extinction cross section of the two-layered
sphere containing left-handed media is rather large.

(ii) If the following conditions $h^{(1)}_{n}(N_{2}\rho_{1})=0$,
$\left[h^{(1)}_{n}(N_{2}\rho_{1})\right]'=0$,
$h^{(1)}_{n}(N_{2}\rho_{2})=j_{n}(\rho_{2})$,
$\left[h^{(1)}_{n}(N_{2}\rho_{2})\right]'=\left[j_{n}(\rho_{2})\right]'$,
$N_{2}=\mu_{2}=\epsilon_{2}=-1$ and $\mu_{0}=+1$ are satisfied,
then it is easily verified that

\begin{equation}
{\rm det}A_{\rm r}=0,   \quad     {\rm det}B_{\rm r}=0     \quad
{\rm and} \quad a^{\rm r}_{n}=0,    \quad     b^{\rm r}_{n}=0.
\end{equation}
Thus in this case the extinction cross section of the two-layered
sphere containing left-handed media is negligibly small.

(iii) The case with $N_{1}=\mu_{1}=\epsilon_{1}=-1$,
$N_{2}=\mu_{2}=\epsilon_{2}=0$, $N_{0}=\mu_{0}=\epsilon_{0}=+1$ is
of physical interest, which deserves consideration by using the
Mie coefficients presented above.

Based on the calculation of Mie coefficients, one can treat the
scattering problem of double-layered sphere irradiated by a plane
wave. The scattering and absorption properties of double-layered
sphere containing left-handed media can thus be discussed in
detail, which are now under consideration and will be submitted
elsewhere for the publication.

\section{Three kinds of compact thin subwavelength cavity resonators made of left-handed media: rectangular, cylindrical, spherical}
Engheta suggested that a slab of metamaterial with negative
electric permittivity and magnetic permeability (and hence
negative optical refractive index) can act as a phase
compensator/conjugator and, therefore, by combining such a slab
with another slab fabricated from a conventional (ordinary)
dielectric material one can, in principle, have a 1-D cavity
resonator whose dispersion relation may not depend on the sum of
thicknesses of the interior materials filling this cavity, but
instead it depends on the ratio of these thicknesses. Namely, one
can, in principle, conceptualize a 1-D compact, subwavelength,
thin cavity resonator with the total thickness far less than the
conventional $\frac{\lambda}{2}$\cite{Engheta}.

Engheta's idea for the 1-D compact, subwavelength, thin cavity
resonator is the two-layer rectangular structure (the left layer
of which is assumed to be a conventional lossless dielectric
material with permittivity and permeability being positive
numbers, and the right layer is taken to be a lossless
metamaterial with negative permittivity and permeability)
sandwiched between the two reflectors ({\it e.g.}, two perfectly
conducting plates)\cite{Engheta}. For the pattern of the 1-D
subwavelength cavity resonator readers may be referred to the
figures of reference\cite{Engheta}. Engheta showed that with the
appropriate choice of the ratio of the thicknesses $d_{1}$ to
$d_{2}$, the phase acquired by the incident wave at the left
(entrance) interface to be the same as the phase at the right
(exit) interface, essentially with no constraint on the total
thickness of the structure. The mechanism of this effect may be
understood as follows: as the planar electromagnetic wave exits
the first slab, it enters the rectangular slab of metamaterial and
finally it leaves this second slab. In the first slab, the
direction of the Poynting vector is parallel to that of phase
velocity, and in the second slab, however, these two vectors are
antiparallel with each other. Thus the wave vector $k_{2}$ is
therefore in the opposite direction of the wave vector $k_{1}$. So
the total phase difference between the front and back faces of
this two-layer rectangular structure is
$k_{1}d_{1}-|k_{2}|d_{2}$\cite{Engheta}. Therefore, whatever phase
difference is developed by traversing the first rectangular slab,
it can be decreased and even cancelled by traversing the second
slab. If the ratio of $d_{1}$ and $d_{2}$ is chosen to be
$\frac{d_{1}}{d_{2}}=\frac{|k_{2}|}{k_{1}}$, then the total phase
difference between the front and back faces of this two-layer
structure becomes zero ({\it i.e.}, the total phase difference is
not $2n\pi$, but instead of zero)\cite{Engheta}. As far as the
properties and phenomena in the subwavelength cavity resonators is
concerned, Tretyakov {\it et al.} investigated the evanescent
modes stored in cavity resonators with backward-wave
slabs\cite{Tretyakov}.

\subsection{A rectangular slab 1-D thin subwavelength cavity
resonator}

To consider the 1-D wave propagation in a compact, subwavelength,
thin cavity resonator, we first take into account a slab cavity of
three-layer structure, where the regions 1 and 2 are located on
the left- and right- handed sides, and the plasmon-type medium (or
a superconductor material) is between the regions 1 and 2. The
above three-layer structure is assumed to be sandwiched between
the two reflectors (or two perfectly conducting
plates)\cite{Engheta}.  Assume that the wave vector of the
electromagnetic wave is parallel to the third component of
Cartesian coordinate. The electric and magnetic fields in the
region 1 (with the permittivity being $\epsilon_{1}$ and the
permeability being $\mu_{1}$) are written in the form
\begin{equation}
E_{x1}=E_{01}\sin \left(n_{1}k_{0}z\right),  \quad
H_{y1}=\frac{n_{1}k_{0}}{i\omega \mu_{1}}E_{01}\cos
\left(n_{1}k_{0}z\right)   \quad  (0\leq z\leq d_{1}),
\label{resoeq1}
\end{equation}
where $k_{0}$ stands for the wave vector of the electromagnetic
wave under consideration in the free space, {\it i.e.},
$k_{0}=\frac{\omega}{c}$, and in the region 2, where $d_{1}+a\leq
z\leq d_{1}+d_{2}+a$, the electric and magnetic fields are of the
form
\begin{equation}
E_{x2}=E_{02}\sin
\left[n_{2}k_{0}\left(z-d_{1}-d_{2}-a\right)\right],   \quad
H_{y2}=\frac{n_{2}k_{0}}{i\omega \mu_{2}}E_{02}\cos
\left[n_{2}k_{0}\left(z-d_{1}-d_{2}-a\right)\right],
\label{resoeq2}
\end{equation}
where $d_{1}$, $d_{2}$ and $a$ denote the thicknesses of the
regions 1, 2 and the plasmon (or superconducting) region,
respectively. The subscripts 1 and 2 in the present paper denote
the physical quantities in the regions 1 and 2. Note that here the
optical refractive indices $n_{1}$ and $n_{2}$ are defined to be
$n_{1}=\sqrt{\epsilon_{1}\mu_{1}}$ and
$n_{2}=\sqrt{\epsilon_{2}\mu_{2}}$. Although in the present paper
we will consider the wave propagation in the negative refractive
index media, the choice of the signs for $n_{1}$ and $n_{2}$ will
be irrelevant in the final results. So, we choose the plus signs
for $n_{1}$ and $n_{2}$ no matter whether the materials 1 and 2
are of left-handedness or not. The choice of the solutions
presented in (\ref{resoeq1}) and (\ref{resoeq2}) guarantees the
satisfaction of the boundary conditions at the perfectly
conducting plates at $z=0$ and $z=d_{1}+d_{2}+a$.

The electric and magnetic fields in the plasmon (or
superconducting) region (with the resonant frequency being
$\omega_{\rm p}$) take the form
\begin{equation}
E_{xs}=A\exp (\beta z)+B\exp (-\beta z),   \quad
H_{ys}=\frac{\beta}{i\omega}\left[A\exp (\beta z)-B\exp (-\beta
z)\right],                  \label{resoeq3}
\end{equation}
where the subscript $s$ represents the quantities in the plasmon
(or superconducting) region, and $\beta=\frac{\sqrt{\omega_{\rm
p}^{2}-\omega^{2}}}{c}$.

To satisfy the boundary conditions
\begin{equation}
E_{x1}|_{z=d_{1}}=E_{xs}|_{z=d_{1}},   \quad
H_{y1}|_{z=d_{1}}=H_{ys}|_{z=d_{1}}
\end{equation}
at the interface ($z=d_{1}$) between the region 1 and the plasmon
region, we should have
\begin{equation}
E_{01}\sin \left(n_{1}k_{0}d_{1}\right)=A\exp (\beta d_{1})+B\exp
(-\beta d_{1}),   \quad   \frac{n_{1}k_{0}}{\beta
\mu_{1}}E_{01}\cos \left(n_{1}k_{0}d_{1}\right)=A\exp (\beta
d_{1})-B\exp(-\beta d_{1}).                  \label{resoeq5}
\end{equation}
It follows that the parameters $A$ and $B$ in Eq.(\ref{resoeq3})
are given as follows
\begin{eqnarray}
A&=&\frac{1}{2}\exp \left(-\beta d_{1}\right)E_{01}\left[\sin\left(n_{1}k_{0}d_{1}\right)+\frac{n_{1}k_{0}}{\beta \mu_{1}}\cos \left(n_{1}k_{0}d_{1}\right)\right],   \nonumber  \\                \nonumber \\
B&=&-\frac{1}{2}\exp \left(\beta
d_{1}\right)E_{01}\left[\frac{n_{1}k_{0}}{\beta \mu_{1}}\cos
\left(n_{1}k_{0}d_{1}\right)-\sin \left(n_{1}k_{0}d_{1}\right)
\right].
\label{resoeq6}
\end{eqnarray}
In the similar fashion, to satisfy the boundary conditions
\begin{equation}
E_{x2}|_{z=d_{1}+a}=E_{xs}|_{z=d_{1}+a},   \quad
H_{y2}|_{z=d_{1}+a}=H_{ys}|_{z=d_{1}+a}
\end{equation}
at the interface ($z=d_{1}+a$) between the region 2 and the
plasmon region, one should arrive at
\begin{eqnarray}
& & E_{02}\sin
\left(-n_{2}k_{0}d_{2}\right)=A\exp\left[\beta(d_{1}+a)\right]+B\exp\left[-\beta(d_{1}+a)\right],
\nonumber \\
& &    \frac{n_{2}k_{0}}{\beta \mu_{2}}E_{02}\cos
\left(-n_{2}k_{0}d_{2}\right)=A\exp\left[\beta(d_{1}+a)\right]-B\exp\left[-\beta(d_{1}+a)\right].
\label{resoeq7}
\end{eqnarray}
It follows that the parameters $A$ and $B$ in Eq.(\ref{resoeq3})
are given as follows
\begin{eqnarray}
A&=&-\frac{1}{2}E_{02}\exp \left[-\beta (d_{1}+a)\right]\left[\sin\left(n_{2}k_{0}d_{2}\right)-\frac{n_{2}k_{0}}{\beta \mu_{2}}\cos \left(n_{2}k_{0}d_{2}\right)\right],   \nonumber  \\                \nonumber \\
B&=&-\frac{1}{2}\exp \left[-\beta
(d_{1}+a)\right]E_{02}\left[\frac{n_{2}k_{0}}{\beta \mu_{2}}\cos
\left(n_{2}k_{0}d_{2}\right)+\sin \left(n_{2}k_{0}d_{2}\right)
\right].                 \label{resoeq8}
\end{eqnarray}
Thus, according to Eq.(\ref{resoeq6}) and (\ref{resoeq8}), we can
obtain the following conditions
\begin{eqnarray}
& & E_{01}\left[\sin\left(n_{1}k_{0}d_{1}\right)+\frac{n_{1}k_{0}}{\beta \mu_{1}}\cos\left(n_{1}k_{0}d_{1}\right)\right]+E_{02}\exp(-\beta a)\left[\sin\left(n_{2}k_{0}d_{2}\right)-\frac{n_{2}k_{0}}{\beta \mu_{2}}\cos\left(n_{2}k_{0}d_{2}\right)\right]=0,   \nonumber  \\
                 \nonumber \\
& &  E_{01}\left[\frac{n_{1}k_{0}}{\beta
\mu_{1}}\cos\left(n_{1}k_{0}d_{1}\right)-\sin
\left(n_{1}k_{0}d_{1}\right)\right]-E_{02}\exp(\beta
a)\left[\frac{n_{2}k_{0}}{\beta \mu_{2}}\cos
\left(n_{2}k_{0}d_{2}\right)+\sin\left(n_{2}k_{0}d_{2}\right)\right]=0.
\label{resoeq10}
\end{eqnarray}
In order to have a nontrivial solution, {\it i.e.}, to have
$E_{01}\neq 0$ and $E_{02}\neq 0$, the determinant in
Eq.(\ref{resoeq10}) must vanish. Thus we obtain the following
restriction condition
\begin{eqnarray}
& &  \left[\exp(\beta a)+\exp(-\beta
a)\right]\left[\frac{n_{1}}{\mu_{1}}\tan
\left(n_{2}k_{0}d_{2}\right)+\frac{n_{2}}{\mu_{2}}\tan
\left(n_{1}k_{0}d_{1}\right)\right]
                                       \nonumber \\
& &   +\frac{\beta}{k_{0}}\left[\exp(\beta a)-\exp(-\beta
a)\right]\left[\tan \left(n_{1}k_{0}d_{1}\right)\tan
\left(n_{2}k_{0}d_{2}\right)+\frac{n_{1}n_{2}k_{0}^{2}}{\beta^{2}\mu_{1}\mu_{2}}\right]=0
\label{resoeq11}
\end{eqnarray}
for the electromagnetic wave in the three-layer-structure
rectangular cavity.

If the thickness, $a$, of the plasmon region is vanishing ({\it
i.e.}, there exists no plasmon region), then the restriction
equation (\ref{resoeq11}) is simplified to
\begin{equation}
\frac{n_{1}}{\mu_{1}}\tan\left(n_{2}k_{0}d_{2}\right)+\frac{n_{2}}{\mu_{2}}\tan
\left(n_{1}k_{0}d_{1}\right)=0.
\end{equation}
In what follows we will demonstrate why the introduction of
left-handed media will give rise to the novel design of the
compact thin cavity resonator. If the material in region 1 is a
regular medium while the one in region 2 is the left-handed
medium, it follows that
\begin{equation}
\frac{\tan\left(n_{1}k_{0}d_{1}\right)}{\tan\left(n_{2}k_{0}d_{2}\right)}=\frac{-n_{1}\mu_{2}}{n_{2}\mu_{1}}.
\end{equation}
According to Engheta\cite{Engheta}, this relation does not show
any constraint on the sum of thicknesses of $d_{1}$ and $d_{2}$.
It rather deals with the ratio of tangent of these thicknesses
(with multiplicative constants). If we assume that $\omega$,
$d_{1}$ and $d_{2}$ are chosen such that the small-argument
approximation can be used for the tangent function, the above
relation can be simplified as
\begin{equation}
\frac{d_{1}}{d_{2}}\simeq -\frac{\mu_{2}}{\mu_{1}}.
\end{equation}
This relation shows even more clearly how $d_{1}$ and $d_{2}$
should be related in order to have a nontrivial 1-D solution with
frequency $\omega$ for this cavity. So conceptually, what is
constrained here is $\frac{d_{1}}{d_{2}}$, not $d_{1}+d_{2}$.
Therefore, in principle, one can have a thin subwavelength cavity
resonator for a given frequency\cite{Engheta}.

 So, one of the most exciting ideas is the possibility to
design the so-called compact thin subwavelength cavity resonators.
It was shown that a pair of plane waves travelling in the system
of two planar slabs positioned between two metal planes can
satisfy the boundary conditions on the walls and on the interface
between two slabs even for arbitrarily thin layers, provided that
one of the slabs has negative material parameters\cite{Tretyakov}.

In the following let us take account of two interesting cases:

(i) If $\beta a\rightarrow 0$, then the restriction equation
(\ref{resoeq11}) is simplified to
\begin{equation}
\frac{n_{1}}{\mu_{1}}\tan
\left(n_{2}k_{0}d_{2}\right)+\frac{n_{2}}{\mu_{2}}\tan
\left(n_{1}k_{0}d_{1}\right)+\frac{2\beta^{2}a}{k_{0}}\left[\tan
\left(n_{1}k_{0}d_{1}\right)\tan
\left(n_{2}k_{0}d_{2}\right)+\frac{n_{1}n_{2}k_{0}^{2}}{\beta^{2}\mu_{1}\mu_{2}}\right]=0,
\end{equation}
which yields
\begin{equation}
\tan
\left(n_{1}k_{0}d_{1}\right)=\frac{k_{0}}{\beta}\frac{n_{1}}{\mu_{1}},
\quad           \tan
\left(n_{2}k_{0}d_{2}\right)=-\frac{k_{0}}{\beta}\frac{n_{2}}{\mu_{2}}.
\label{resoeq214}
\end{equation}
If both $n_{1}k_{0}d_{1}$ and $n_{2}k_{0}d_{2}$ are very small,
then one can arrive at $d_{1}\doteq \frac{1}{\beta \mu_{1}}$,
$d_{2}\doteq -\frac{1}{\beta \mu_{2}}$ from Eq.(\ref{resoeq214}),
which means that the thicknesses $d_{1}$ and $d_{2}$ depend upon
the plasmon parameter $\beta$.

(ii) If the resonant frequency $\omega_{\rm p}$ is very large (and
hence $\beta$), then it follows from Eq.(\ref{resoeq10}) and
(\ref{resoeq11}) that
\begin{equation}
\frac{n_{1}}{\mu_{1}}\tan \left(n_{2}k_{0}d_{2}\right)=0,   \quad
\frac{n_{2}}{\mu_{2}}\tan \left(n_{1}k_{0}d_{1}\right)=0,
\end{equation}
namely, regions 1 and 2 are isolated from each other., which is a
result familiar to us.

In conclusion, as was shown by Engheta, it is possible that when
one of the slab has a negative permeability, electromagnetic wave
in two adjacent slabs bounded by two metal walls can satisfy the
boundary conditions even if the distance between the two walls is
much smaller than the wavelength\cite{Engheta}.

\subsection{A cylindrical thin subwavelength cavity resonator}

Here we will consider the restriction equation for a cylindrical
 cavity to be a thin subwavelength cavity resonator containing left-handed media.
 It is well known that the Helmholtz equation $\nabla^{2}{\bf E}+k^{2}{\bf E}=0$
 in an axially symmetric cylindrical cavity (with the 2-D polar coordinates $\rho$ and $\varphi$) can be rewritten as
\begin{eqnarray}
& &  \nabla^{2}E_{\rho}-\frac{1}{\rho^{2}}E_{\rho}-\frac{2}{\rho^{2}}\frac{\partial E_{\varphi}}{\partial \varphi}+k^{2}E_{\rho}=0,                \nonumber \\
& &   \nabla^{2}E_{\varphi}-\frac{1}{\rho^{2}}E_{\varphi}+\frac{2}{\rho^{2}}\frac{\partial E_{\rho}}{\partial \varphi}+k^{2}E_{\varphi}=0,               \nonumber \\
& &  \nabla^{2}E_{z}+k^{2}E_{z}=0.       \label{resoeq201}
\end{eqnarray}
One can obtain the electromagnetic field distribution, $E_{\rho}$,
$E_{\varphi}$ and $H_{\rho}$, $H_{\varphi}$, in the above axially
symmetric cylindrical cavity via Eq.(\ref{resoeq201}). But here we
will adopt another alternative way to get the solutions of
electromagnetic fields in the cylindrical cavity. If the
electromagnetic fields are time-harmonic, {\it i.e.}, ${\bf
E}(\rho,\varphi,z,t)=\vec{{\mathcal
E}}(\rho,\varphi)\exp[i(hz-kct)]$ and ${\bf
H}(\rho,\varphi,z,t)=\vec{{\mathcal
H}}(\rho,\varphi)\exp[i(hz-kct)]$, then it follows from Maxwell
equations that
\begin{eqnarray}
& &  -ikc{\mathcal E}_{\rho}=\frac{1}{\epsilon}\left(\frac{1}{\rho}\frac{\partial {\mathcal H}_{z}}{\partial \varphi}-ih{\mathcal H}_{\varphi}\right),                \nonumber \\
& &  -ikc{\mathcal E}_{\varphi}=\frac{1}{\epsilon}\left(ih{\mathcal H}_{\rho}-\frac{\partial {\mathcal H}_{z}}{\partial \rho}\right),                \nonumber \\
& &  -ikc{\mathcal E}_{z}=\frac{1}{\epsilon}\left(\frac{\partial
{\mathcal H}_{\varphi}}{\partial \rho}+\frac{1}{\rho}{\mathcal
H}_{\varphi}-\frac{1}{\rho}\frac{\partial {\mathcal
H}_{\rho}}{\partial \varphi}\right),
\end{eqnarray}
and
\begin{eqnarray}
& &  ikc{\mathcal H}_{\rho}=\frac{1}{\mu}\left(\frac{1}{\rho}\frac{\partial {\mathcal E}_{z}}{\partial \varphi}-ih{\mathcal E}_{\varphi}\right),                \nonumber \\
& &  ikc{\mathcal H}_{\varphi}=\frac{1}{\mu}\left(ih{\mathcal E}_{\rho}-\frac{\partial {\mathcal E}_{z}}{\partial \rho}\right),                \nonumber \\
& &  ikc{\mathcal H}_{z}=\frac{1}{\mu}\left(\frac{\partial
{\mathcal E}_{\varphi}}{\partial \rho}+\frac{1}{\rho}{\mathcal
E}_{\varphi}-\frac{1}{\rho}\frac{\partial {\mathcal
E}_{\rho}}{\partial \varphi}\right).
\end{eqnarray}

Thus it is demonstrated that the electromagnetic fields ${\mathcal
E}_{\rho}$, ${\mathcal E}_{\varphi}$ and ${\mathcal H}_{\rho}$,
${\mathcal H}_{\varphi}$ can be expressed in terms of ${\mathcal
E}_{z}$ and ${\mathcal H}_{z}$, {\it i.e.},
\begin{equation}
{\mathcal E}_{\rho}=\frac{i}{k^{2}-h^{2}}\left(h\frac{\partial
{\mathcal E}_{z}}{\partial \rho}+\frac{k^{2}}{\epsilon
\rho}\frac{\partial {\mathcal H}_{z}}{\partial \varphi}\right),
\quad {\mathcal
E}_{\varphi}=\frac{i}{k^{2}-h^{2}}\left(h\frac{1}{\rho}\frac{\partial
{\mathcal E}_{z}}{\partial
\varphi}-\frac{k^{2}}{\epsilon}\frac{\partial {\mathcal
H}_{z}}{\partial \rho}\right), \label{resoeq2036}
\end{equation}
and
\begin{equation}
{\mathcal H}_{\rho}=\frac{i}{k^{2}-h^{2}}\left(h\frac{\partial
{\mathcal H}_{z}}{\partial \rho}-\frac{k^{2}}{\mu
\rho}\frac{\partial {\mathcal E}_{z}}{\partial \varphi}\right),
\quad {\mathcal
H}_{\varphi}=\frac{i}{k^{2}-h^{2}}\left(h\frac{1}{\rho}\frac{\partial
{\mathcal H}_{z}}{\partial
\varphi}+\frac{k^{2}}{\mu}\frac{\partial {\mathcal
E}_{z}}{\partial \rho}\right).      \label{resoeq2037}
\end{equation}
As an illustrative example, in what follows, we will consider only
the TM wave ({\it i.e.}, ${\mathcal H}_{z}=0$) in the axially
symmetric double-layer cylindrical thin subwavelength cavity
resonator. Assume that the permittivity, permeability and radius
of media in the interior and exterior layers of this double-layer
cavity resonator are $\epsilon_{1}$, $\mu_{1}$, $R_{1}$ and
$\epsilon_{2}$, $\mu_{2}$, $R_{2}$, respectively. According to the
Helmholtz equation (with the boundary material being the perfectly
conducting medium, ${\mathcal E}_{2z}|_{R_{2}}=0$), we can obtain
${\it
E}_{1z}=J_{m}\left(\sqrt{k_{1}^{2}-h_{1}^{2}}\rho\right)\left\{
{\begin{array}{*{20}c}
   {\cos m\varphi }  \\
   {\sin m\varphi}  \\
\end{array}}\right\}$ and ${\it
E}_{2z}=\left[AJ_{m}\left(\sqrt{k_{2}^{2}-h_{2}^{2}}\rho\right)+BN_{m}\left(\sqrt{k_{2}^{2}-h_{2}^{2}}\rho\right)\right]\left\{
{\begin{array}{*{20}c}
   {\cos m\varphi }  \\
   {\sin m\varphi}  \\
\end{array}}\right\}$. Thus it follows from Eq.(\ref{resoeq2036}) and
(\ref{resoeq2037}) that the electromagnetic fields in both
interior and exterior layers are of the form
\begin{eqnarray}
{\it
E}_{1\rho}&=&\frac{ih_{1}}{\sqrt{k_{1}^{2}-h_{1}^{2}}}J'_{m}\left(\sqrt{k_{1}^{2}-h_{1}^{2}}\rho\right)\left\{
{\begin{array}{*{20}c}
   {\cos m\varphi }  \\
   {\sin m\varphi}  \\
\end{array}}\right\},                \nonumber \\
{\it
E}_{1\varphi}&=&\frac{imh_{1}}{\left(k_{1}^{2}-h_{1}^{2}\right)\rho}J_{m}\left(\sqrt{k_{1}^{2}-h_{1}^{2}}\rho\right)\left\{
{\begin{array}{*{20}c}
   {\sin m\varphi }  \\
   {-\cos m\varphi}  \\
\end{array}}\right\},
                          \nonumber \\
   {\it
E}_{1z}&=&J_{m}\left(\sqrt{k_{1}^{2}-h_{1}^{2}}\rho\right)\left\{
{\begin{array}{*{20}c}
   {\cos m\varphi }  \\
   {\sin m\varphi}  \\
\end{array}}\right\},
\end{eqnarray}

\begin{eqnarray}
{\mathcal
H}_{1\rho}&=&\frac{imk_{1}^{2}}{\left(k_{1}^{2}-h_{1}^{2}\right)\rho}\frac{1}{\mu_{1}}J_{m}\left(\sqrt{k_{1}^{2}-h_{1}^{2}}\rho\right)\left\{
{\begin{array}{*{20}c}
   {-\sin m\varphi }  \\
   {\cos m\varphi}  \\
\end{array}}\right\},                  \nonumber \\
{\mathcal
H}_{1\varphi}&=&\frac{ik_{1}^{2}}{\sqrt{k_{1}^{2}-h_{1}^{2}}}\frac{1}{\mu_{1}}J'_{m}\left(\sqrt{k_{1}^{2}-h_{1}^{2}}\rho\right)\left\{
{\begin{array}{*{20}c}
   {\cos m\varphi }  \\
   {\sin m\varphi}  \\
\end{array}}\right\}, \nonumber \\
{\mathcal H}_{1z}&=& 0,
\end{eqnarray}
and
\begin{eqnarray}
{\it
E}_{2\rho}&=&\frac{ih_{2}}{\sqrt{k_{2}^{2}-h_{2}^{2}}}\left[AJ'_{m}\left(\sqrt{k_{2}^{2}-h_{2}^{2}}\rho\right)+BN'_{m}\left(\sqrt{k_{2}^{2}-h_{2}^{2}}\rho\right)\right]\left\{
{\begin{array}{*{20}c}
   {\cos m\varphi }  \\
   {\sin m\varphi}  \\
\end{array}}\right\},                \nonumber \\
{\it
E}_{2\varphi}&=&\frac{imh_{2}}{\left(k_{2}^{2}-h_{2}^{2}\right)\rho}\left[AJ_{m}\left(\sqrt{k_{2}^{2}-h_{2}^{2}}\rho\right)+BN_{m}\left(\sqrt{k_{2}^{2}-h_{2}^{2}}\rho\right)\right]\left\{
{\begin{array}{*{20}c}
   {\sin m\varphi }  \\
   {-\cos m\varphi}  \\
\end{array}}\right\},
                          \nonumber \\
   {\it
E}_{2z}&=&\left[AJ_{m}\left(\sqrt{k_{2}^{2}-h_{2}^{2}}\rho\right)+BN_{m}\left(\sqrt{k_{2}^{2}-h_{2}^{2}}\rho\right)\right]\left\{
{\begin{array}{*{20}c}
   {\cos m\varphi }  \\
   {\sin m\varphi}  \\
\end{array}}\right\},
\end{eqnarray}

\begin{eqnarray}
{\mathcal
H}_{2\rho}&=&\frac{imk_{2}^{2}}{\left(k_{2}^{2}-h_{2}^{2}\right)\rho}\frac{1}{\mu_{2}}\left[AJ_{m}\left(\sqrt{k_{2}^{2}-h_{2}^{2}}\rho\right)+BN_{m}\left(\sqrt{k_{2}^{2}-h_{2}^{2}}\rho\right)\right]\left\{
{\begin{array}{*{20}c}
   {-\sin m\varphi }  \\
   {\cos m\varphi}  \\
\end{array}}\right\},                  \nonumber \\
{\mathcal
H}_{2\varphi}&=&\frac{ik_{2}^{2}}{\sqrt{k_{2}^{2}-h_{2}^{2}}}\frac{1}{\mu_{2}}\left[AJ'_{m}\left(\sqrt{k_{2}^{2}-h_{2}^{2}}\rho\right)+BN'_{m}\left(\sqrt{k_{2}^{2}-h_{2}^{2}}\rho\right)\right]\left\{
{\begin{array}{*{20}c}
   {\cos m\varphi }  \\
   {\sin m\varphi}  \\
\end{array}}\right\}, \nonumber \\
{\mathcal H}_{2z}&=& 0,
\end{eqnarray}
where $J_{m}$ and $N_{m}$ denote Bessel functions and Norman
functions, respectively, and
$k_{1}=\sqrt{\epsilon_{1}\mu_{1}}\frac{\omega}{c}$,
$k_{2}=\sqrt{\epsilon_{2}\mu_{2}}\frac{\omega}{c}$.

By using the boundary conditions ${\mathcal
E}_{2z}|_{\rho=R_{2}}=0$, ${\mathcal
E}_{2\varphi}|_{\rho=R_{2}}=0$ (due to the perfectly conducting
medium at the boundary $\rho=R_{2}$), one can determine the
relationship between $A$ and $B$, {\it i.e.},
\begin{equation}
 B=-A\frac{J_{m}\left(\sqrt{k_{2}^{2}-h_{2}^{2}}R_{2}\right)}{N_{m}\left(\sqrt{k_{2}^{2}-h_{2}^{2}}R_{2}\right)}.
 \label{resoeq2010}
\end{equation}
By using the boundary conditions ${\mathcal
E}_{1z}|_{R1}={\mathcal E}_{2z}|_{R1}$, ${\mathcal
E}_{1\varphi}|_{R1}={\mathcal E}_{2\varphi}|_{R1}$, one can obtain
\begin{eqnarray}
& & J_{m}\left(\sqrt{k_{1}^{2}-h_{1}^{2}}R_{1}\right)=AJ_{m}\left(\sqrt{k_{2}^{2}-h_{2}^{2}}R_{1}\right)+BN_{m}\left(\sqrt{k_{2}^{2}-h_{2}^{2}}R_{1}\right),                 \nonumber \\
& &
\frac{h_{1}}{k_{1}^{2}-h_{1}^{2}}J_{m}\left(\sqrt{k_{1}^{2}-h_{1}^{2}}R_{1}\right)=\frac{h_{2}}{k_{2}^{2}-h_{2}^{2}}\left[AJ_{m}\left(\sqrt{k_{2}^{2}-h_{2}^{2}}R_{1}\right)+BN_{m}\left(\sqrt{k_{2}^{2}-h_{2}^{2}}R_{1}\right)\right].
\label{resoeq2011}
\end{eqnarray}
The second equation in Eq.(\ref{resoeq2011}) is employed to
determine the relation between $h_{1}$ and $h_{2}$. With the help
of the boundary conditions ${\mathcal H}_{1z}|_{R1}={\mathcal
H}_{2z}|_{R1}$ , ${\mathcal H}_{1\varphi}|_{R1}={\mathcal
H}_{2\varphi}|_{R1}$, one can arrive at
\begin{equation}
\frac{n_{1}}{\sqrt{k_{1}^{2}-h_{1}^{2}}}\frac{1}{\mu_{1}}J'_{m}\left(\sqrt{k_{1}^{2}-h_{1}^{2}}R_{1}\right)=\frac{n_{2}}{\sqrt{k_{2}^{2}-h_{2}^{2}}}\frac{1}{\mu_{2}}\left[AJ'_{m}\left(\sqrt{k_{2}^{2}-h_{2}^{2}}R_{1}\right)+BN'_{m}\left(\sqrt{k_{2}^{2}-h_{2}^{2}}R_{1}\right)\right].
\label{resoeq2012}
\end{equation}
Combination of the first equation in Eq.(\ref{resoeq2011}) and
(\ref{resoeq2012}), we have
\begin{equation}
\frac{n_{1}}{\mu_{1}\sqrt{k_{1}^{2}-h_{1}^{2}}}\frac{J'_{m}\left(\sqrt{k_{1}^{2}-h_{1}^{2}}R_{1}\right)}{J_{m}\left(\sqrt{k_{1}^{2}-h_{1}^{2}}R_{1}\right)}=\frac{n_{2}}{\mu_{2}\sqrt{k_{2}^{2}-h_{2}^{2}}}\frac{AJ'_{m}\left(\sqrt{k_{2}^{2}-h_{2}^{2}}R_{1}\right)+BN'_{m}\left(\sqrt{k_{2}^{2}-h_{2}^{2}}R_{1}\right)}{AJ_{m}\left(\sqrt{k_{2}^{2}-h_{2}^{2}}R_{1}\right)+BN_{m}\left(\sqrt{k_{2}^{2}-h_{2}^{2}}R_{1}\right)},
\end{equation}
which may be viewed as the restriction condition for the
cylindrical cavity resonator. If we choose a typical case with
$h_{1}=h_{2}=0$, then the obtained restriction condition is
simplified to be
\begin{equation}
\frac{1}{\mu_{1}}\frac{J'_{m}\left(k_{1}R_{1}\right)}{J_{m}\left(k_{1}R_{1}\right)}=\frac{1}{\mu_{2}}\frac{AJ'_{m}\left(k_{2}R_{1}\right)+BN'_{m}\left(k_{2}R_{1}\right)}{AJ_{m}\left(k_{2}R_{1}\right)+BN_{m}\left(k_{2}R_{1}\right)}.
\label{reso2013}
\end{equation}
Substitution of the relation (\ref{resoeq2010}) into
(\ref{reso2013}) yields
\begin{equation}
\frac{1}{\mu_{1}}\frac{J'_{m}\left(k_{1}R_{1}\right)}{J_{m}\left(k_{1}R_{1}\right)}=\frac{1}{\mu_{2}}\frac{J'_{m}\left(k_{2}R_{1}\right)N_{m}\left(k_{2}R_{2}\right)-J_{m}\left(k_{2}R_{2}\right)N'_{m}\left(k_{2}R_{1}\right)}{J_{m}\left(k_{2}R_{1}\right)N_{m}\left(k_{2}R_{2}\right)-J_{m}\left(k_{2}R_{2}\right)N_{m}\left(k_{2}R_{1}\right)}.
\label{resoeq2015}
\end{equation}
Eq.(\ref{resoeq2015}) is just the simplified restriction condition
for the cylindrical cavity resonator.

Similar to the analysis presented in Sec. I, it is readily shown
that by introducing the left-handed media such cylindrical cavity
will also act as a compact thin subwavelength resonator. The
discussion on this subject will not be performed further in the
present paper.
\subsection{A spherical thin subwavelength cavity resonator}

Here we will consider briefly the restriction equation for a
spherical cavity to be a thin subwavelength cavity resonator
containing left-handed media. Assume that the permittivity,
permeability and radius of media in the interior and exterior
layers of this double-layer cavity resonator are $\epsilon_{1}$,
$\mu_{1}$, $\rho_{1}$ and $\epsilon_{2}$, $\mu_{2}$, $\rho_{2}$,
respectively, and that the boundary medium at $\rho=\rho_{2}$ is
the perfectly conducting material. Note that here the functions,
symbols and quantities are adopted in the paper\cite{Shen}, {\it
e.g.}, $N_{1}$, $N_{2}$ denote the refractive indices of interio
and exterior layers, and $j_{n}$ and $h_{n}^{(1)}$ are the
spherical Bessel and Hankel functions.

According to the paper\cite{Shen}, it follows from the boundary
condition at $\rho=\rho_{2}$ that

\begin{eqnarray}
& & a^{m}_{n}j_{n}(N_{2}\rho_{2})+a_{n}^{\bar{m}}h^{(1)}_{n}(N_{2}\rho_{2})=0,                 \nonumber \\
& &
b^{m}_{n}[N_{2}\rho_{2}j_{n}(N_{2}\rho_{2})]'+b_{n}^{\bar{m}}\left[N_{2}\rho_{2}h^{(1)}_{n}(N_{2}\rho_{2})\right]'=0.
\label{resoeq41}
\end{eqnarray}
The roles of Eqs.(\ref{resoeq41}) is to determine the Mie
coefficients $a_{n}^{\bar{m}}$ and $b_{n}^{\bar{m}}$ in terms of
$a^{m}_{n}$ and $b^{m}_{n}$.

At the boundary $\rho=\rho_{1}$, it follows from the boundary
condition ${\bf i}_{1}\times {\bf E}_{\rm m}={\bf i}_{1}\times
{\bf E}_{\rm t}$ that one can obtain
\begin{eqnarray}
& &   a^{m}_{n}j_{n}(N_{2}\rho_{1})+a_{n}^{\bar{m}}h^{(1)}_{n}(N_{2}\rho_{1})=a_{n}^{t}j_{n}(N_{1}\rho_{1}),               \nonumber \\
& &
N_{1}b^{m}_{n}\left[N_{2}\rho_{1}j_{n}(N_{2}\rho_{1})\right]'+N_{1}b_{n}^{\bar{m}}\left[N_{2}\rho_{1}h^{(1)}_{n}(N_{2}\rho_{1})\right]'=N_{2}b_{n}^{t}\left[N_{1}\rho_{1}j_{n}(N_{1}\rho_{1})\right]'.
\label{resoeq42}
\end{eqnarray}
The role of the first and second equations in Eq.(\ref{resoeq42})
is to obtain the expressions for the Mie coefficients $a_{n}^{t}$
and $b_{n}^{t}$ in terms of $a^{m}_{n}$ and $b^{m}_{n}$,
respectively.

In the same manner, at the boundary $\rho=\rho_{1}$, it follows
from the boundary condition ${\bf i}_{1}\times {\bf H}_{\rm
m}={\bf i}_{1}\times {\bf H}_{\rm t}$ that one can obtain
\begin{eqnarray}
& &   \mu_{1}\left\{a^{m}_{n}\left[N_{2}\rho_{1}j_{n}(N_{2}\rho_{1})\right]'+a_{n}^{\bar{m}}\left[N_{2}\rho_{1}h^{(1)}_{n}(N_{2}\rho_{1})\right]'\right\}=\mu_{2}a_{n}^{t}\left[N_{1}\rho_{1}j_{n}(N_{1}\rho_{1})\right]',               \nonumber \\
& &    N_{2}
\mu_{1}\left[b^{m}_{n}j_{n}(N_{2}\rho_{1})+b_{n}^{\bar{m}}h^{(1)}_{n}(N_{2}\rho_{1})\right]=N_{1}\mu_{2}b_{n}^{t}j_{n}(N_{1}\rho_{1}).
\label{resoeq43}
\end{eqnarray}
Insertion of the expressions for the Mie coefficients
$a_{n}^{\bar{m}}$ and $b_{n}^{\bar{m}}$ and $a_{n}^{t}$ and
$b_{n}^{t}$ in terms of $a^{m}_{n}$ and $b^{m}_{n}$ obtained by
Eqs.(\ref{resoeq41}) and (\ref{resoeq42}) into
Eqs.(\ref{resoeq43}) will lead to a set of equations of the Mie
coefficients $a^{m}_{n}$ and $b^{m}_{n}$. In order to have a
nontrivial solutions of $a^{m}_{n}$ and $b^{m}_{n}$, the
determinant in Eqs.(\ref{resoeq43}) must vanish. Thus we will
obtain a restriction condition for the spherical cavity resonator.

By analogy with the analysis presented in Sec. I, it is easily
verified that by involving the left-handed media such spherical
cavity will also serve as a compact thin subwavelength resonator.
In a word, the possibility to satisfy the boundary conditions for
small distances between metal plates is based on the fact that
plane waves in Veselago media are backward waves, meaning that the
phase shift due to propagation in a usual slab can be compensated
by a negative phase shift inside a backward-wave
slab\cite{Engheta,Tretyakov}. We will not discuss further this
topic in this paper.

\section{A quantum mechanical problem: cone-angle-independent geometric phases and photon frequency shift in biaxially
anisotropic left-handed media}

Berry's discovery\cite{Berry} that there exists a topological
(geometric) phase (in addition to the dynamical phase that is
familiar to physicists) in quantum mechanical wavefunction of an
adiabatic process opens up new opportunities for investigating the
global and topological properties of quantum evolution. Geometric
phases thereby gained particular and considerable attention of
many researchers in various fields such as quantum
mechanics\cite{AA}, differential geometry\cite{Simon}, gravity
theory\cite {Furtado,Shen2}, atomic and molecular physics\cite
{Kuppermann,Kuppermann2,Levi}, nuclear physics\cite {Wagh},
quantum optics\cite{Gong}, condensed matter
physics\cite{Taguchi,Falci}, molecular structures and chemical
reaction\cite{Kuppermann} as well. More recently, many authors
concentrated on their special attention on the potential
applications of geometric phases to the geometric (topological)
quantum computation, quantum decoherence and related
topics\cite{Wangzd,Wangxb}. It is well known that, due to its
global and topological feature, geometric phases depends upon the
solid angle subtended at the parameter space of Hamiltonian. For
example, the adiabatic cyclic geometric phases of photons
propagating inside a noncoplanar optical fiber is $2\pi
\sigma(1-\cos \theta)$ with $\sigma$ denoting the photon helicity
eigenvalue, where $2\pi (1-\cos \theta)$ is just the expression
for the solid angle (with $2\theta$ being the cone angle)
subtended by a curve traced by the direction of wave vector of
light, at the center of photon momentum space\cite{Chiao,Tomita}.
In contrast, in this paper, we will propose a new geometric phase
that is independent of cone angle, by taking into account the
light propagation in a curved fiber made of biaxially anisotropic
left-handed media.

It should be noted that most of the recent theoretical works
discussed mainly the characteristics of electromagnetic wave
propagation through {\it isotropic} left-handed media, but up to
now, the left-handed media that have been prepared successfully
experimentally are actually {\it anisotropic} in nature, and it
may be very difficult to prepare an isotropic left-handed
medium\cite{Smith,Hu}. Here we consider the wave propagation
inside such biaxially anisotropic left-handed medium the
permittivity and permeability tensors of which are written as
follows
\begin{equation}
(\hat{\epsilon})_{ik}=\left(\begin{array}{cccc}
\epsilon  & 0 & 0 \\
0 &   -\epsilon & 0  \\
 0 &  0 &  \epsilon_{3}
 \end{array}
 \right),                 \qquad          (\hat{\mu})_{ik}=\left(\begin{array}{cccc}
-\mu  & 0 & 0 \\
0 &   \mu & 0  \\
 0 &  0 &  \mu_{3}
 \end{array}
 \right) .              \label{queq1}
\end{equation}
If the propagation vector of time-harmonic electromagnetic wave is
${\bf k}=(0, 0, k)$, then according to the Maxwellian Equations,
one can arrive at ${\bf k}\times {\bf E}=(-kE_{2}, kE_{1}, 0)$,
$[(\hat{\mu})_{ik}H_{k}]=(-\mu H_{1}, \mu H_{2}, 0)$. It follows
from the Faraday's electromagnetic induction law $\nabla
\times{\bf E}=-\frac{\partial{\bf B}}{\partial t}$ that
$H_{1}=\frac{kE_{2}}{\mu\mu_{0}\omega}$ and
$H_{2}=\frac{kE_{1}}{\mu\mu_{0}\omega}$. Thus, the third component
of Poynting vector of this time-harmonic wave is obtained
\begin{equation}
S_{3}=E_{1}H_{2}-E_{2}H_{1}=\frac{k}{\mu\mu_{0}\omega}(E_{1}^{2}-E_{2}^{2}),
\label{queq2}
\end{equation}
which implies that the Poynting vectors corresponding to the
$E_{1}$- and $E_{2}$- fields are of the form
\begin{equation}
{\bf S}^{(1)}=\frac{E_{1}^{2}}{\mu\mu_{0}\omega}{\bf k},    \quad
{\bf S}^{(2)}=-\frac{E_{2}^{2}}{\mu\mu_{0}\omega}{\bf k},
\label{queq3}
\end{equation}
respectively. It is apparently seen from (\ref{queq3}) that the
direction of wave vector ${\bf S}^{(2)}$ is opposite to that of
${\bf S}^{(1)}$. This, therefore, means that if $\mu>0$, then for
the $E_{1}$ field, this biaxially anisotropic medium characterized
by (\ref{queq1}) is like a right-handed material (regular
material) whereas for the $E_{2}$ field, it serves as a
left-handed one.

In view of above discussions, it is concluded that inside the
above biaxially anisotropic medium, the wave vectors of $E_{1}$-
and $E_{2}$- fields of propagating planar wave are apposite to
each other. Consider a hypothetical optical fiber that is made of
this biaxially anisotropic left-handed medium, inside which the
wave vector of $E_{1}$ field propagating is assumed to be ${\bf
k(t)}=k(\sin\theta \cos\varphi, \sin\theta \sin \varphi ,
\cos\theta)$. If both $\theta$ and $\varphi$ are nonvanishing,
then this fiber is noncoplanarly curved and in consequence the
geometric phases of light will arise. It is readily verified from
(\ref{queq3}) that the wave vector of $E_{2}$- field is $-{\bf
k(t)}=k(\sin\theta' \cos\varphi', \sin\theta' \sin \varphi' ,
\cos\theta')$ with $\theta'=\pi-\theta$ and
$\varphi'=\varphi+\pi$. Note that for the latter case ({\it i.e.},
$E_{2}$- field), the expression for the time-dependent coefficient
in photon geometric phases changes from
$\int_{0}^{t}\dot{\varphi}(t^{^{\prime }})\left[1-\cos \theta
(t^{^{\prime }})\right]{\rm d}t^{^{\prime }}$ to
$\int_{0}^{t}\dot{\varphi}(t^{^{\prime }})\left[1+\cos \theta
(t^{^{\prime }})\right]{\rm d}t^{^{\prime }}$ (because of
$\theta\rightarrow\pi-\theta$ and $\varphi\rightarrow\varphi+\pi$
). In the next subsection these results will be useful in
calculating the cone angle independent geometric phases of
circularly polarized light in biaxially anisotropic left-handed
media.

Note that such biaxially anisotropic medium can be fabricated by
current technology\cite{Smith,Hu}. So, in what follows we will
consider the geometric phases inside a noncoplanarly curved
optical fiber made of the biaxially anisotropic left-handed media
mentioned above. The time-dependent Schr\"{o}dinger equation that
governs the time development of the photon wavefunction in a
curved fiber (composed of regular isotropic media) is written in
the form
\begin{equation}
i\frac{\partial \left| \sigma ,{\bf{k}}(t),
n_{\sigma}\right\rangle }{\partial t}=\frac{ {\bf{k}}(t)\times
\dot{\bf{k}}(t)}{k^{2}}\cdot {\bf{S}}\left| \sigma ,{\bf{k}}(t),
n_{\sigma} \right\rangle      \label{queq12}
\end{equation}
with ${\bf{S}}$ being the spin operator of photon fields. By
making use of the Lewis-Riesenfeld invariant theory and the
invariant-related unitary transformation
formulation\cite{Riesenfeld,Gao1,Jap}, we obtain the exact
particular solutions $ \left| \sigma ,{\bf{k}}(t), n_{\sigma}
\right\rangle=\exp \left[\frac{1}{i}\phi _{\sigma }^{\rm
(g)}(t)\right]V(t)\left| n_{\sigma} \right\rangle$ to
Eq.(\ref{queq12}), where $\left| n_{\sigma}
\right\rangle\equiv\left| \sigma,{\bf{k}}(t=0), n_{\sigma}
\right\rangle$ is the initial polarized photon state,
$V(t)=\exp[\beta(t){\bf S_{+}}-\beta^{\ast}(t){\bf S_{-}}]$ with
$\beta(t)=-[\theta(t)/2]\exp[-i\varphi(t)]$,
$\beta^{\ast}(t)=-[\theta(t)/2]\exp[i\varphi(t)]$\cite{Zhu,Shenpla}
and ${\bf S}_{\pm}={\bf S}_{1}\pm i{\bf S}_{2}$. The noncyclic
nonadiabatic geometric phase is given as follows

\begin{equation}
\phi _{\sigma }^{\rm
(g)}(t)=\left\{{\int_{0}^{t}\dot{\varphi}(t^{^{\prime }})[1-\cos
\theta (t^{^{\prime }})]{\rm d}t^{^{\prime }}}\right\}\left\langle
n_{\sigma} \right|S_{3}\left|n_{\sigma} \right\rangle.
\label{queq4}
\end{equation}
Note here that the photon states $\left|n_{\sigma} \right\rangle$
corresponding to the right- and left- handed polarized light with
the photon occupation numbers being $n_{R}$ and $n_{L}$ are
respectively defined to be
$|n_{R}\rangle=\frac{\left(a^{\dagger}_{R}\right)^{n_{R}}}{\sqrt{n_{R}!}}|0_{R}\rangle$
and
$|n_{L}\rangle=\frac{\left(a^{\dagger}_{L}\right)^{n_{L}}}{\sqrt{n_{L}!}}|0_{L}\rangle$.
The creation operators of left- and right- handed circularly
polarized light are constructed in terms of $a^{\dagger}_{1}$ and
$a^{\dagger}_{2}$, {\it i.e.},
$a^{\dagger}_{L}=\frac{a^{\dagger}_{1}+ia^{\dagger}_{2}}{\sqrt{2}}$
and
$a^{\dagger}_{R}=\frac{a^{\dagger}_{1}-ia^{\dagger}_{2}}{\sqrt{2}}$,
respectively\cite{Bjorken}, where $a^{\dagger}_{1}$ and
$a^{\dagger}_{2}$ are the creation operators corresponding to the
two mutually perpendicular polarization vectors, which are also
orthogonal to the wave vector ${\bf k}$ of the time harmonic
electromagnetic wave.

Note, however, that according to Eq.(\ref{queq3}), in the above
biaxially anisotropic left-handed media the wave vector of light
corresponding to the two mutually perpendicular polarization
vectors are antiparallel to each other. Since the wave vector of
$E_{1}$-field in such anisotropic left-handed media is
antiparallel to that of $E_{2}$-field, we should first calculate
the following expectation value $\langle n_{R}
|a_{1}^{\dagger}a_{1}|n_{R}\rangle$, $\langle n_{R}
|a_{2}^{\dagger}a_{2}|n_{R}\rangle$, $\langle n_{L}
|a_{1}^{\dagger}a_{1}|n_{L}\rangle$ and $\langle n_{L}
|a_{2}^{\dagger}a_{2}|n_{L}\rangle$ in order to obtain the
expressions for geometric phases of left- and right- handed
circularly polarized light in this peculiar biaxially anisotropic
left-handed medium. With the help of
$a_{1}|n_{R}\rangle=\sqrt{\frac{n_{R}}{2}}|n_{R}-1\rangle$,
$\langle n_{R} |a_{1}^{\dagger}=\sqrt{\frac{n_{R}}{2}}\langle
n_{R}-1|$,
$a_{2}|n_{R}\rangle=i\sqrt{\frac{n_{R}}{2}}|n_{R}-1\rangle$ and $
\langle n_{R} |a_{2}^{\dagger}=-i\sqrt{\frac{n_{R}}{2}}\langle
n_{R}-1|$, we obtain $\langle n_{R}
|a_{1}^{\dagger}a_{1}|n_{R}\rangle=\frac{n_{R}}{2}$ and $\langle
n_{R} |a_{2}^{\dagger}a_{2}|n_{R}\rangle=\frac{n_{R}}{2}$. Hence
the nonadiabatic noncyclic geometric phases of right-handed
polarized photons corresponding to $E_{1}$- and $E_{2}$- fields
are
\begin{equation}
\phi_{R}^{(1)}(t)=\frac{n_{R}}{2}\left\{{\int_{0}^{t}\dot{\varphi}(t)(t^{^{\prime
}})\left[1-\cos \theta (t^{^{\prime }})\right]{\rm d}t^{^{\prime
}}}\right\},   \quad
\phi_{R}^{(2)}(t)=\frac{n_{R}}{2}\left\{{\int_{0}^{t}\dot{\varphi}(t)(t^{^{\prime
}})\left[1+\cos \theta (t^{^{\prime }})\right]{\rm d}t^{^{\prime
}}}\right\},
\end{equation}
respectively, and their sum is
\begin{equation}
\phi_{R}(t)=\phi_{R}^{(1)}(t)+\phi_{R}^{(2)}(t)=n_{R}\int_{0}^{t}\dot{\varphi}(t)(t^{^{\prime
}}){\rm d}t^{^{\prime }},
\end{equation}
which is independent of the cone angle $\theta (t)$ of photon
momentum $\bf k$ space.

In the same fashion, we obtain $
a_{1}|n_{L}\rangle=\sqrt{\frac{n_{L}}{2}}|n_{L}-1\rangle$, $
\langle n_{L} |a_{1}^{\dagger}=\sqrt{\frac{n_{L}}{2}}\langle
n_{L}-1|$, $\langle n_{L}
|a_{1}^{\dagger}a_{1}|n_{L}\rangle=\frac{n_{L}}{2}$; and
$a_{2}|n_{L}\rangle=-i\sqrt{\frac{n_{L}}{2}}|n_{L}-1\rangle$,
$\langle n_{L} |a_{2}^{\dagger}=i\sqrt{\frac{n_{L}}{2}}\langle
n_{L}-1|$, $\langle n_{L}
|a_{2}^{\dagger}a_{2}|n_{L}\rangle=\frac{n_{L}}{2}$. Hence the
nonadiabatic noncyclic geometric phases of left-handed polarized
photons corresponding to $E_{1}$- and $E_{2}$- fields are
\begin{equation}
\phi_{L}^{(1)}(t)=-\frac{n_{L}}{2}\left\{{\int_{0}^{t}\dot{\varphi}(t)(t^{^{\prime
}})\left[1-\cos \theta (t^{^{\prime }})\right]{\rm d}t^{^{\prime
}}}\right\},         \quad
\phi_{L}^{(2)}(t)=-\frac{n_{L}}{2}\left\{{\int_{0}^{t}\dot{\varphi}(t)(t^{^{\prime
}})\left[1+\cos \theta (t^{^{\prime }})\right]{\rm d}t^{^{\prime
}}}\right\},
\end{equation}
respectively, and their sum is
\begin{equation}
\phi_{L}(t)=\phi_{L}^{(1)}(t)+\phi_{L}^{(2)}(t)=-n_{L}\int_{0}^{t}\dot{\varphi}(t)(t^{^{\prime
}}){\rm d}t^{^{\prime }},
\end{equation}
which is also independent of the cone angle $\theta (t)$.

Thus the total geometric phases of left- and right- handed
polarized photons is given by
\begin{equation}
\phi_{\rm tot}^{\rm (g)
}(t)=\phi_{R}(t)+\phi_{L}(t)=\left(n_{R}-n_{L}\right)\int_{0}^{t}\dot{\varphi}(t)(t^{^{\prime
}}){\rm d}t^{^{\prime }},                    \label{queqq}
\end{equation}
which differs from the total geometric phases of circularly light
in the regular curved fiber only by a cone angle $\theta (t)$ of
photon momentum $\bf k$ space.

In the Chiao-Wu-Tomita fiber experiment\cite{Chiao,Tomita}, the
geometric phases of photons is related close to the the geometric
nature of the pathway (expressed by $\theta$ and ${\varphi}(t)$)
along which quantum systems evolve. However, in our case where the
light propagating inside a biaxially anisotropic left-handed
media, the geometric phase (\ref{queqq}) is related only to the
precessional frequency $\dot{\varphi}$ of wave propagation on the
helicoid inside the curved fiber.

Consider the Chiao-Wu adiabatic cyclic case in which the rotating
angular frequency ({\it i.e.}, the precessional frequency) of
photon moving on the helicoid reads $\dot{\varphi}=\Omega $ with
$\Omega =\frac{2\pi c}{\sqrt{d^{2}+(4\pi a)^{2}}}$\cite{Shenpla}
where $d$ and $a$ respectively denote the pitch length and the
radius of the helix, and $c$ is the speed of light. It follows
from (\ref{queqq}) that the photon frequency shift, $\Delta$, is
$\sigma \Omega$, namely, for the right-handed circularly polarized
photon, the frequency shift $\Delta$ is $\Omega$ (due to
$\sigma=+1$), while for the left-handed circularly polarized
photon, $\Delta$ is $-\Omega$ (due to $\sigma=-1$). Generally
speaking, one can obtain the photon frequency shift via some
nonlinear optical effects such as Raman effect and left-right
coupling of circularly polarized light in biaxially gyrotropic
left-handed media\cite{arxiv}, which may all be viewed as {\it
dynamical} scheme. However, here we propose a so-called {\it
geometric} scheme to achieve photon frequency shift, since the
precessional frequency $\Omega$ depends on both the pitch length
$d$ and the radius $a$ of the fiber helix. This, therefore, means
that this photon frequency shift is controllable by manipulating
the spatial shape and helix radius of the curved optical fiber.

For the present, physicists' control over the behavior of photons
has spread to include the photon number, phase\cite{Duan} and
polarization\cite{Muller} of light wave. If we could engineer all
the degrees of freedom of photons, our technology would benefit.
Already, the fiber communication, which simply guides light, has
revolutionized the telecommunications industry. Apparently, it is
of essential significance to control and utilize the degrees of
freedom of photons (photon number, polarization, helicity,
geometric phase, etc.) in information science and technology. In
this paper, we suggest a {\it geometric} scheme, which can control
the photon frequency shift by altering the curvature (and pitch
length and helix radius) of the noncoplarnarly curved optical
fiber.

\section{Extra phases of light at the interfaces between left- and right- handed media}
In this section, we will consider the effects of light appearing
at the interfaces between left- and right- handed media. In order
to treat this problem conveniently, we study the wave propagation
inside an optical fiber which is periodically modulated by
altering regular and negative media. Although it is doubtful
whether such periodically modulated fibers could be designed and
realized or not in experiments at the optical scale, it could be
argued that the work presented here can be considered only a
speculative one. But the method and results obtained via the use
of this optical fiber system composed by such sequences of right-
and left- handed materials can also be applied to the light
propagation at the interfaces between left- and right- handed
media in arbitrary geometric shapes of optical materials. In this
periodical optical fiber, helicity inversion (or the transitions
between helicity states) of photons may be easily caused by the
interaction of light field with media near both sides of the
interfaces between LRH materials. Since photon helicity inversion
at the interfaces mentioned above is a {\it time-dependent}
process, this new geometric phase arises during the light
propagates through the interfaces (in the following we will call
them the LRH interfaces) between left- and right- handed media,
where the anomalous refraction occurs when the incident lightwave
travels to the LRH interfaces. we think that, in the literature it
gets less attention than it deserves. In what follows we calculate
the photon wavefunction and corresponding extra phases (including
the geometric phases) in this physical process, and emphasize that
we should attach importance to this geometric phases when
considering the wave propagation near the LRH interfaces.
\subsection{Model Hamiltonian}
We now treat the helicity reversal problem of light wave adjacent
to the interfaces of left- and right- handed media. For
convenience, let us consider a hypothetical optical fiber that is
fabricated periodically from both left- and right- handed (LRH)
media with the optical refractive indices being $-n$ and $n$,
respectively. Thus the wave vector of photon moving along the
fiber is respectively $-n\frac{\omega }{c}$ in left-handed (LH)
section and $n\frac{\omega }{c}$ in right-handed (RH) section,
where $\omega$ and $c$ respectively denote the frequency and the
speed of light in a vacuum. For simplicity, we assume that the
periodical length, $b$, of LH is equal to that of RH in the fiber.
If the eigenvalue of photon helicity is $\sigma $ in right-handed
sections, then, according to the definition of helicity,
$h=\frac{\bf k}{\left| {\bf k}\right| }\cdot{\bf J}$ with ${\bf
J}$ denoting the total angular momentum of the photon, the
eigenvalue of helicity acquires a minus sign in left-handed
sections. We assume that at $t=0$ the light propagates in the
right-handed section and the initial eigenvalue of photon helicity
is $\sigma $. So, in the wave propagation inside the
LRH-periodical optical fiber, the helicity eigenvalue of $h$ is
then $\left(-\right) ^{m}\sigma $ with $m=\left[
\frac{ct}{nb}\right] $, where $\left[\frac{ct}{nb}\right] $
represents the integer part of $\frac{ct}{nb}$. It is clearly seen
that $\left(-\right) ^{m}$ stands for the switching on and off of
the helicity reversal, {\it i.e.}, the positive and negative value
of $\left(-\right) ^{m}$ alternate in different time intervals.
This, therefore, means that if $2k\left(\frac{nb}{c}\right) <t\leq
\left(2k+1\right) \left(\frac{nb}{c}\right) $, then
$\left(-\right) ^{m} =+1$, and if $\left(2k+1\right)
\left(\frac{nb}{c}\right) <t\leq (2k+2)\left(\frac{nb}{c}\right)
$, then $\left(-\right) ^{m} =-1$, where $k$ is zero or a positive
integer. It follows that the incidence of lightwave on the LRH
interfaces in the fiber gives rise to the transitions between the
photon helicity states ($\left| +\right\rangle $ and $\left|
-\right\rangle $). This enables us to construct a time-dependent
effective Hamiltonian
\begin{equation}
H\left(t\right) =\frac{1}{2}\omega \left(t\right) \left(
S_{+}+S_{-}\right)                     \label{exeq1}
\end{equation}
in terms of $\left| +\right\rangle $ and $\left| -\right\rangle $
to describes this instantaneous transition process of helicity
states at the LRH interfaces, where $S_{+}=\left| +\right\rangle
\left\langle -\right| ,S_{-}=\left| -\right\rangle \left\langle
+\right| ,S_{3}=\frac{1}{2}\left(\left| +\right\rangle
\left\langle +\right| -\left| -\right\rangle \left\langle -\right|
\right) $ satisfying the following SU(2) Lie algebraic commuting
relations $\left[ S_{+},S_{-}\right] =2S_{3}$ and $\left[
S_{3},S_{\pm }\right] =\pm S_{\pm }$. The time-dependent frequency
parameter $\omega \left(t\right) $ may be taken to be $\omega
\left(t\right) =\varsigma \frac{\rm d}{{\rm d}t}p\left(t\right) $,
where $p\left(t\right) =\left(-\right) ^{m}$ with $m=\left[
\frac{ct}{nb}\right] $, and $\varsigma$ is the coupling
coefficient, which can, in principle, be determined by the
physical mechanism of interaction between light fields and media.
Since $p\left(t\right) $ is a periodical function, by using the
analytical continuation procedure, it can be rewritten as the
following linear combinations of analytical functions
\begin{equation}
p\left(t\right) =\sum_{k=1}^{\infty }\frac{2}{k\pi }\left[
1-\left(-\right) ^{k}\right] \sin \left(\frac{k\pi c}{nb}t\right).
\label{exeq2}
\end{equation}

 In what follows, we solve the time-dependent Schr\"{o}dinger
equation (in the unit $\hbar =1$)
\begin{equation}
H\left(t\right) \left| \Psi _{\sigma }\left(t\right) \right\rangle
=i\frac{\partial }{\partial t}\left| \Psi _{\sigma }\left(
t\right) \right\rangle     \label{exeq3}
\end{equation}
governing the propagation of light in the LRH- periodical fiber.
According to the Lewis-Riesenfeld invariant theory\cite{Lewis},
the exact particular solution $\left| \Psi _{\sigma
}\left(t\right) \right\rangle$ of the time-dependent
Schr\"{o}dinger equation (\ref{exeq3}) is different from the
eigenstate of the invariant $I(t)$ only by a time-dependent $c$-
number factor $\exp \left[ \frac{1}{i}\phi _{\sigma
}\left(t\right) \right]$, where
\begin{equation}
\phi _{\sigma }\left(t\right)=\int_{0}^{t}\left\langle \Phi
_{\sigma }\left(t^{\prime }\right) \right|[H(t^{\prime
})-i\frac{\partial }{\partial t^{\prime }}]\left| \Phi _{\sigma
}\left(t^{\prime }\right) \right\rangle {\rm d}t^{\prime }
 \label{exeq4}
\end{equation}
with $\left| \Phi _{\sigma }\left(t\right) \right\rangle $ being
the eigenstate of the invariant $I(t)$ (corresponding to the
particular eigenvalue $\sigma$) and satisfying the eigenvalue
equation $I\left(t\right) \left| \Phi _{\sigma }\left(t\right)
\right\rangle =\sigma \left| \Phi _{\sigma }\left(t\right)
\right\rangle$, where the eigenvalue $\sigma$ of the invariant
$I(t)$ is {\it time-independent}. Thus we have
\begin{equation}
\left| \Psi _{\sigma }\left(t\right) \right\rangle =\exp \left[
\frac{1}{i}\phi _{\sigma }\left(t\right) \right] \left| \Phi
_{\sigma }\left(t\right) \right\rangle.    \label{exeq51}
\end{equation}
In order to obtain $\left| \Psi _{\sigma }\left(t\right)
\right\rangle$, we should first obtain the eigenstate $\left| \Phi
_{\sigma }\left(t\right) \right\rangle $ of the invariant $I(t)$.
\subsection{Photon geometric phases due to helicity inversions inside a periodical fiber made of left-handed media}
Here we investigate the time evolution of photon wavefunctions and
extra phases due to photon helicity inversion at the LRH
interfaces. As has been stated above, for convenience, we consider
the wave propagation inside a hypothetical optical fiber which is
composed periodically of left- and right- handed media. Now we
solve the time-dependent Schr\"{o}dinger equation. In accordance
with the Lewis-Riesenfeld theory, the invariant $I(t)$ is a
conserved operator ({\it i.e.}, it possesses {\it
time-independent} eigenvalues) and agrees with the following
Liouville-Von Neumann equation

\begin{equation}
\frac{\partial I(t)}{\partial t}+\frac{1}{i}[I(t),H(t)]=0.
                 \label{exeq5}
\end{equation}
It follows from Eq.(\ref{exeq5}) that the invariant $I(t)$ may
also be constructed in terms of $S_{\pm }$ and $S_{3}$, {\it
i.e.},

\begin{equation}
I\left(t\right) =2\left\{\frac{1}{2}\sin \theta \left(t\right)
\exp \left[ -i\varphi  \left(t\right)\right] S_{+}+\frac{1}{2}\sin
\theta \left(t\right) \exp \left[ i\varphi \left(t\right)\right]
S_{-}+\cos \theta \left(t\right) S_{3}\right\}. \label{exeq6}
\end{equation}
Inserting Eq.(\ref{exeq1}) and (\ref{exeq6}) into
Eq.(\ref{exeq5}), one can arrive at a set of auxiliary equations

\begin{eqnarray}
\exp \left[ -i\varphi \right] \left(\dot{\theta}\cos \theta
-i\dot{\varphi}\sin \theta \right) -i\omega \cos \theta =0,  \nonumber \\
\quad \dot{\theta}+\omega \sin \varphi =0,
 \label{exeq7}
\end{eqnarray}
which are used to determine the time-dependent parameters,
$\theta\left(t\right)$ and $\varphi\left(t\right)$, of the
invariant $I(t)$\cite{Lewis}.

It should be noted that we cannot easily solve the eigenvalue
equation $I\left(t\right) \left| \Phi _{\sigma }\left(t\right)
\right\rangle =\sigma \left| \Phi _{\sigma }\left(t\right)
\right\rangle$, for the time-dependent parameters $\theta\left(
t\right)$ and $\varphi\left(t\right)$ are involved in the
invariant (\ref{exeq6}). If, however, we could find (or construct)
a unitary transformation operator $V(t)$ to make $V^{\dagger
}(t)I(t)V(t)$ be {\it time-independent}, then the eigenvalue
equation problem of $I(t)$ is therefore easily resolved. According
to our experience for utilizing the invariant-related unitary
transformation formulation\cite{Shen2}, we suggest a following
unitary transformation operator

\begin{equation}
V\left(t\right) =\exp \left[ \beta \left(t\right) S_{+}-\beta
^{\ast }\left(t\right) S_{-}\right] ,
 \label{exeq8}
\end{equation}
where $\beta (t)$ and $\beta ^{\ast }(t)$ will be determined by
calculating $I_{\rm V}=V^{\dagger }(t)I(t)V(t)$ in what follows.

Calculation of $I_{\rm V}=V^{\dagger }(t)I(t)V(t)$ yields

\begin{equation}
I_{\rm V}=V^{\dagger }\left(t\right) I\left(t\right) V\left(
t\right) =2S_{3},
 \label{exeq9}
\end{equation}
if $\beta $ and $\beta ^{\ast }$ are chosen to be $\beta \left(
t\right) =-\frac{\theta \left(t\right) }{2}\exp \left[ -i\varphi
\left(t\right) \right]$, $\beta ^{\ast }\left(t\right)
=-\frac{\theta \left(t\right) }{2}\exp \left[ i\varphi \left(
t\right) \right] $. This, therefore, means that we can change the
{\it time-dependent} $I(t)$ into a {\it time-independent} $I_{\rm
V}$, and the result is $I_{\rm V}=2S_{3}$. Thus, the eigenvalue
equation of $I_{\rm V}$ is $I_{\rm V}\left| \sigma \right\rangle
=\sigma \left| \sigma \right\rangle $ with $\sigma=\pm 1$, and
consequently the eigenvalue equation of $I(t)$ is written $I\left(
t\right) V\left(t\right) \left| \sigma \right\rangle =\sigma
V\left(t\right) \left| \sigma \right\rangle $. So, we obtain the
eigenstate $\left| \Phi _{\sigma }\left(t\right) \right\rangle$ of
$I(t)$, {\it i.e.}, $\left| \Phi _{\sigma }\left(t\right)
\right\rangle=V\left(t\right)\left| \sigma \right\rangle$.

Correspondingly, $H(t)$ is transformed into

\begin{eqnarray}
H_{\rm V}\left(t\right) =V^{\dagger }\left(t\right) \left[ H\left(
t\right) -i\frac{\partial }{\partial t}\right] V\left(t\right)
\label{exeq11}
\end{eqnarray}
and the time-dependent Schr\"{o}dinger equation (\ref{exeq3}) is
rewritten
\begin{equation}
H_{\rm V}\left(t\right) \left| \Psi _{\sigma }\left(t\right)
\right\rangle_{\rm V} =i\frac{\partial }{\partial t}\left| \Psi
_{\sigma }\left(t\right) \right\rangle_{\rm V}   \label{exeq12}
\end{equation}
under the unitary transformation $V(t)$, where $ \left| \Psi
_{\sigma }\left(t\right) \right\rangle_{\rm V}=V^{\dagger }\left(
t\right) \left| \Psi _{\sigma }\left(t\right) \right\rangle$.

Further analysis shows that the exact particular solution $ \left|
\Psi _{\sigma }\left(t\right) \right\rangle_{\rm V}$ of the
time-dependent Schr\"{o}dinger equation (\ref{exeq12}) is
different from the eigenstate $\left| \sigma \right\rangle$ of the
{\it time-independent} invariant $I_{\rm V}$ only by a
time-dependent $c$- number factor $\exp \left[ \frac{1}{i}\phi
_{\sigma }\left(t\right) \right]$\cite{Lewis}, which is now
rewritten as $\exp \left \{ \int_{0}^{t}\left\langle \sigma\right|
H_{\rm V}\left(t^{\prime }\right)\left|\sigma \right\rangle {\rm
d}t^{\prime }\right\}$.

By using the auxiliary equations (\ref{exeq7}), the Glauber
formula and the Baker-Campbell-Hausdorff formula, it is verified
that $H_{\rm V}(t)$ depends only on the operator $S_{3}$, {\it
i.e.},

\begin{equation}
H_{\rm V}\left(t\right) =\left\{ \omega \left(t\right)\sin \theta
\left(t\right) \cos \varphi \left(t\right)+\dot{\varphi}\left(
t\right) \left[ 1-\cos \theta \left(t\right) \right] \right\}
S_{3}
 \label{exeq13}
\end{equation}
and the time-dependent $c$- number factor $\exp \left[
\frac{1}{i}\phi _{\sigma }\left(t\right) \right] $ is therefore
$\exp \left\{ \frac{1}{i}\left[ \phi _{\sigma }^{\left({\rm
d}\right) }\left(t\right) +\phi _{\sigma }^{\left({\rm g}\right)
}\left(t\right) \right] \right\} $, where the dynamical phase is

\begin{equation}
\phi _{\sigma }^{\left({\rm d}\right) }\left(t\right) =\sigma
\int_{0}^{t}\omega \left(t^{\prime }\right)\sin \theta \left(
t^{\prime }\right) \cos \varphi \left(t^{\prime }\right){\rm
d}t^{\prime } \label{exeq14}
\end{equation}
and the geometric phase is
\begin{equation}
\phi _{\sigma }^{\left({\rm g}\right) }\left(t\right) =\sigma
\int_{0}^{t}\dot{\varphi}\left(t^{\prime }\right) \left[ 1-\cos
\theta \left(t^{\prime }\right) \right] {\rm d}t^{\prime }.
 \label{exeq15}
\end{equation}

Hence the particular exact solution of the time-dependent
Schr\"{o}dinger equation (\ref{exeq3}) corresponding to the
particular eigenvalue, $\sigma$, of the invariant $I(t)$ is of the
form

\begin{equation}
\left| \Psi _{\sigma }\left(t\right) \right\rangle =\exp \left\{
\frac{1}{i}\left[ \phi _{\sigma }^{\left({\rm d}\right) }\left(
t\right) +\phi _{\sigma }^{\left({\rm g}\right) }\left(t\right)
\right] \right\} V\left(t\right) \left| \sigma \right\rangle.
 \label{exeq16}
\end{equation}

It follows from the obtained expression (\ref{exeq15}) for
geometric phase of photons that, if the frequency parameter
$\omega$ is small ({\it i.e.}, the adiabatic quantum process) and
then according to the auxiliary equations (\ref{exeq7}),
$\dot{\theta}\simeq 0$, the Berry phase (adiabatic geometric phase
) in a cycle ({\it i.e.}, one round trip, $T\simeq
\frac{2\pi}{\omega}$) of parameter space of invariant $I(t)$ is

\begin{equation}
\phi _{\sigma }^{\left({\rm g}\right) }\left(T\right)= 2\pi \sigma
(1-\cos \theta),
 \label{exeq17}
\end{equation}
where $2\pi (1-\cos \theta)$ is a solid angle over the parameter
space of the invariant $I(t)$, which means that the geometric
phase is related only to the geometric nature of the pathway along
which quantum systems evolve. Expression (\ref{exeq17}) is
analogous to the magnetic flux produced by a monopole of strength
$\sigma$ existing at the origin of the parameter space. This,
therefore, implies that geometric phases differ from dynamical
phases and involve the global and topological properties of the
time evolution of quantum systems.
\subsection{In biaxially anisotropic left-handed media}
Note that in the previous subsection, we treat the helicity
reversals of single photon in electromagnetic media made of
isotropic left- and right- handed materials. Now we consider this
problem in a system composed by a sequence of right-handed
(isotropic) and biaxially anisotropic left-handed media, the
permittivity and permeability of the latter is given
\begin{equation} (\hat{\epsilon})_{ik}=\left(\begin{array}{cccc}
\epsilon  & 0 & 0 \\
0 &   -\epsilon & 0  \\
 0 &  0 &  \epsilon_{3}
 \end{array}
 \right),                 \qquad          (\hat{\mu})_{ik}=\left(\begin{array}{cccc}
-\mu  & 0 & 0 \\
0 &   \mu & 0  \\
 0 &  0 &  \mu_{3}
 \end{array}
 \right) .
\end{equation}
It has been verified in Sec.III that for the $E_{1}$- field this
biaxially anisotropic medium can be regarded as a right-handed
material while for the $E_{2}$- field it can be considered a
left-handed one. This, therefore, implies that only the $E_{2}$-
field propagating through this biaxially anisotropic left-handed
medium will acquire an additional phase due to its helicity
inversion at the LRH interfaces. But since the $E_{2}$- field is
not the eigenmode of the photon helicity, we cannot obtain the
extra phases immediately by using the formula (\ref{exeq4}).
According to the treatment in Sec.III, one can arrive at the
expression for the total extra phases of circularly polarized
light (with the occupation numbers of polarized photons being
$n_{L}$ and $n_{R}$) propagating through the system made of a
sequence of isotropic regular media and biaxially anisotropic
left-handed one, and the result is written as
\begin{equation}
\phi_{\rm tot}(t)=\frac{n_{R}-n_{L}}{2}\left[\phi^{\rm
(d)}(t)+\phi^{\rm (g)}(t)\right]
\end{equation}
with $\phi^{\rm (d)}(t)+\phi^{\rm
(g)}(t)=\int_{0}^{t}\left\{\omega \left(t^{\prime }\right)\sin
\theta \left(t^{\prime }\right) \cos \varphi \left(t^{\prime
}\right)+\dot{\varphi}\left(t^{\prime }\right) \left[ 1-\cos
\theta \left(t^{\prime }\right) \right]\right\}{\rm d}t^{\prime
}$. It follows that if $n_{R}=n_{L}$, the added phase due to
photon helicity reversal on the interfaces between left- and
right- handed media vanishes.

Based on the restriction\cite{german}
\begin{equation}
n^{4}(2\omega+\Omega_{L})(2\omega+\Omega_{R})\Omega_{L}\Omega_{R}+\zeta^{4}(\omega+\Omega_{L})^{2}(\omega+\Omega_{R})^{2}=0
\label{exeqq27}
\end{equation}
imposed on the frequency shifts of circularly polarized light, we
now consider a possibility that only one of the polarized light,
say $E_{R}$, can be propagated in biaxially gyrotropic left-handed
media. If the frequency shift $\Omega_{L}$ of left-handed
polarized light is $-\omega-i\Gamma$, {\it i.e.},
$\omega+\Omega_{L}=-i\Gamma$, where $\Gamma$ is a positive real
number, then according to Eq.(\ref{exeqq27}), one can arrive at
\begin{equation}
n^{4}(\Gamma^{2}+\omega^{2})(2\omega+\Omega_{R})\Omega_{R}+\zeta^{4}\Gamma^{2}(\omega+\Omega_{R})^{2}=0.
\label{exeq419}
\end{equation}
It is easy to obtain $\Omega_{R}$ from Eq.(\ref{exeq419}), and the
result is given
$\Omega_{R}=-\omega+\gamma^{\frac{1}{2}}(1+\gamma)^{-\frac{1}{2}}\omega$
with
$\gamma=\frac{n^{4}(\Gamma^{2}+\omega^{2})}{\zeta^{4}\Gamma^{2}}$.
Thus, the frequency of right-handed polarized light is
\begin{equation}
\omega+\Omega_{R}=\gamma^{\frac{1}{2}}(1+\gamma)^{-\frac{1}{2}}\omega,
\label{exeq420}
\end{equation}
which means that the frequency of $E_{R}$ is modified by a factor
$\gamma^{\frac{1}{2}}(1+\gamma)^{-\frac{1}{2}}$. If no L-R
coupling exists, {\it i.e.}, $\zeta=0$, then $\gamma$ tends to
infinity and the factor
$\gamma^{\frac{1}{2}}(1+\gamma)^{-\frac{1}{2}}$ approaches unity,
and then the frequency shift $\Omega_{R}$ of right-handed
polarized light is vanishing, which can be easily seen in the
expression (\ref{exeq420}).

Note that in the case discussed above, the left-handed polarized
light in this type of media exponentially decreases (due to the
imaginary frequency $\omega+\Omega_{L}$, which is $-i\Gamma$)
while the right-handed one can be propagated, {\it i.e.}, only one
wave can be present in this media. So, in this case the additional
phase acquired by photon wavefunction due to helicity inversions
on LRH interfaces is
\begin{equation}
\phi_{\rm tot}(t)=\frac{n_{R}}{2}\left[\phi^{\rm (d)}(t)+\phi^{\rm
(g)}(t)\right].
\end{equation}

Additionally, it is of interest to show that the above scheme is
applicable to the detection of quantum-vacuum geometric
phases\cite{Shenpla}. Since the left-handed polarized light
(including the quantum vacuum fluctuation corresponding to the
left-handed polarized light) cannot be propagated in this
media\cite{Klimov,Veselago}, the quantum-vacuum geometric phase of
right-handed polarized light will not be cancelled by that of
left-handed one, namely, the only retained quantum-vacuum
geometric phase is that of right-handed circularly polarized light
and therefore it is possible for the nonvanishing quantum-vacuum
geometric phases to be detected in experiments.

\subsection{Discussion: physical significance and potential applications}

It is worthwhile to point out that the geometric phase of photons
caused by helicity inversion presented here is of quantum level.
However, whether the Chiao-Wu geometric phase due to the spatial
geometric shape of fiber is of quantum level or not is not
apparent (see, for example, the arguments between Haldane and
Chiao {\it et al.} about this problem \cite{Haldane2,Chiao2}),
since the expression for the Chiao-Wu geometric phase can be
derived by using both the classical Maxwell's electromagnetic
theory, differential geometry and quantum
mechanics\cite{Haldane1,Haldane2,Chiao2}. However, the geometric
phase in this paper can be considered only by Berry's adiabatic
quantum theory and Lewis-Riesenfeld invariant theory, namely, the
classical electrodynamics cannot predict this geometric phase.
Although many investigators have taken into account the boundary
condition problem and anomalous refraction in left-handed media by
using the classical Maxwell's theory\cite{Shelby,Veselago}, less
attention is paid to this geometric phase due to helicity
inversion. It is believed that this geometric phase originates at
the quantum level, but survives the correspondence-principle limit
into the classical level. So, We emphasize that it may be
essential to take into consideration this geometric phase in
investigating the anomalous refraction at the LRH interfaces.

It is well known that geometric phases arise only in
time-dependent quantum systems. In the present problem, the
transitions between helicity states on the LRH interfaces, which
is a time-dependent process, results in the geometric phase of
photons. This may be viewed from two aspects: (i) it is apparently
seen in Eq.(\ref{exeq7}) that if the frequency parameter $\omega$
in the Hamiltonian (\ref{exeq1}) vanishes, then $\dot{\varphi}=0$
and the geometric phase (\ref{exeq15}) is therefore vanishing;
(ii) it follows from (\ref{exeq2}) that the frequency coefficient
$\omega \left(t\right)$ of Hamiltonian (\ref{exeq1}) is

\begin{equation}
\omega \left(t\right) =\varsigma \frac{\rm d}{{\rm d}t}p\left(
t\right) =\frac{2c}{nb}\varsigma\sum_{k=1}^{\infty }\left[
1-\left(-\right) ^{k}\right] \cos \left(\frac{k\pi c}{nb}t\right).
 \label{exeq18}
\end{equation}
Since ${\left| \cos \left(\frac{k\pi c}{nb}t\right)\right|\leq
1}$, the frequency coefficient, the transition rates between
helicity states, and the consequent time-dependent phase ($\varphi
_{\sigma }^{\left({\rm g}\right) }\left(t\right)+\varphi _{\sigma
}^{\left({\rm d}\right) }\left(t\right)$) greatly decrease
correspondingly as the periodical optical path $nb$ increases.
Thus we can conclude that the interaction of light fields with
media near the LRH interfaces gives rise to this topological
quantum phase.

In addition to obtaining the expression (\ref{exeq15}) for
geometric phase, we obtain the wavefunction (\ref{exeq16}) of
photons in the LRH- optical fiber by solving the time-dependent
Schr\"{o}dinger equation (\ref{exeq3}) based on the
Lewis-Riesenfeld invariant theory\cite{Lewis} and the
invariant-related unitary transformation
formulation\cite{Gao,Gao2}. We believe that this would enable us
to consider the propagation of light fields inside the optical
fiber in more detail.

In the above treatment, we constructed an effective Hamiltonian
(\ref{exeq1}) to describe the time evolution of helicity states of
photons. It should be noted that the method presented here is only
a phenomenological description of propagation of lightwave in the
LRH-periodical fiber. This phenomenological description is based
on the assumption that the direction of wave vector ${\bf k}$
becomes opposite nearly instantaneously on the LRH interfaces.
This assumption holds true so long as the periodical length $b$ is
much larger than the wavelength of lightwave in the fiber.

\section{Wave propagation in generalized gyrotropic media}

In this section, we investigate the wave propagation in a
generalized gyrotropic medium with the following permittivity and
permeability
\begin{equation}
(\epsilon)_{ik}=\left(\begin{array}{cccc}
a_{\rm e}+d_{\rm e}  & b_{\rm e}-ic_{\rm e} & 0 \\
b_{\rm e}+ic_{\rm e} & a_{\rm e}-d_{\rm e} & 0  \\
 0 &  0 &  \epsilon_{3}
 \end{array}
 \right),         \quad    (\mu)_{ik}=\left(\begin{array}{cccc}
a_{\rm h}+d_{\rm h}  & b_{\rm h}-ic_{\rm h} & 0 \\
b_{\rm h}+ic_{\rm h} & a_{\rm h}-d_{\rm h} & 0  \\
 0 &  0 &  \mu_{3}
 \end{array}
 \right).        \label{geneq1}
\end{equation}
This type of materials is a generalization of the ordinary
gyrotropic medium with
\begin{equation}
(\epsilon)_{ik}=\left(\begin{array}{cccc}
\epsilon_{1}  & i\epsilon_{2} & 0 \\
-i\epsilon_{2} &   \epsilon_{1} & 0  \\
 0 &  0 &  \epsilon_{3}
 \end{array}
 \right),                 \qquad          (\mu)_{ik}=\left(\begin{array}{cccc}
\mu_{1}  & i\mu_{2} & 0 \\
-i\mu_{2} &   \mu_{1} & 0  \\
 0 &  0 &  \mu_{3}
 \end{array}
 \right)              \label{geneq2}
\end{equation}
and can encompass the electromagnetic parameters of
$\epsilon_{ik}$ and $\mu_{ik}$ in (\ref{geneq1}) material
responses experimentally obtained.

Assume that the wave vector ${\bf k}$ of electromagnetic wave
propagating inside this medium is parallel to the the third
component of the Cartesian coordinate system. According to
Maxwellian Equations, one can arrive at
\begin{equation}
-\nabla^{2}{
E}_{1}=\mu_{0}\left(\mu_{21}\frac{\partial^{2}}{\partial
t^{2}}D_{2}-\mu_{22}\frac{\partial^{2}}{\partial
t^{2}}D_{1}\right),   \quad   -\nabla^{2}{
E}_{2}=-\mu_{0}\left(\mu_{11}\frac{\partial^{2}}{\partial
t^{2}}D_{2}-\mu_{12}\frac{\partial^{2}}{\partial
t^{2}}D_{1}\right),
\end{equation}
where the electric displacement vector $D_{1}$ and $D_{2}$ are of
the form
\begin{equation}
D_{1}=\epsilon_{0}\left(\epsilon_{11}E_{1}+\epsilon_{12}E_{2}\right),
\quad
D_{2}=\epsilon_{0}\left(\epsilon_{21}E_{1}+\epsilon_{22}E_{2}\right).
\end{equation}
Due to the transverse nature of planar electromagnetic wave, the
following acquirements are satisfied
\begin{equation}
E_{3}=0,\quad  H_{3}=0, \quad  {\bf k}\cdot{\bf E}=0,  \quad  {\bf
k}\cdot{\bf H}=0,  \quad  {\bf k}\cdot{\bf D}=0,    \quad  {\bf
k}\cdot{\bf B}=0.
\end{equation}
Thus with the help of above equations, it is verified that
\begin{eqnarray}
\nabla^{2}{
E}_{1}=-\epsilon_{0}\mu_{0}\left[\left(\mu_{21}\epsilon_{21}-\mu_{22}\epsilon_{11}\right)\frac{\partial^{2}}{\partial
t^{2}}E_{1}+\left(\mu_{21}\epsilon_{22}-\mu_{22}\epsilon_{12}\right)\frac{\partial^{2}}{\partial
t^{2}}E_{2}\right],
\nonumber \\
\nabla^{2}{
E}_{2}=\epsilon_{0}\mu_{0}\left[\left(\mu_{11}\epsilon_{21}-\mu_{12}\epsilon_{11}\right)\frac{\partial^{2}}{\partial
t^{2}}E_{1}+\left(\mu_{11}\epsilon_{22}-\mu_{12}\epsilon_{12}\right)\frac{\partial^{2}}{\partial
t^{2}}E_{2}\right].
\end{eqnarray}
So,
\begin{eqnarray}
\nabla^{2}\left(\frac{E_{1}\pm
iE_{2}}{\sqrt{2}}\right)&=&\frac{1}{\sqrt{2}}\epsilon_{0}\mu_{0}\left[\left(\mu_{22}\epsilon_{11}-\mu_{21}\epsilon_{21}\right)\pm
i\left(\mu_{11}\epsilon_{21}-\mu_{12}\epsilon_{11}\right)\right]\frac{\partial^{2}}{\partial
t^{2}}E_{1}           \nonumber \\
&+&\frac{1}{\sqrt{2}}\epsilon_{0}\mu_{0}\left[\left(\mu_{22}\epsilon_{12}-\mu_{21}\epsilon_{22}\right)\pm
i\left(\mu_{11}\epsilon_{22}-\mu_{12}\epsilon_{12}\right)\right]\frac{\partial^{2}}{\partial
t^{2}}E_{2}.
\end{eqnarray}
It follows from (\ref{geneq1}) that
\begin{eqnarray}
\left(\mu_{22}\epsilon_{11}-\mu_{21}\epsilon_{21}\right)+
i\left(\mu_{11}\epsilon_{21}-\mu_{12}\epsilon_{11}\right)=A_{+}+iB_{+},
 \nonumber \\
\left(\mu_{22}\epsilon_{12}-\mu_{21}\epsilon_{22}\right)+
i\left(\mu_{11}\epsilon_{22}-\mu_{12}\epsilon_{12}\right)=iA_{+}+B_{+},
\end{eqnarray}
where
\begin{eqnarray}
A_{+}=a_{\rm e}a_{\rm h}-a_{\rm e}c_{\rm h}-d_{\rm e}d_{\rm
h}-id_{\rm e}b_{\rm h}-b_{\rm e}b_{\rm h}+ib_{\rm e}d_{\rm
h}+c_{\rm e}c_{\rm h}-c_{\rm e}a_{\rm h},
 \nonumber \\
 B_{+}=-a_{\rm e}b_{\rm h}-id_{\rm e}a_{\rm h}+id_{\rm e}c_{\rm
 h}-b_{\rm e}c_{\rm h}+b_{\rm e}a_{\rm h}-c_{\rm e}b_{\rm h}+ic_{\rm e}d_{\rm
 h}+ia_{\rm e}d_{\rm h}.
\end{eqnarray}
In the same fashion, one can arrive at
\begin{eqnarray}
\left(\mu_{22}\epsilon_{11}-\mu_{21}\epsilon_{21}\right)-
i\left(\mu_{11}\epsilon_{21}-\mu_{12}\epsilon_{11}\right)=A_{-}+iB_{-},
 \nonumber \\
 \left(\mu_{22}\epsilon_{12}-\mu_{21}\epsilon_{22}\right)-
i\left(\mu_{11}\epsilon_{22}-\mu_{12}\epsilon_{12}\right)=-iA_{-}-B_{-},
\end{eqnarray}
where
\begin{eqnarray}
A_{-}=a_{\rm e}a_{\rm h}+a_{\rm e}c_{\rm h}-d_{\rm e}d_{\rm
h}+id_{\rm e}b_{\rm h}-b_{\rm e}b_{\rm h}-ib_{\rm e}d_{\rm
h}+c_{\rm e}c_{\rm h}+c_{\rm e}a_{\rm h},
 \nonumber \\
 B_{-}=a_{\rm e}b_{\rm h}-id_{\rm e}a_{\rm h}-id_{\rm e}c_{\rm
 h}-b_{\rm e}c_{\rm h}-b_{\rm e}a_{\rm h}-c_{\rm e}b_{\rm h}-ic_{\rm e}d_{\rm
 h}+ia_{\rm e}d_{\rm h}.
\end{eqnarray}
Hence, the wave equations of left- and right- handed circularly
polarized light\footnote{For simplicity, without loss of
generality, it is assumed that the two mutually perpendicular real
unit polarization vectors ${\vec\varepsilon}(k,1)$ and
${\vec\varepsilon}(k,2)$ can be taken to be as follows:
$\varepsilon_{1}(k,1)=\varepsilon_{2}(k,2)=1$,
$\varepsilon_{1}(k,2)=\varepsilon_{2}(k,1)=0$ and
$\varepsilon_{3}(k,1)=\varepsilon_{3}(k,2)=0$. Thus by the aid of
the formula ${\bf E}=-\frac{\partial {\bf A}}{\partial t}$ for the
electric field strength, in the second-quantization framework one
can arrive at\cite{Shen}
\begin{eqnarray}
E_{R}=\frac{E_{1}+ iE_{2}}{\sqrt{2}}=i\int{\rm d}^{3}{\bf
k}\sqrt{\frac{\omega}{2(2\pi)^{3}}}[a_{L}(k)\exp(-ik\cdot
x)-a_{R}^{\dagger}(k)\exp(ik\cdot x)],   \nonumber  \\
E_{L}=\frac{E_{1}-iE_{2}}{\sqrt{2}}=i\int{\rm d}^{3}{\bf
k}\sqrt{\frac{\omega}{2(2\pi)^{3}}}[a_{R}(k)\exp(-ik\cdot
x)-a_{L}^{\dagger}(k)\exp(ik\cdot x)].
\end{eqnarray} } is written
\begin{eqnarray}
 \nabla^{2}\left(\frac{E_{1}-
iE_{2}}{\sqrt{2}}\right)=\epsilon_{0}\mu_{0}A_{-}\frac{\partial^{2}}{\partial
t^{2}}\left(\frac{E_{1}-
iE_{2}}{\sqrt{2}}\right)+i\epsilon_{0}\mu_{0}B_{-}\frac{\partial^{2}}{\partial
t^{2}}\left(\frac{E_{1}+iE_{2}}{\sqrt{2}}\right),
 \nonumber \\
 \nabla^{2}\left(\frac{E_{1}+
iE_{2}}{\sqrt{2}}\right)=\epsilon_{0}\mu_{0}A_{+}\frac{\partial^{2}}{\partial
t^{2}}\left(\frac{E_{1}+
iE_{2}}{\sqrt{2}}\right)+i\epsilon_{0}\mu_{0}B_{+}\frac{\partial^{2}}{\partial
t^{2}}\left(\frac{E_{1}-iE_{2}}{\sqrt{2}}\right). \label{geneq13}
\end{eqnarray}

It should be pointed out that the interaction between left- and
right-handed polarized light arises in Eq.(\ref{geneq13}), which
is analogous to the Josephson's effect and may therefore be of
physical interest.

The coupling of left-handed polarized light to the right-handed
one can be characterized by the frequency shift $\Omega_{L}$ and
$\Omega_{R}$, namely, the amplitudes of left- and right- handed
circularly polarized light propagating along the
$\hat{z}$-direction are written as follows
\begin{equation}
E_{L}\sim\exp\left\{\frac{1}{i}\left[\left(\omega+\Omega_{L}\right)t-\sqrt{A_{-}}\frac{\omega}{c}
z+\phi_{L}\right]\right\},  \quad
E_{R}\sim\exp\left\{\frac{1}{i}\left[\left(\omega+\Omega_{R}\right)t-\sqrt{A_{+}}\frac{\omega}{c}
z+\phi_{R}\right]\right\}.     \label{geneq15}
\end{equation}
Insertion of Eq.(\ref{geneq15}) into Eq.(\ref{geneq13}) yields
 \begin{equation}
A_{-}\omega^{2}E_{L}=A_{-}\left(\omega+\Omega_{L}\right)^{2}E_{L}+iB_{-}\left(\omega+\Omega_{R}\right)^{2}E_{R},
\quad
A_{+}\omega^{2}E_{R}=A_{+}\left(\omega+\Omega_{R}\right)^{2}E_{R}+iB_{+}\left(\omega+\Omega_{L}\right)^{2}E_{L},
     \label{geneq16}
\end{equation}
and consequently
 \begin{equation}
A_{+}A_{-}\left(2\omega\Omega_{R}+\Omega_{R}^{2}\right)\left(2\omega\Omega_{L}+\Omega_{L}^{2}\right)+B_{+}B_{-}\left(\omega+\Omega_{L}\right)^{2}\left(\omega+\Omega_{R}\right)^{2}=0,
     \label{geneq17}
\end{equation}
which is a restricted condition regarding the frequency-shift
relation between left- and right- handed circularly polarized
light.

Note that in the case of conventional gyrotropic media
characteristic of such permittivity and permeability tensors
(\ref{geneq2}), where $d_{\rm e}=b_{\rm e}=d_{\rm h}=b_{\rm h}=0$,
$a_{\rm e }=\epsilon_{1}$, $c_{\rm e }=-\epsilon_{2}$, $a_{\rm h
}=\mu_{1}$, $c_{\rm h}=-\mu_{2}$, the optical refractive index
squared is of the form
\begin{equation}
A_{\pm}=\left(\epsilon_{1}\pm\epsilon_{2}\right)\left(\mu_{1}\pm\mu_{2}\right),
\end{equation}
and the coupling coefficients $B_{\pm}$ are vanishing.
\\ \\

\textbf{Acknowledgements}  This project was supported partially by
the National Natural Science Foundation of China under Project No.
$90101024$ and $60378037$, and also by Zhejiang Provincial Natural
Science Foundation of China under Key Project No. ZD0002. The
author was grateful to Prof. Sai Ling He for the discussion on the
wave propagation in left-handed media.
\\ \\

\textbf{Appendix A.}

In this Appendix, we will briefly consider the permittivity of the
metal thin wire array medium. For a nearly free electron in the
structures of {\it array of long metallic wires}
(ALMWs)\cite{Pendry2}, the equation of motion is $\frac{\partial
{\bf v} }{\partial t}=\frac{e}{m}{\bf E}$, and the electric
current density ${\bf j}$ therefore satisfies the equation
$\frac{\partial {\bf j} }{\partial t}=\frac{ne^{2}}{m}{\bf E}$,
where ${\bf j}=ne{\bf v}$ with $n$ being the electron number
density in ALMWs media. If the differential operator
$\frac{\partial}{\partial t}$ can be replaced with $i\omega$,
where $\omega$ is the frequency of incident electromagnetic wave,
then we can obtain
\begin{equation}
{\bf j}=\frac{ne^{2}}{i\omega m}{\bf E}.       \label{appendix a1}
\end{equation}
Thus with the help of ${\bf j}=\frac{\partial {\bf P} }{\partial
t}=i\omega {\bf P}$, one can arrive at
\begin{equation}
{\bf P}=-\frac{ne^{2}}{\omega^{2}m}{\bf E}.     \label{appendix
a2}
\end{equation}
According to the definition of $\epsilon$, {\it i.e.}, ${\bf
P}=(\epsilon-1)\epsilon_{0}{\bf E}$, it follows that
\begin{equation}
\epsilon=1-\frac{\omega_{\rm p}^{2}}{\omega^{2}}   \quad  {\rm
with} \quad      \omega_{\rm p}^{2}=\frac{ne^{2}}{m\epsilon_{0}}.
\end{equation}

\textbf{Appendix B.}

SRR essentially behaves as a LC circuit, the equation of which may
be as follows
\begin{equation}
L\frac{{\rm d}^{2}}{{\rm d}t^{2}}q+R\frac{\rm d}{{\rm
d}t}q+\frac{q}{C}=-\frac{\rm d}{{\rm d}t}B\pi r^{2}_{0}
\end{equation}
according to the electromagnetic induction law. By replacing the
derivative operator $\frac{\rm d}{{\rm d}}$ with $i\omega$, and
using the electric current $I=\frac{\rm d}{{\rm d}t}q=i\omega q$
(hence $q=\frac{I}{i\omega}$), one can arrive at
\begin{equation}
I=-\frac{\frac{B\pi
r^{2}_{0}}{L}\omega^{2}}{\omega^{2}-i\frac{R}{L}\omega-\frac{1}{LC}}.
\end{equation}
Thus we have the magnetization (total moments per unit volume)
\begin{equation}
M=NI\pi r^{2}_{0}=-\frac{\frac{NB\left(\pi
r^{2}_{0}\right)^{2}}{L}\omega^{2}}{\omega^{2}-i\frac{R}{L}\omega-\frac{1}{LC}},
\end{equation}
which is equal to $\frac{1}{\mu_{0}}\frac{\mu-1}{\mu}B$. So, we
may obtain
\begin{eqnarray}
\mu
&=&\frac{\omega^{2}-i\frac{R}{L}\omega-\frac{1}{LC}}{\omega^{2}\left(1+\frac{N\mu_{0}\left(\pi
r^{2}_{0}\right)^{2}}{L}\right)-i\frac{R}{L}\omega-\frac{1}{LC}}                 \nonumber \\
&=& 1-\frac{{\frac{\frac{N\mu_{0}\left(\pi
r^{2}_{0}\right)^{2}}{L}}{1+\frac{N\mu_{0}\left(\pi
r^{2}_{0}\right)^{2}}{L}}\omega^{2}}}{{\omega^{2}-i\omega
\frac{\frac{R}{L}}{1+\frac{N\mu_{0}\left(\pi
r^{2}_{0}\right)^{2}}{L}}-\frac{1}{LC\left(1+\frac{N\mu_{0}\left(\pi
r^{2}_{0}\right)^{2}}{L}\right)}}},
\end{eqnarray}
which is of the form
\begin{equation}
\mu=1-\frac{F\omega^{2}}{\omega^{2}+i\omega \Gamma-\omega^{2}_{0}
}.
\end{equation}


\end{document}